\documentclass[aps, pre, twocolumn, a4paper, groupedaddress, floatfix]{revtex4-1}
\usepackage{graphicx}
\usepackage{amsmath}
\usepackage{placeins}
\usepackage{color}
\usepackage{soul} 
\usepackage{hyperref}
\usepackage{textcomp} 
\usepackage{url}
\usepackage{xurl}
\usepackage[T1]{fontenc}

\begin{document}

\title{Spatial Structures of Wind Farms: Correlation Analysis of the Generated Electrical Power}

\author{Edgar Jungblut\footnote{now at Forschungszentrum Jülich GmbH,\\Email: edgar.jungblut@uni-due.de}, Henrik M. Bette, Thomas Guhr}
\affiliation{Fakultät für Physik, Universität Duisburg-Essen, Duisburg, Germany}

\begin{abstract}

We investigate the interaction of many wind turbines in a wind farm with a focus on their electrical power production. The operational data of two offshore wind farms with a ten minute and a ten second time resolution, respectively, are analyzed.
For the correlations of the active power between turbines over the entire wind farms, we find a dominant collective behavior. We manage to subtract the collective behavior and find a significant dependence of the correlation structure on the spatial structure of the wind farms.
We further show a connection between the observed correlation structures and the prevailing wind direction. We attribute the differences between the two wind farms to the differences in the turbine spacing within the two wind farms.

\end{abstract}
 
\maketitle

\section{Introduction}

The field of wind energy research ranges from physical and technical issues to sociological and economic ones \cite{wes-1-1-2016}. Regarding the physical and technical aspects, the aims are a better understanding of atmospheric flows and the interactions of wind turbines within a wind farm, an optimal design (including materials) of wind turbines, an optimized control strategy for wind turbines within a wind farm, and a strategy for the integration of wind energy into the power grid \cite{Veerseaau2027}. With our research we want to contribute to statistical analysis of wind turbine data of whole wind farms, utilizing Supervisory Control and Data Acquisition (SCADA) data from already installed SCADA systems as a simple and cost-effective data source.

Wake effects and turbulences generated by the wind turbines are important for the layout of wind farms. Many sophisticated efforts are undertaken to describe the flows and incorporate them into wind farm design \cite{PorteAgel2020, Parada2017}. Detailed physics based Computational Fluid Dynamics (CFD) models are needed to accurately describe the air flows within the wind farms. Models have been developed to calculate wake effects and turbulence generated by the wind turbines themselves \cite{jensen1983note, frandsen2006analytical}. However, these models are often computationally costly and need detailed information on atmospheric flows as well as wind turbine and wind farm geometries. Therefore, we focus on a statistical analysis that is input-free beyond the measured data itself. Previously, data on wind conditions within the wind farm were often only based on single measured values from separate measuring masts and model calculations \cite{barthelmie2007modelling, gao2019investigation}. Due to the availability of SCADA data, it is now possible to determine electrical power production, wind speed and orientation individually for each wind turbine \cite{castellani2017analyzing, mckay2013wake, barthelmie2007modellingpower}. Especially for large wind farms this might be a significant improvement for investigating wake effects, because the wind conditions may vary at different locations and the measured value of a single separate measuring mast may not be representative for the wind farm as a whole \cite{mechali2006wake}.

The interactions between wind turbines are relevant not only for the design but also for control and monitoring of wind farms. To ensure optimal operation, current operating data are evaluated and appropriate adjustments are made. For this, it is essential to have fast analysis procedures that can process real-time data. Computing time can be shortened, for example, by reducing the complexity of the data. This can be done by aggregating several wind turbines, based on wind speed, wind direction and wind farm layout, taking into account wake effects \cite{ali2013wind, castro1999aggregated, marinopoulos2011investigating}. Another approach is to use correlation analyses by determining the correlation structure for the wind farm at one time and comparing it to the structure at other times. This was used in the examination of correlations for return time series in stock markets. Recurring, consistent structures were identified, that can be understood as states \cite{munnix2012identifying, stepanov2015stability, rinn2015dynamics}. The dynamics can then be represented by the state time series. This significantly reduces the complexity of the system and allows for rapid analysis. Likewise, states in the correlation structure were found in analyses of highway traffic flow \cite{Wang_2020, Wang_2021, GARTZKE2022127367}.

Our goal is to investigate the correlation structure of wind farms. We apply statistical methods to see which information can be gained on the interactions in the wind farm from the SCADA data. We do this without additional input or modeling assumptions. Our focus is on the electrical power production of individual wind turbines and the collective behavior within the wind farm. First, we calculate correlation matrices for the active power. Second, we apply clustering methods and find different states in the correlation structure.

Our study is organized as follows: In Sec.~\ref{Data} we describe the data used. In Sec.~\ref{Method} we introduce our methods for the calculation of the correlation matrices and the further analysis. In Sec.~\ref{Data analysis} we show our results for the correlation analysis for the active power of the wind turbines for each wind farm. In Sec.~\ref{Conclusion} we summarize our results and provide an outlook.

\section{Data}
\label{Data}

We analyze SCADA operational data of the German offshore wind farm \textsc{Riffgat} and the British offshore wind farm \textsc{Thanet}. We introduce the two wind farms and datasets briefly in Sec.~\ref{Des Farms}, in Sec.~\ref{Data Pre} we describe our data pre-processesing.

\begin{figure}[htb]
	\begin{center}
		\includegraphics[width=0.9\columnwidth]{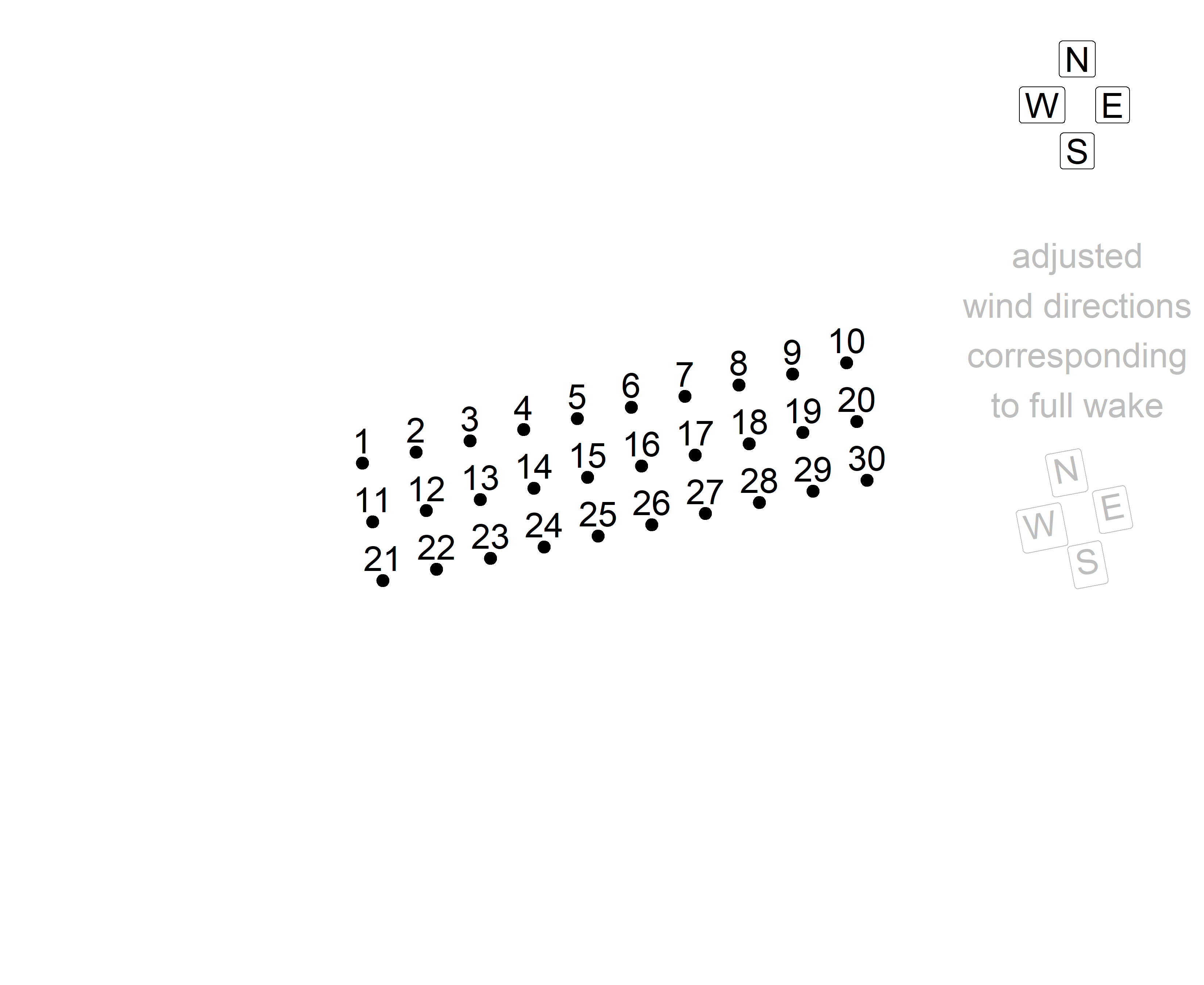}
		\includegraphics[width=0.9\columnwidth]{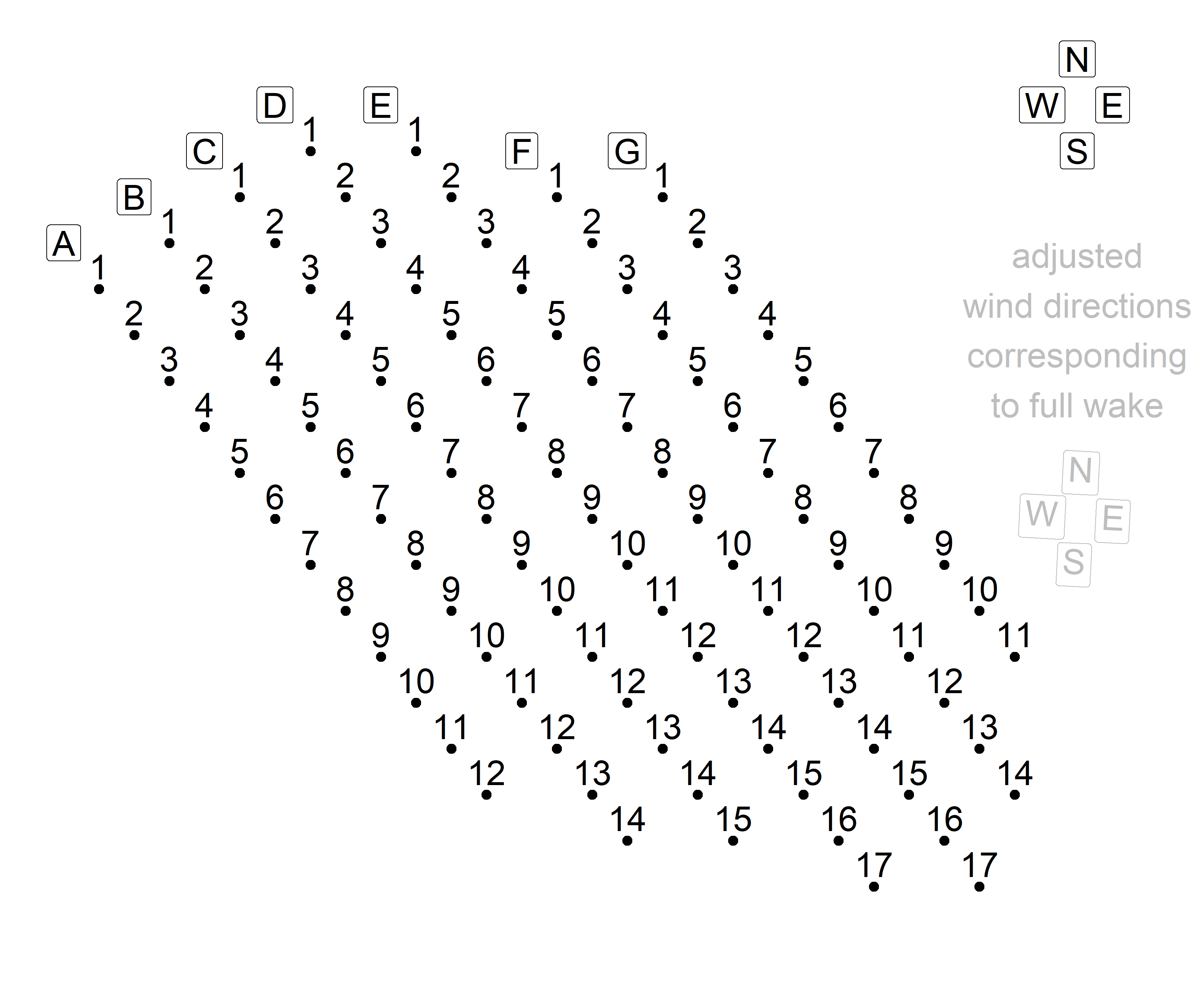}
		\caption{Top: Schematic layout of the wind farm \textsc{Riffgat}. Bottom: Schematic layout of the wind farm \textsc{Thanet}. The adjusted wind directions are used in the later analysis to align the wind farms with the eight wind directions (N, NE, E, $\dots$). For the wind farm \textsc{Riffgat} wind direction north corresponds to compass direction 349°N. For the wind farm \textsc{Thanet} wind direction north corresponds to compass direction 3°N.}
		\label{fig:Layout_Parks}
	\end{center}
\end{figure}

\subsection{Description of the wind farms and data}
\label{Des Farms}

The wind farm \textsc{Riffgat} was the first commercial North Sea wind farm when it was built in August 2013 \cite{ewe_riffgat}. It is located about $15\,$km north-west of the island of Borkum \cite{tennet}. The wind farm consists of $30$ \textit{\textsc{Siemens} SWT-3.6-120} wind turbines with a rotor diameter of $120\,$m and a rated electrical power of $3600\,$kW each. Thus, the wind farm has a total rated electrical power of $108\,$MW.
The wind farm layout is shown schematically in Fig.~\ref{fig:Layout_Parks}, top. It consists of three rows running from west to east. Each row consists of ten wind turbines. The wind turbine spacing within the rows is approximately $550\,$m ($4.6$ rotor diameters). The spacing between rows is approximately $600\,$m ($5$ rotor diameters).
The analyzed operational data span a twelve month period from March 01, 2014 (one month after wind farm launch) to February 28, 2015.
The data includes measurements at the wind turbines (\textit{e.g.} wind speed, nacelle direction, pitch angle, generator speed), grid characteristics (\textit{e.g.} active and reactive power of the wind turbine, grid voltage, current) as well as status values. The operational data are measured at high frequency, but recorded as the mean values, the standard deviations and minimum and maximum values of ten minute intervals.

The \textsc{Thanet} wind farm was built in 2010 and was one of the largest offshore wind farms in the world \cite{thanet_bbc} at the time. It is located about $12\,$km off the coast of Kent in the south-east of the United Kingdom \cite{thanet_vattenfall}. The wind farm consists of $100$ \textit{\textsc{Vestas} V90-3.0 MW} wind turbines with a rotor diameter of $90\,$m and a rated electrical power of $3000\,$kW each. In total, therefore, the wind farm has a rated electrical power of $300\,$MW.
The wind farm layout is shown schematically in Fig.~\ref{fig:Layout_Parks}, bottom. It consists of seven rows running from north-west to south-east. The wind turbine number per row varies between $11$ and $17$. The wind turbine spacing within the rows is approxiamtely $470\,$m ($5.2$ rotor diameters). The spacing between rows is approximately $720\,$m ($8$ rotor diameters). However, spacing may vary slightly for individual wind turbines.
The various measured variables have different time resolutions. Measured variables such as wind speed and active power are measured in approximately $5\,$s increments. For synchronization, we average them to $10\,$s increments for our studies. Temperature measurements, \textit{e.g.}, of the generator, are made in $10\,$min increments. Measured variables such as the nacelle direction are transmitted in each case as changes occur.
The \textsc{Thanet} dataset is very large due to its high temporal resolution. This has a significant impact on the handling of the dataset and the required computation times. Therefore, the analysis in this paper is limited to the operational data for February 2017.

For both datasets, the data do not have entries for all observables at all times. The times without measurement vary per wind turbine and observable. Thus, there may be a wind speed measurement for a wind turbine at one point in time, but not its associated electrical power production. Missing points in the data are replaced by \textit{NA} (Not Available).

\subsection{Data pre-processing}
\label{Data Pre}

The data analysis requires preparation. We sort out and correct erroneous values as well as implausible data. We use different preparation steps for the low-resolution \textsc{Riffgat} dataset and the high-resolution \textsc{Thanet} dataset.

The preparation of the \textsc{Riffgat} data is based on Refs.~\cite{schlechtingen2011comparative, schlechtingen2013using, godwin2013classification, zheng2014raw} and is done according to the following steps:
First, we discard data (replace it by \textit{NA}) for which the exact same measurement value was recorded for two consecutive measurements. Two exactly equal measured values are highly unlikely for a specification with such a high resolution (five decimal places). Therefore, we interpret this as a measurement error.
Second, we discard all data where the standard deviation for the respective measurement interval has a value of zero.
Last, we discard wind speeds above $30\,$m/s because they are well above the operating range of the wind turbines and are likely to represent measurement errors.

For the \textsc{Thanet} dataset, data preparation does not involve any steps other than transforming it into the format used for our analysis, because of its high temporal resolution.

For the calculation of the correlation matrices, missing points in the data are a major problem. Hence, they have to be handled in an appropriate manner.
For the \textsc{Riffgat} dataset all missing points in the data are replaced by the value $0\,$kW. We do this because the overall number of missing values is low ($2.25\,\%$) and the impact of this filling is therefore minimal.
For the \textsc{Thanet} dataset, the handling of missing points in the data is more complicated. First of all, the number of missing points is much higher ($13.03\,\%$). Furthermore, missing points in the data are not evenly (randomly) distributed across all wind turbines and the entire time period. We observe two main patterns. First, there are long periods with missing data points for individual wind turbines. This is due to failures of these individual wind turbines and is referred to below as \textit{failed wind turbines}. Second, missing data points occur more frequently for multiple wind turbines at the same time when electrical power production and wind speed are low. This is due to shutdowns of the wind turbines due to insufficient wind speeds and is referred to as \textit{wind turbines shut down} in the following.

Failures and shutdowns differ in particular in the fact that the former are individual effects and the latter are collective effects. Therefore, to characterize failures and shutdowns, we first consider the densities of missing data points in the time series of individual wind turbines, $\textit{NA}_{\textrm{dens}}$, and the number of simultaneous missing data points in the time series of all wind turbines, $\textit{NA}_{\textrm{farm}}$.
To ensure proper characterization of failures and shutdowns, the following measures are taken:
First, the density of missing data points in the time series of individual wind turbines, $\textit{NA}_{\textrm{dens}}$, is determined over a large, sliding time window of twelve hours. This gives greater weight to the long duration of failures and less weight to short periods of shutdowns with many missing data points.
Second, to remove collective increases in the density of missing data points, the deviation of the density of missing data points from the mean density of missing data points for all wind turbines is determined and then averaged for each wind turbine over a sliding time window of twelve hours
\begin{equation}
	\textit{NA}_{\textrm{dens,dev}} = \left< \textit{NA}_{\textrm{dens}} - \frac{1}{N_{\textrm{WT}}} \sum_{\textrm{WT}} \textit{NA}_{\textrm{dens}} \right>_{12\,\textrm{h}} \quad ,
\end{equation}
with the number of wind turbines $N_{\textrm{WT}}$.
For the failed wind turbines, we observe another effect in addition to the missing data points. Frequently, individual measured values occur between the missing data points, which deviate strongly from the measured values of all other wind turbines. In our opinion, these are erroneous measured values due to the failure of the respective wind turbine, which should also be replaced.
To determine such values, the deviation, $\Psi$, of the individual wind turbine's electrical power production from the mean electrical power production of all wind turbines is averaged over a ten minute interval for each wind turbine
\begin{equation}
	\Psi_{10} = \left< \Psi \right>_{10\,\textrm{min}} \quad .
\end{equation}
If a wind turbine produces significantly less electrical power on average than all other wind turbines over a time interval of $10$ minutes, its measured values are considered erroneous.
The criteria and thresholds chosen to characterize failures and shutdowns are explicitly summarized as:
\begin{enumerate}
	\item A data point is interpreted as a failure,\\
	if $\textit{NA}_{\textrm{dens}} > 0.6$ and $\textit{NA}_{\textrm{dens,dev}} > 0.1$ or\\
	if $\Psi_{10} < -1000\,$kW.
	\item A data point is interpreted as a shutdown,\\
	if $u<4\,$m/s and $\textit{NA}_{\textrm{farm}}>20$ and\\
	it is not characterized as a failure.
\end{enumerate}
A missing data point in the wind speed measurement is considered as a wind speed $u<4\,$m/s for the second condition.
The thresholds for failure detection were chosen based on how well they were able to identify short- and long-term failures for some specific days. For the shutdowns, the threshold value for wind speed stems from the operations handbook of the wind turbines. The number of wind turbines is chosen as a fifth of the wind farm to reflect shutdowns as a collective phenomenon, the data of the turbines being shutdown the earliest are kept until at least one fifth of the wind farm is shut down.
With this classification, of the $13.03\,\%$ missing data points, $3.85\,\%$ are due to shutdowns and $6.65\,\%$ are due to failures. The remaining $2.53\,\%$ cannot be assigned to either category.

Having identified failures and shutdowns, the question is how to handle them. Failures and shutdowns are fundamentally different and therefore their handling should be different as well.
Shutdowns occur at low wind speeds and low electrical power productions. Consequently, the simplest option is to fill the shutdowns with values of $0\,$kW. This is consistent with the logic that no electrical power is produced during a shutdown. We neglect at this point that the wind turbines themselves need a certain electrical power for their operation and the missing data points would have to be filled with negative values accordingly.
A second possibility is to fill the missing data points with the last existing value in each case. This is especially suitable for short series of missing data points.
Both methods lead to the same qualitative results, but the correlations are stronger in the second case. Hence, for our analysis we use the second method.
Failures occur for individual wind turbines. They do not depend on wind speed or electrical power production. Filling the failures with a fixed value such as $0\,$kW is therefore not useful. Using the last value before the failure is also unsuitable as an option, since on the one hand it is not ensured that this is not already faulty and on the other hand failures usually last for a longer period of time and this value can thus become as unsuitable as a fixed value of $0\,$kW. Thus, the simplest way to fill up the failures with a variable value is to use the respective current average electrical power production of the remaining wind turbines. We will do this in the following. Please note that the mean value of the wind turbines is calculated after the missing data points due to shutdowns have been filled using the aforementioned method.

\begin{figure*}[htb]
	\begin{center}
		\includegraphics[width=0.32\textwidth]{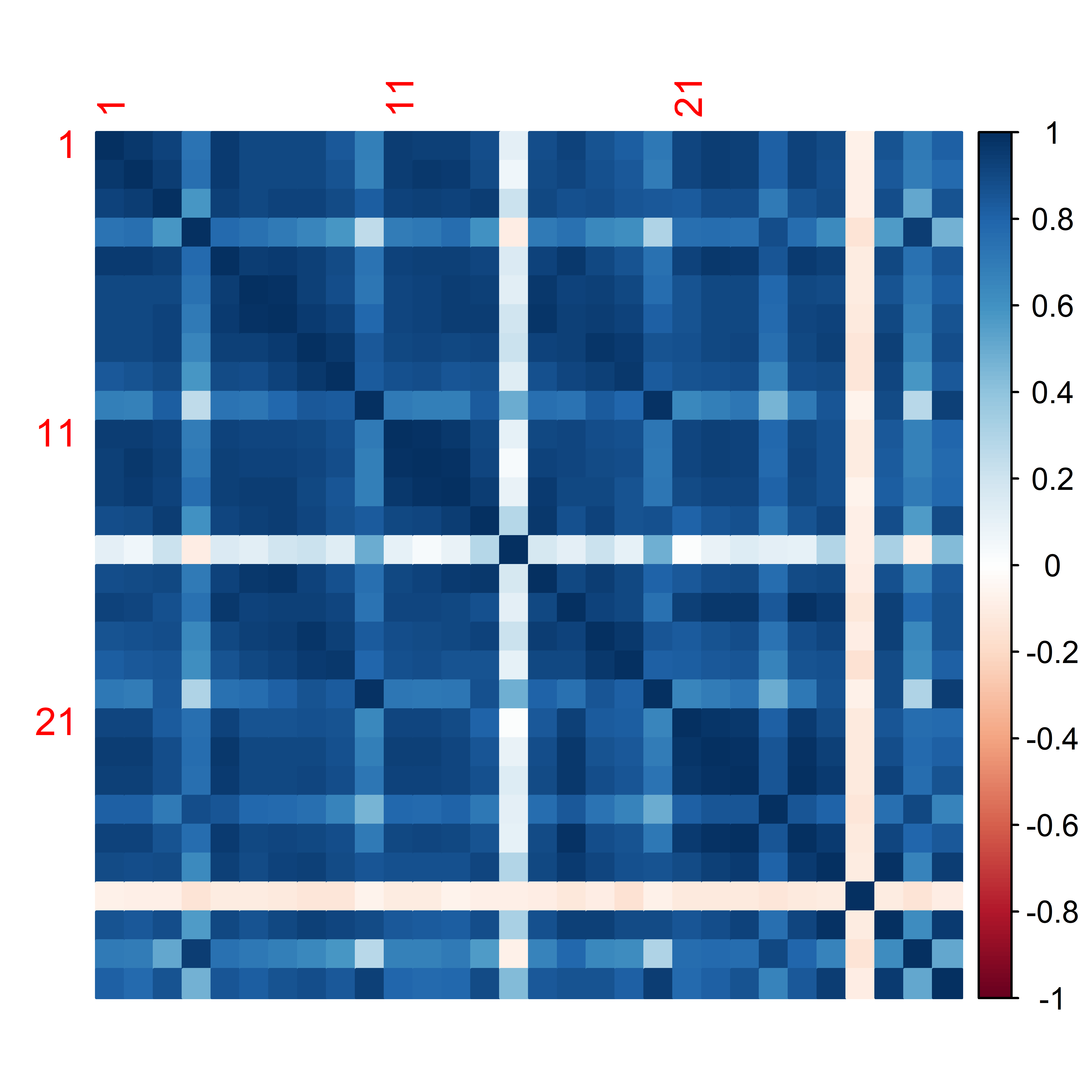}
		\includegraphics[width=0.32\textwidth]{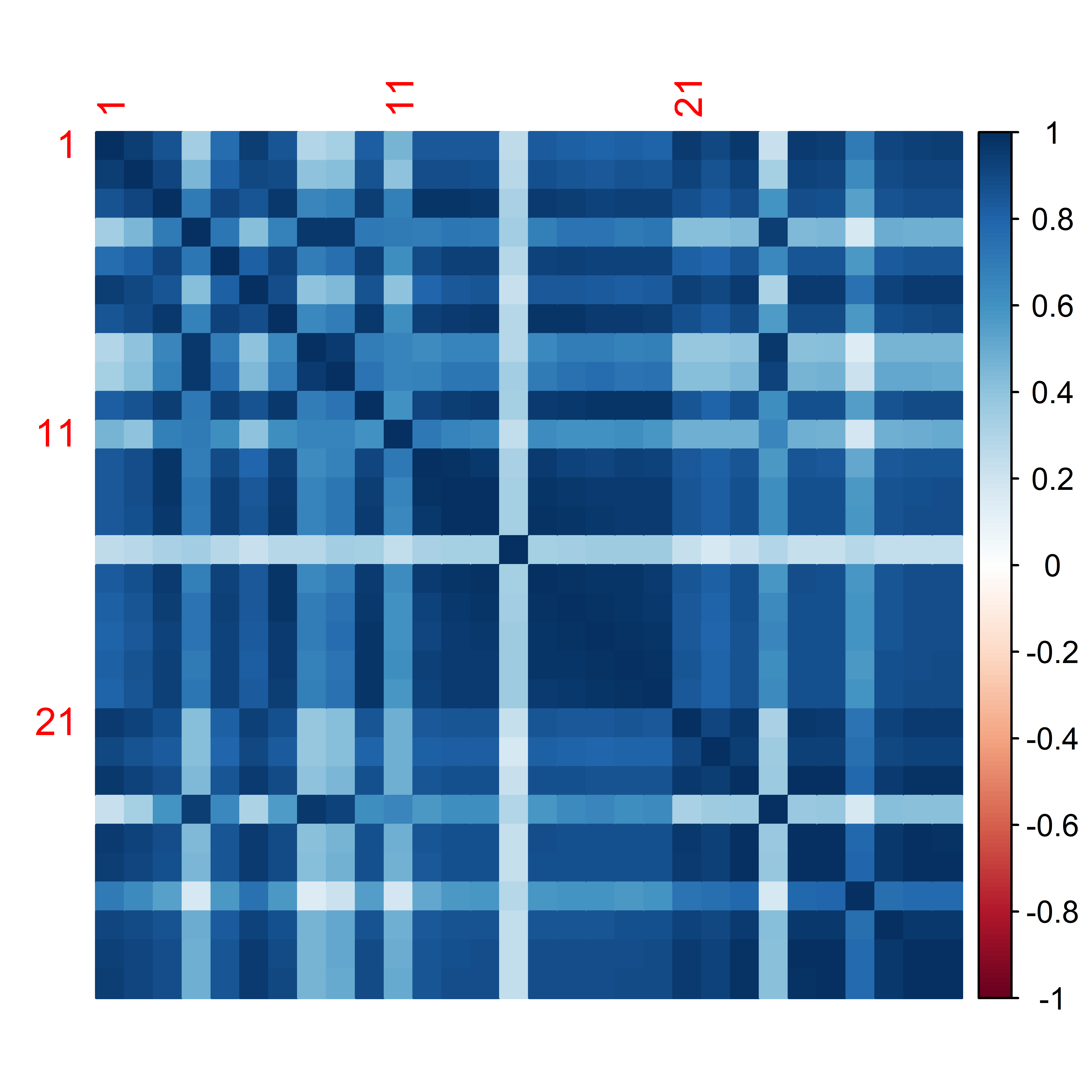}
		\includegraphics[width=0.32\textwidth]{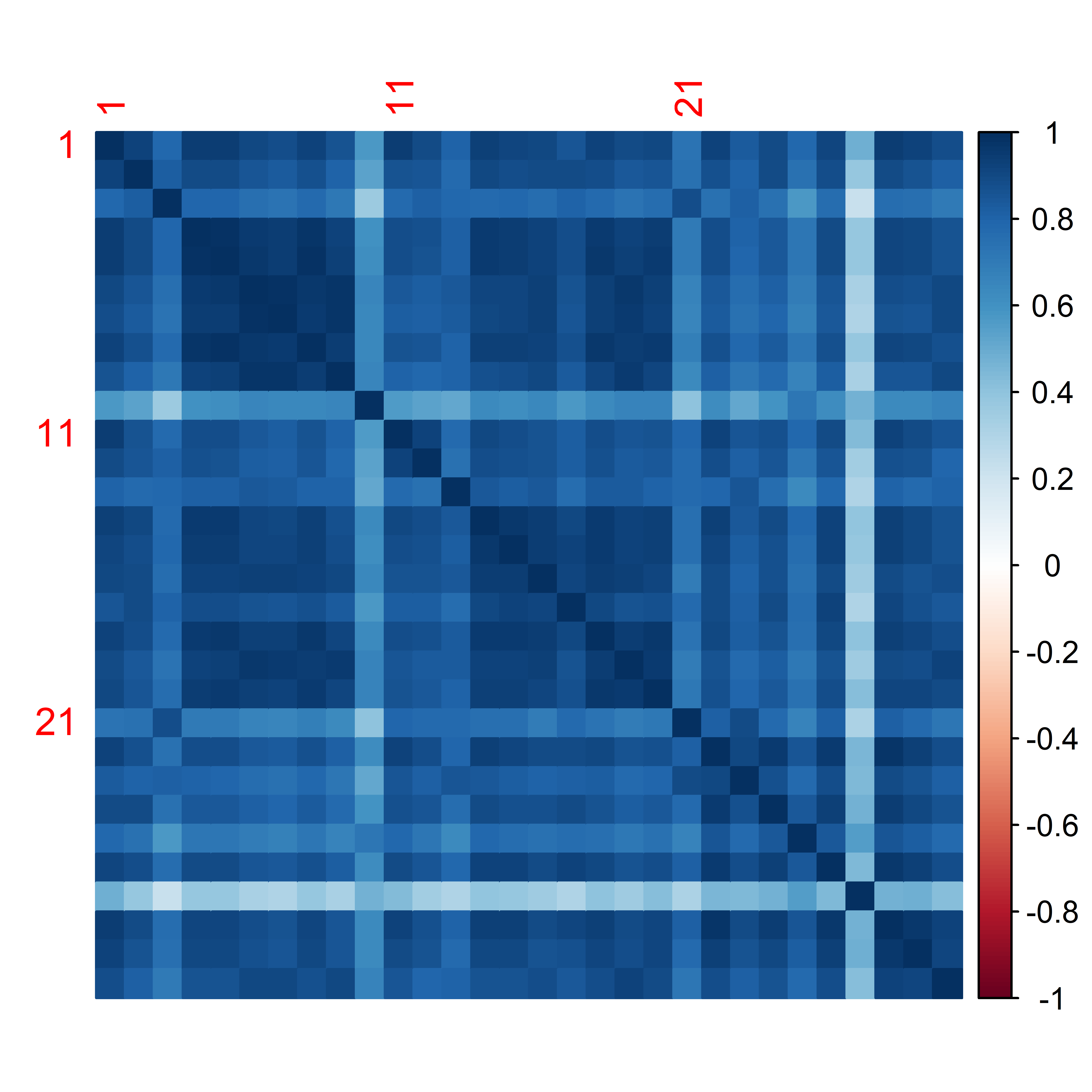}
		\caption{Correlation matrices for the active power over half a day (left), a day (middle) and a week (right) for the wind farm \textsc{Riffgat}.}
		\label{fig:Corr_ex_Riffgat}
	\end{center}
\end{figure*}

\section{Correlation analysis}
\label{Method}

We analyze the correlation of the active power for all wind turbines of a wind farm. To calculate the respective correlation matrices we use the following method:
Let $x_i(t), ~t=1,\dots,T$, be the time series of $T$ measurement of an observable $x$ for a wind turbine $i, ~i=1,\dots,N$, where N is the number of turbines.
Let further $X$ be an $N\times T$ matrix whose rows contain the respective $x_i(t)$.
Now let
\begin{equation}
	m_i(t) = x_i(t) - \langle x_i\rangle_T ,
\end{equation}
with
\begin{equation}
	\langle x_i\rangle_T = \frac{1}{T} \sum_{t = 1}^{T} x_i(t)
\end{equation}
be the time series normalized to zero mean and $M$ be the $N\times T$ matrix of these time series analogous to $X$.
Next, we calculate the $N\times N$ covariance matrix
\begin{equation}
	\label{cov_matrix}
	\Sigma = \frac{1}{T} M M^\dagger .
\end{equation}
We normalize it with the standard deviations $\sigma_i$ of the time series $x_i(t)$ to obtain the $N\times N$ correlation matrix $C$
\begin{equation}
	\label{corr_matrix}
	C = \sigma^{-1} \Sigma \sigma^{-1} \quad ,
\end{equation}
with
\begin{equation}
	\sigma = \textrm{diag}(\sigma_1, \dots , \sigma_N) .
\end{equation}
In the eigenvectors to a given eigenvalue, entries with the same sign and comparable numerical value, indicate coherent, i. e. collective, behavior of the corresponding time series. The largest eigenvalue of $C$ and its eigenvalue measure the collectivity of the system as a whole.  Further large eigenvalues usually represent behavior of subgroups of variables \cite{Guhr2003}. In our analysis we find that the correlation matrices are dominated by collective effects, \textit{i.e.}, we find a large first eigenvalue with an eigenvector that consists of almost equal values in all entries. This is consistent with other complex systems \cite{munnix2012identifying, stepanov2015stability, rinn2015dynamics, Wang_2020, Wang_2021, GARTZKE2022127367, heckens2020uncovering}. To subtract the collective part, we use a singular value decomposition, which always exists \cite{Liesen2015} and has been used in a similar context in Ref.~\cite{heckens2020uncovering}
\begin{equation}
	\label{sing_decomposition}
	M = U S V^\dagger
	~ \mathrm{with} ~ S= \begin{bmatrix}
		s_{1} & & 0 & 0 & & 0 \\
		& \ddots & & & \ddots & \\
		0 & & s_{n} & 0 & & 0
	\end{bmatrix}.
\end{equation}
Here $U$ is an orthogonal $N\times N$ matrix with the eigenvectors of $MM^\dagger$ as columns and $V$ is an orthogonal $T\times T$ matrix with the eigenvectors of $M^\dagger M$ as columns. The matrix $S$ is a $N\times T$ matrix with the singular values $s$ on its main diagonal and all other entries being zero. The number of nonzero singular values corresponds to the rank of $M$. We notice that the $N\times N$ matrix $S S^\dagger$ and the $T \times T$ matrix $S^\dagger S$ contain the (non-negative) eigenvalues of $M M^\dagger$ and $M^\dagger M$, respectively.
By setting one or more singular values to zero in Eq.~(\ref{sing_decomposition}), we generate reduced time series.
For these we calculate the associated reduced-rank correlation matrices with Eq.~(\ref{cov_matrix}) and Eq.~(\ref{corr_matrix}). This procedure is equivalent to the construction of reduced-rank covariance and correlation matrices in Ref. \cite{heckens2020uncovering}. 

\section{Results}
\label{Data analysis}

\begin{figure}[htb]
	\begin{center}
		\includegraphics[width=0.9\columnwidth]{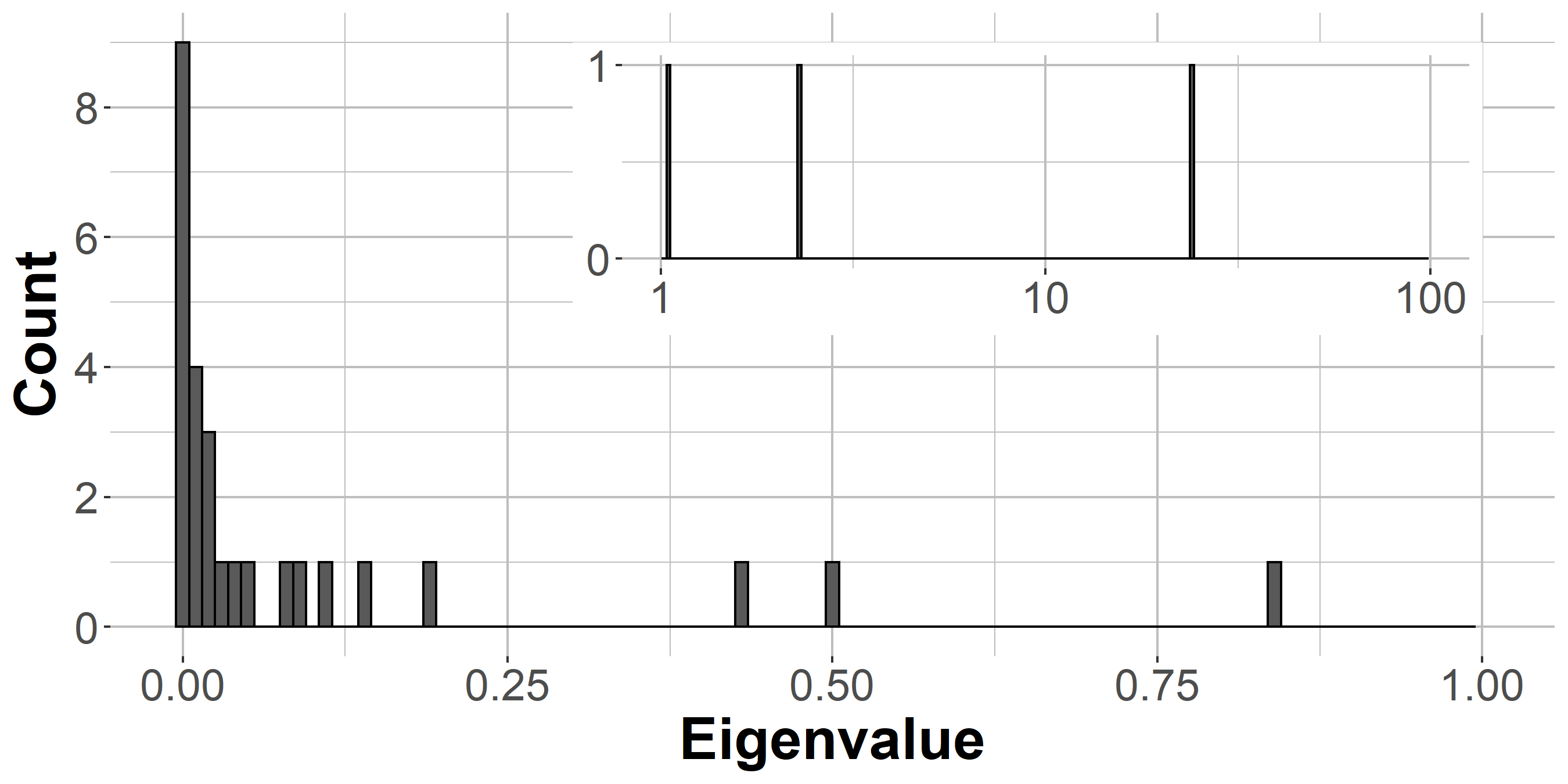}
		\includegraphics[width=0.9\columnwidth]{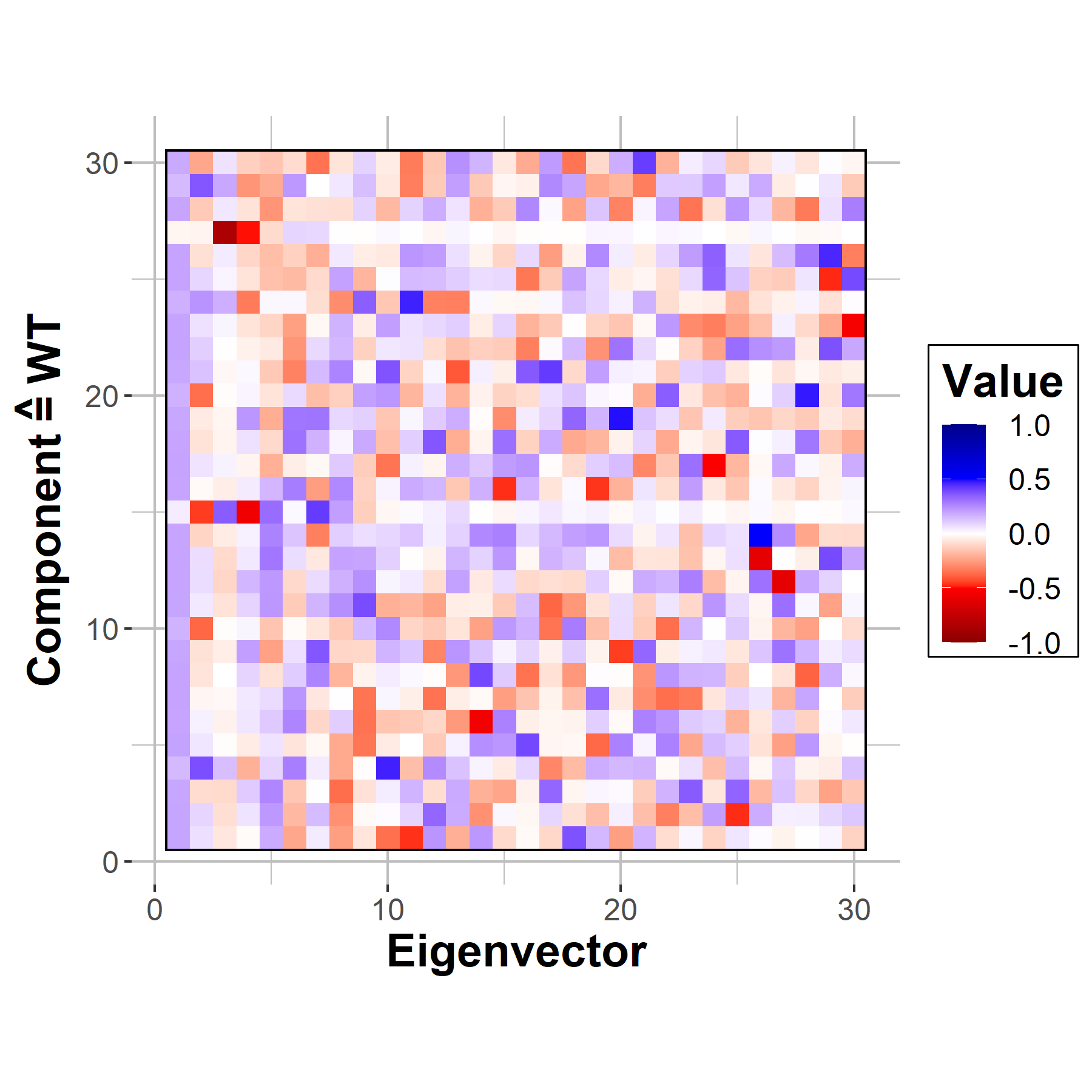}
		\caption{Top: Eigenvalue spectrum as histogram counts of the correlation matrix for the active power for half a day. The inlay shows the large values, while the main plot is zoomed in on small eigenvalues. Bottom: Corresponding eigenvectors of the correlation matrix for the active power. Each column represents an eigenvector. From left to right the corresponding eigenvalue decreases. For numerical values of the entries, see color code.}
		\label{fig:Eigen_Riffgat}
	\end{center}
\end{figure}

\begin{figure*}[htb]
	\begin{center}
		\includegraphics[width=0.32\textwidth]{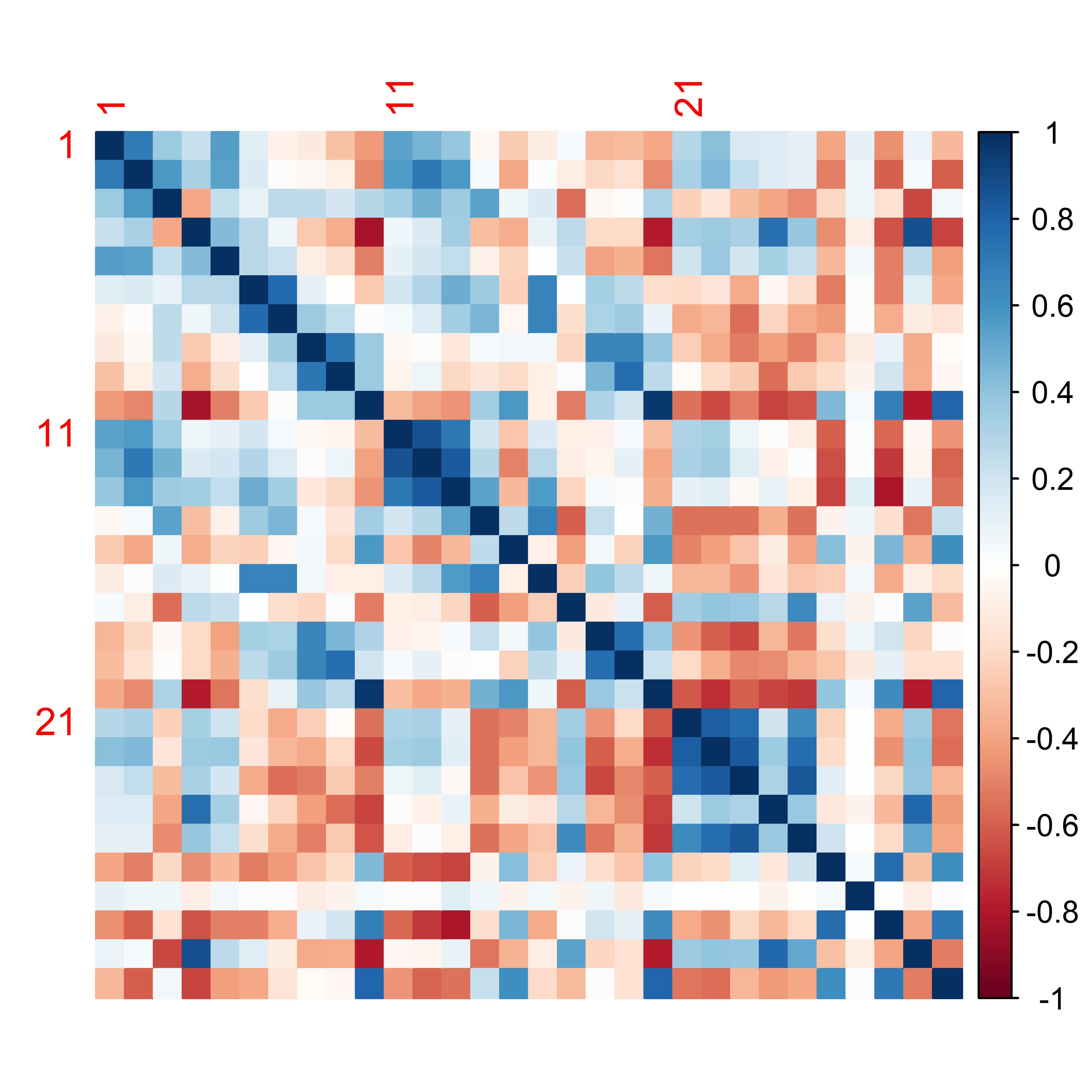}
		\includegraphics[width=0.32\textwidth]{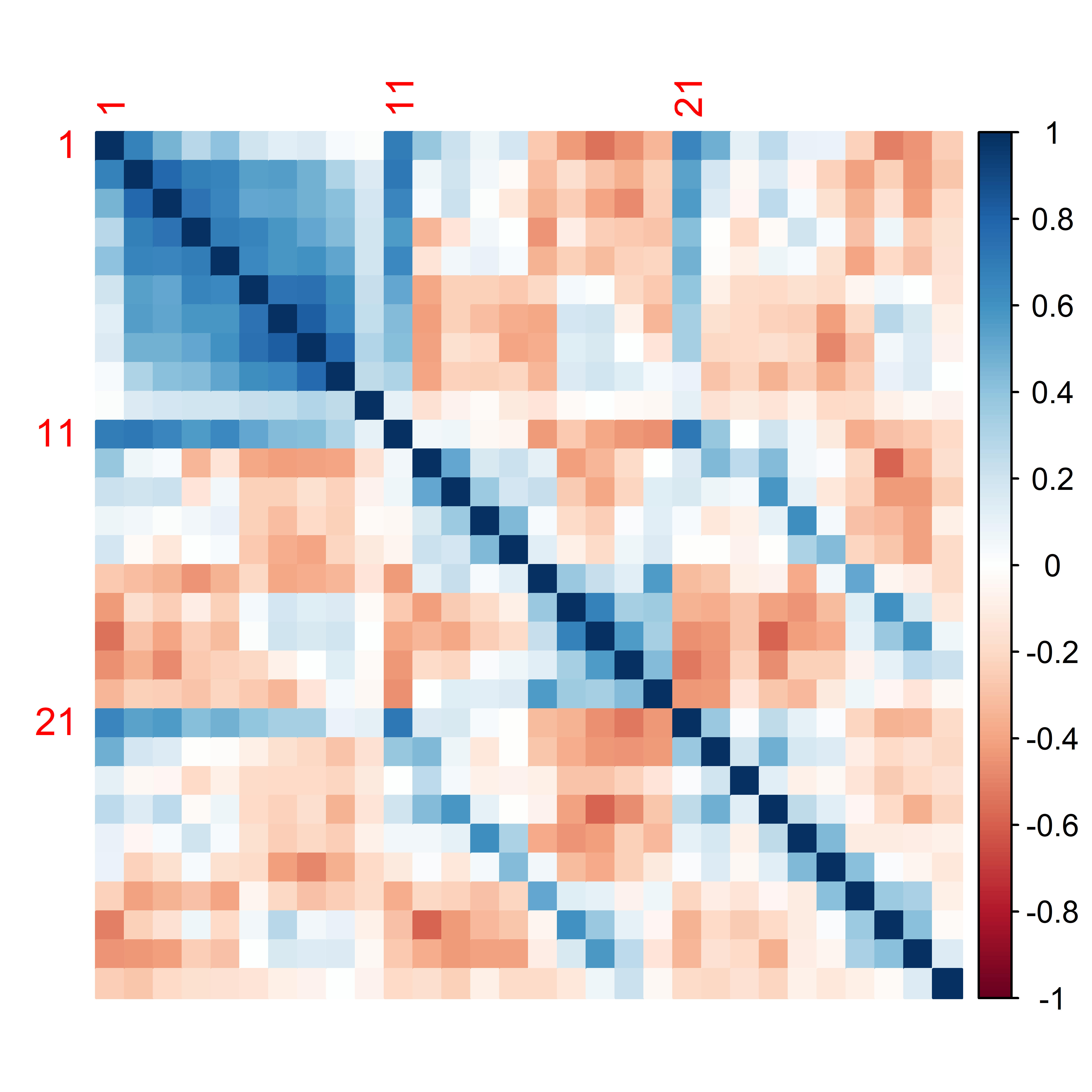}
		\includegraphics[width=0.32\textwidth]{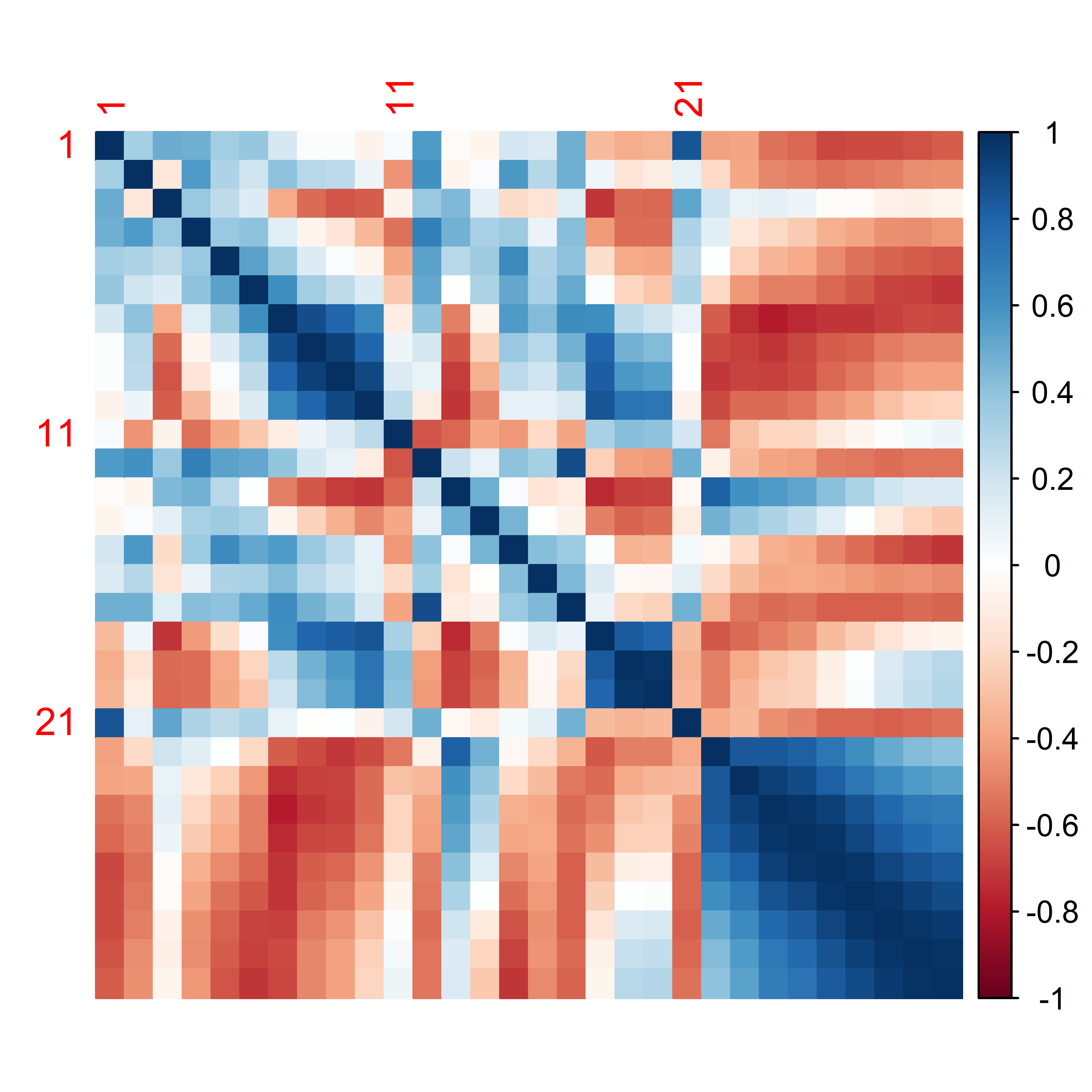}
		\caption{Correlation matrices for the active power without the contribution of the first eigenvalue for three different times for the wind farm \textsc{Riffgat}.}
		\label{fig:Corr_Red_Riffgat}
	\end{center}
\end{figure*}

\begin{figure*}[htb]
	\centering
	\includegraphics[width=0.24\textwidth]{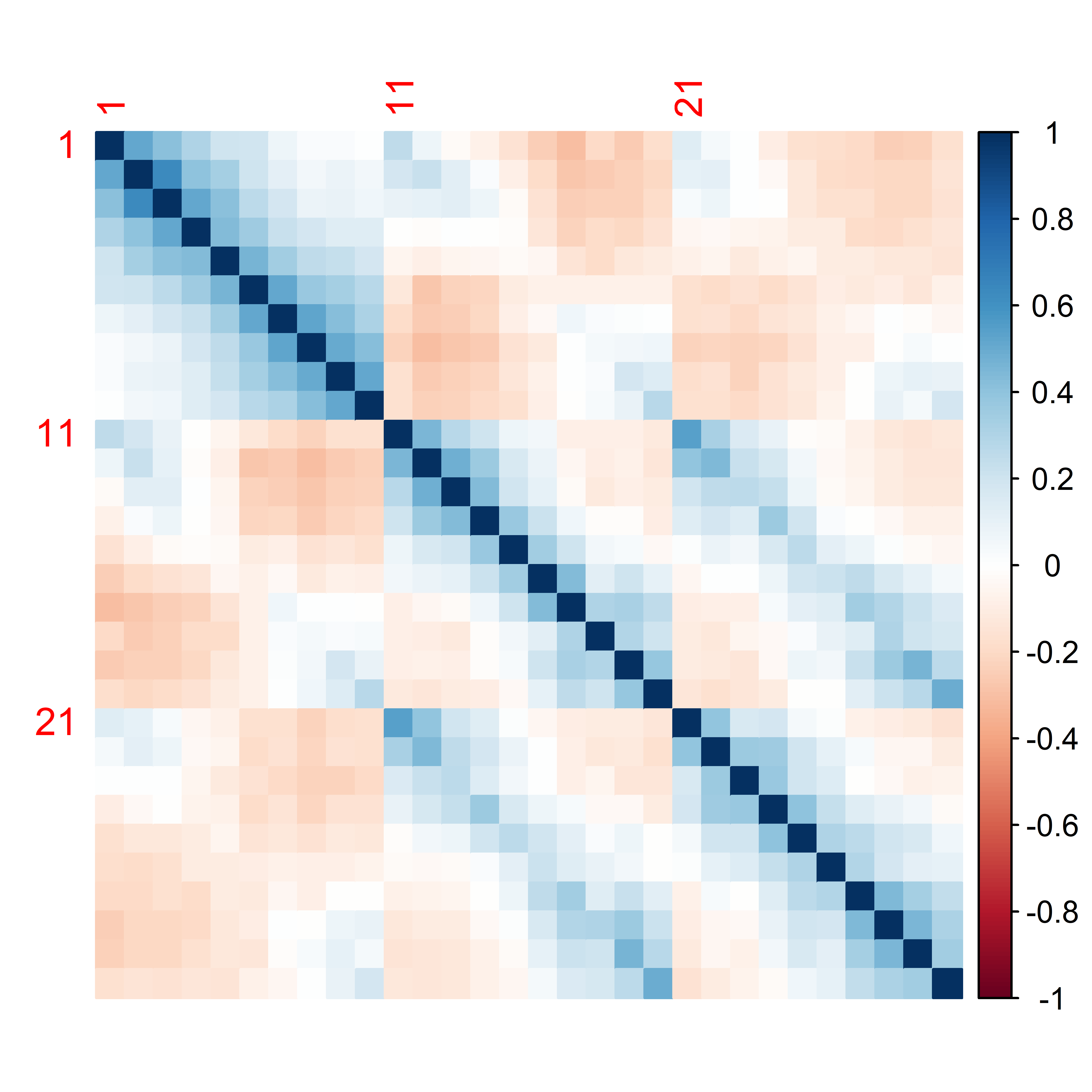}
	\includegraphics[width=0.24\textwidth]{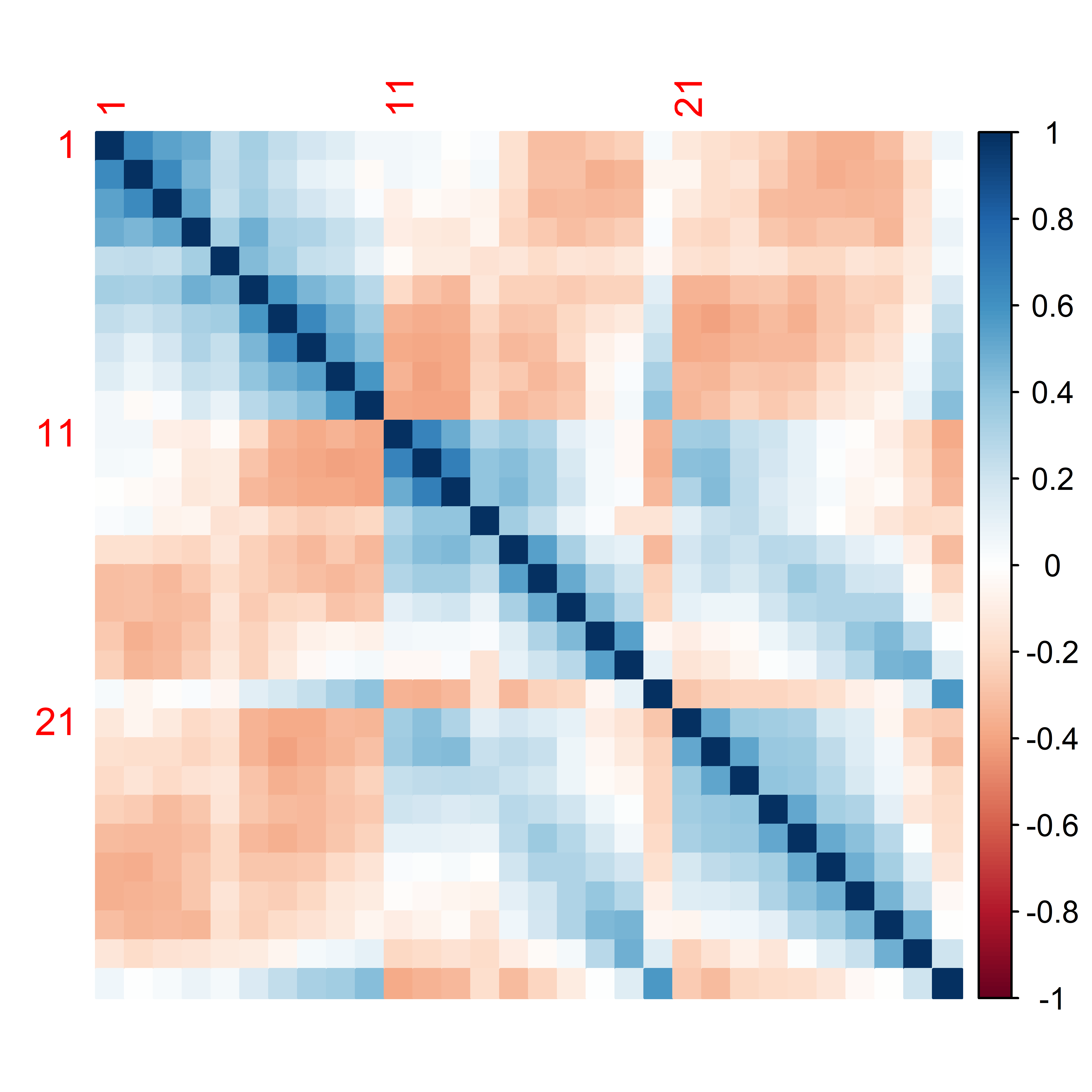}
	\includegraphics[width=0.24\textwidth]{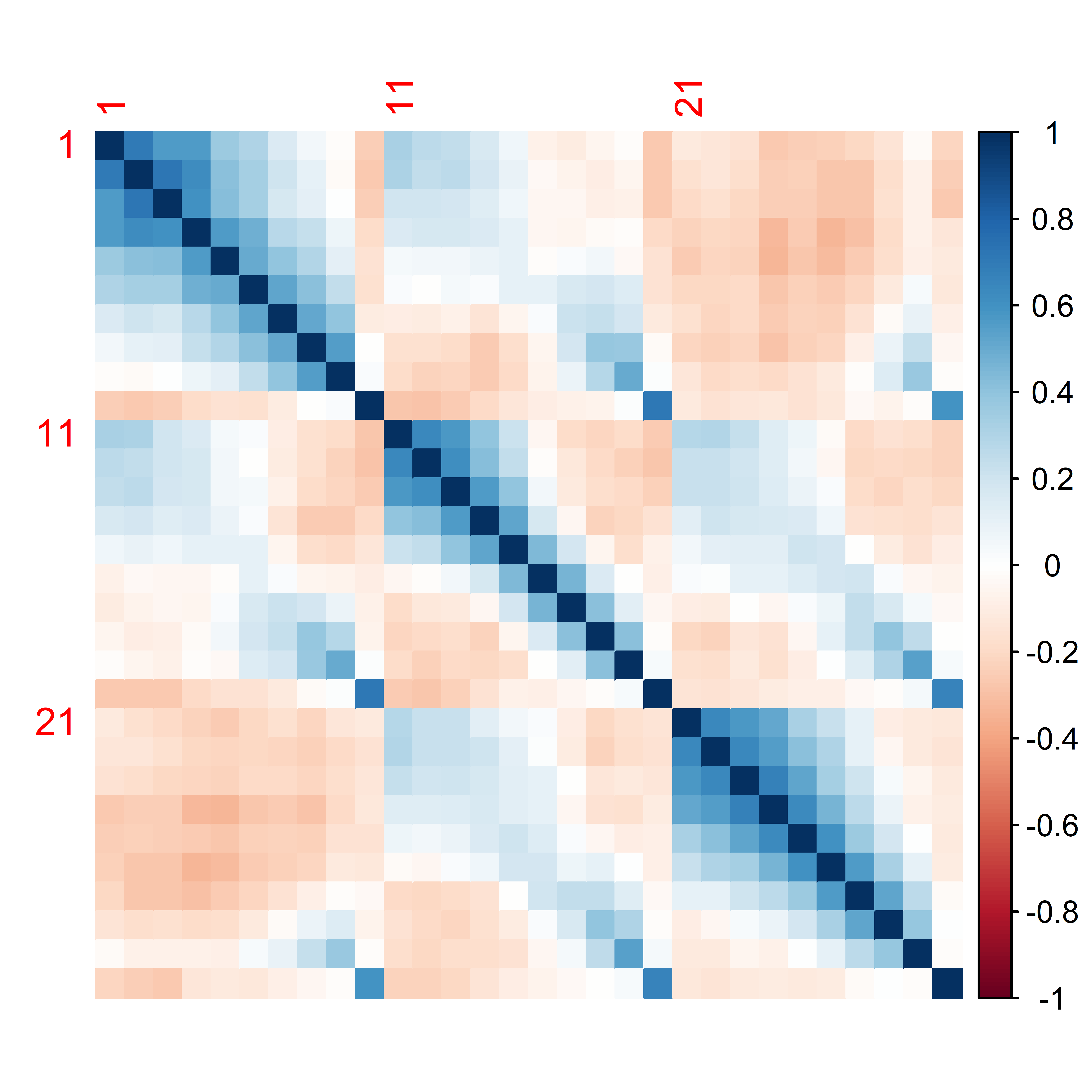}
	\includegraphics[width=0.24\textwidth]{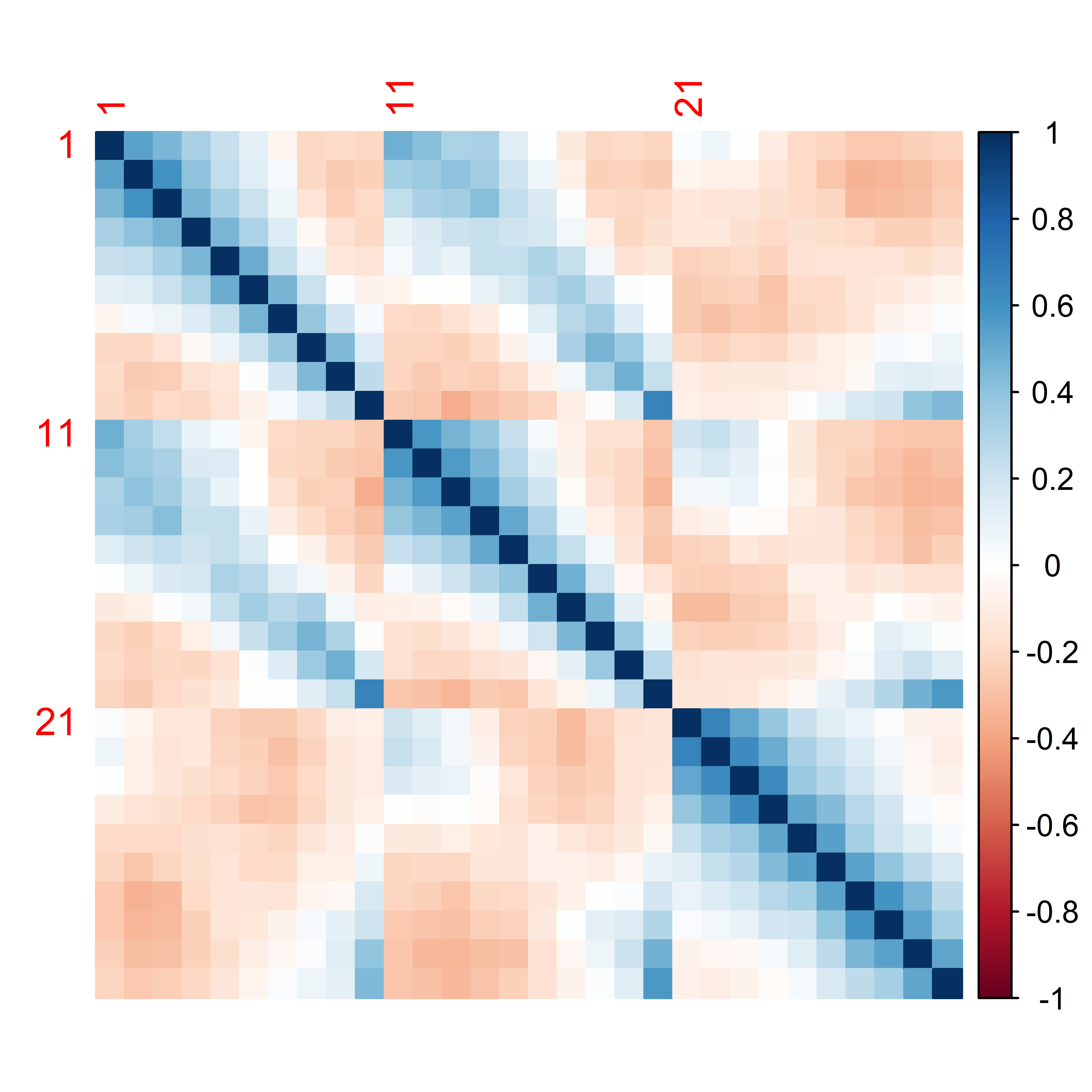}
	\includegraphics[width=0.24\textwidth]{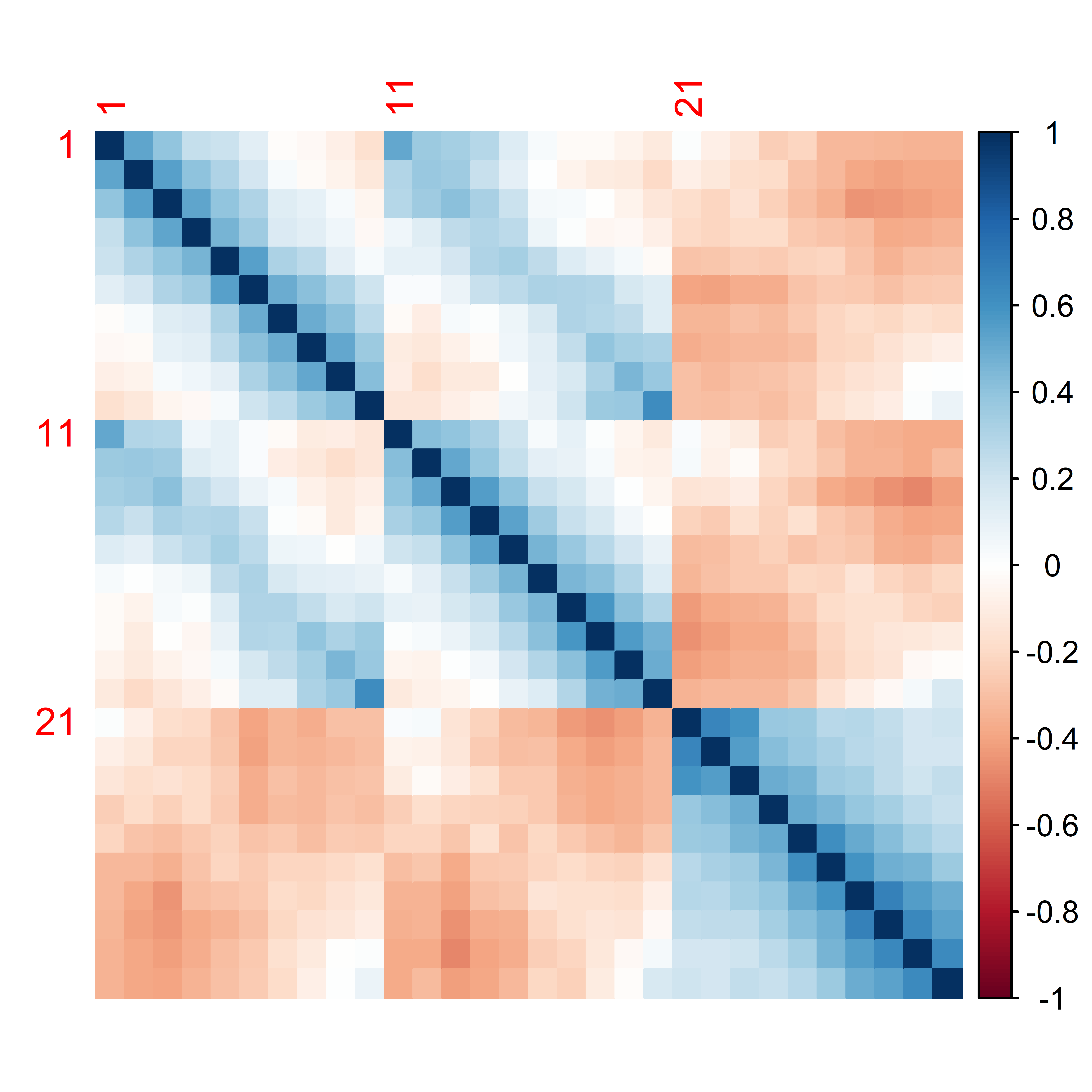}
	\includegraphics[width=0.24\textwidth]{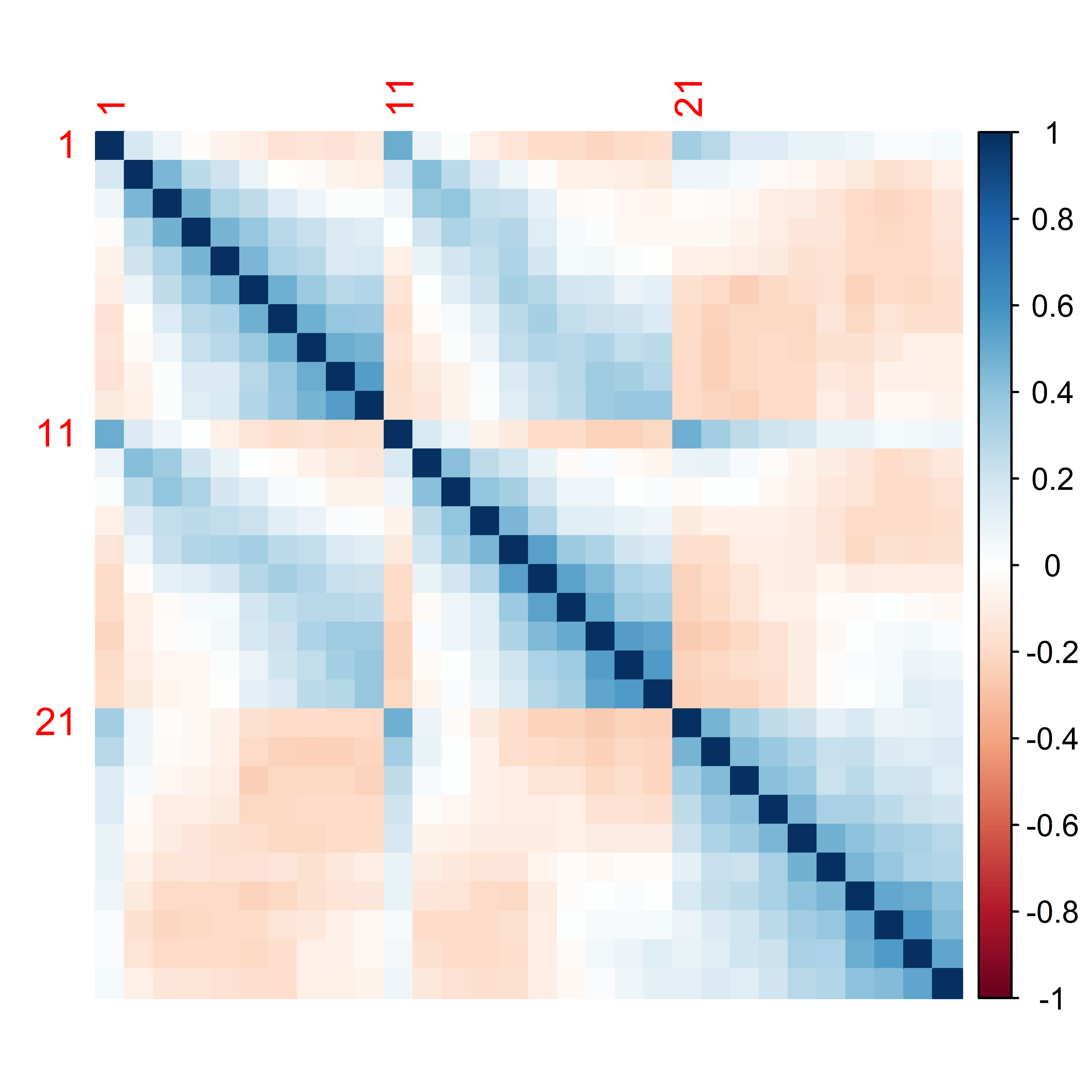}
	\includegraphics[width=0.24\textwidth]{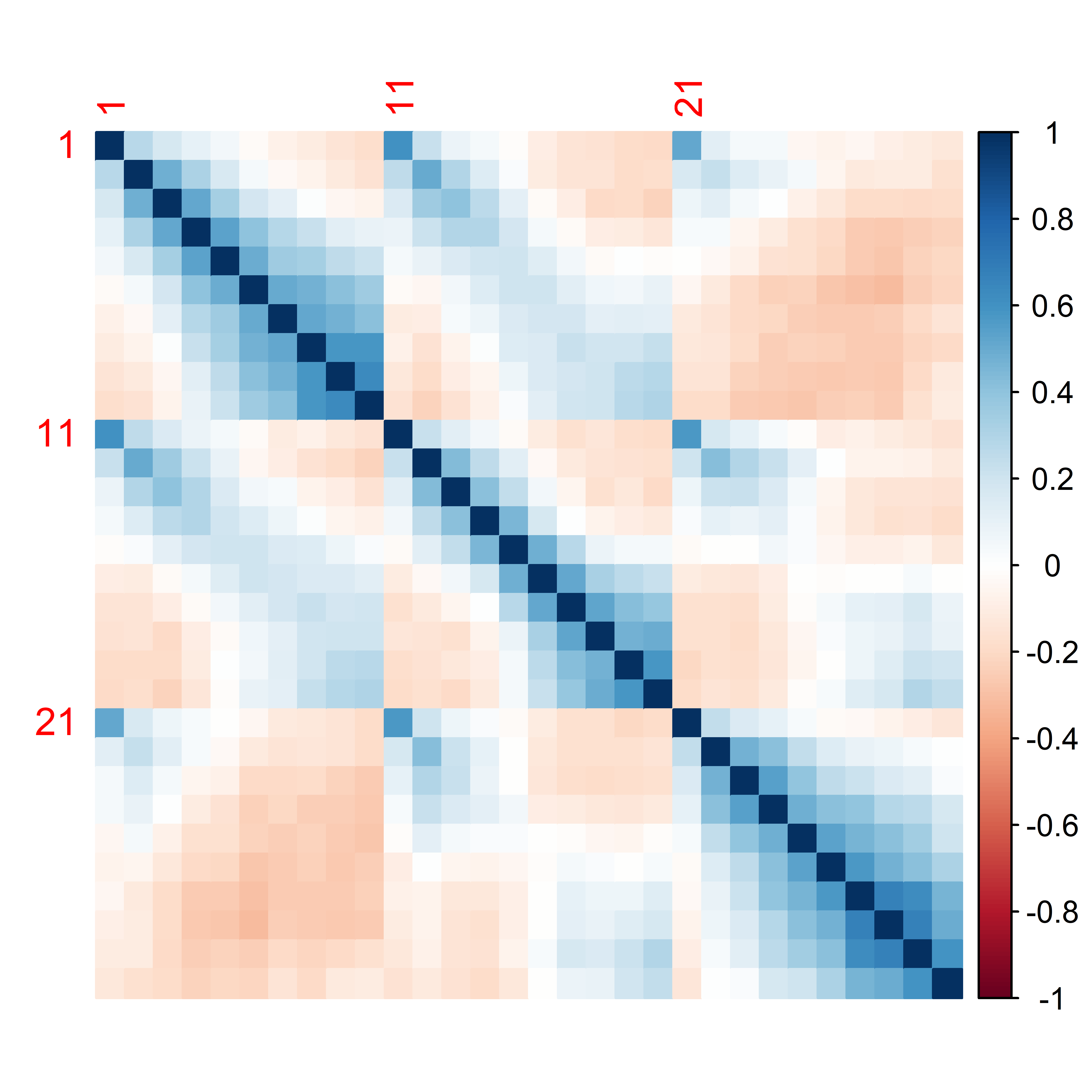}
	\includegraphics[width=0.24\textwidth]{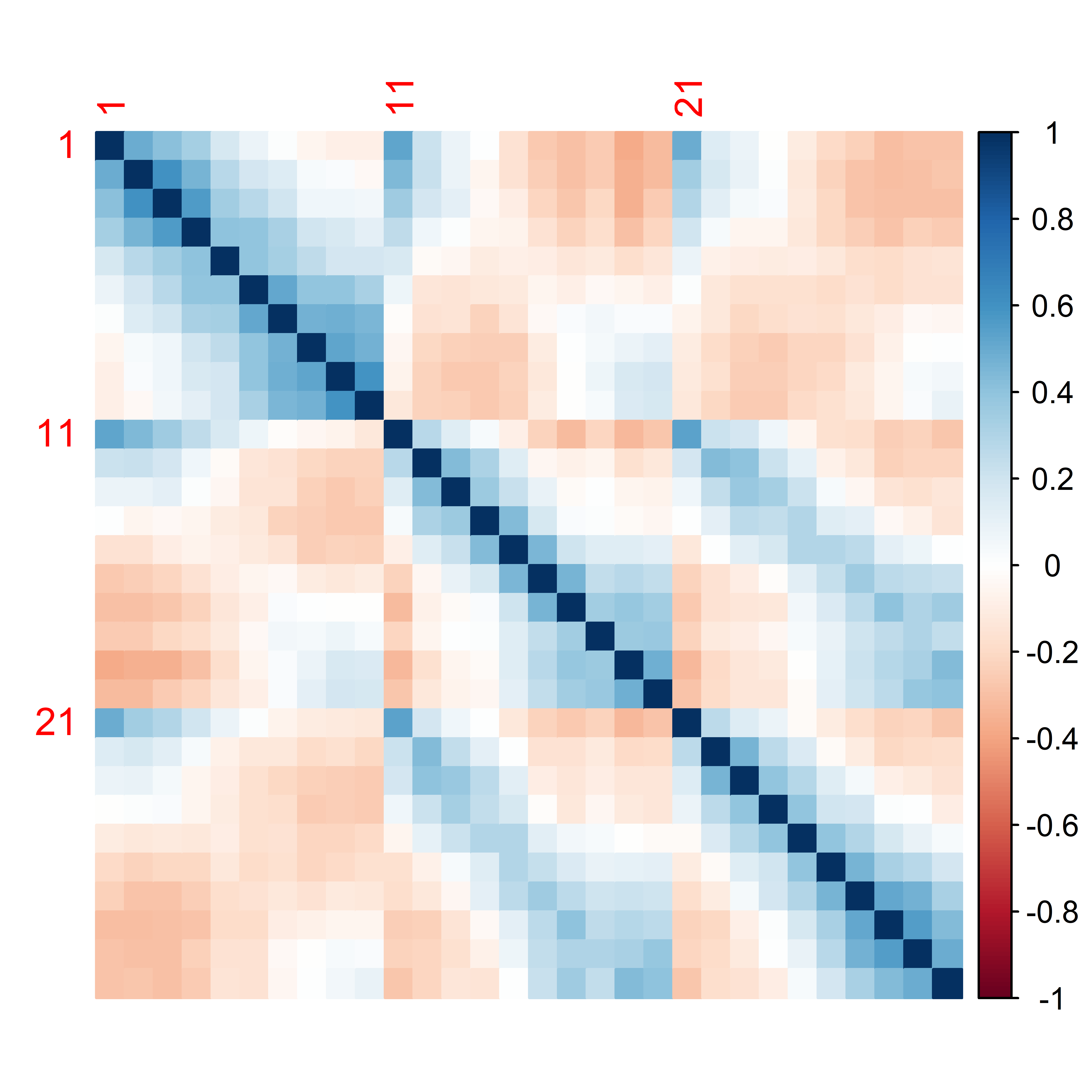}
	\caption{Average correlation matrices for the active power without the contribution of the first eigenvalue for the different $45^\circ$ ranges for the wind farm \textsc{Riffgat}. The wind directions are from left to right N, NE, E, SE and S, SW, W, NW for the two rows.}
	\label{fig:Corr_Red_Dir_Riffgat}
\end{figure*}

\begin{figure*}[htb]
	\begin{center}
		\includegraphics[width=0.32\textwidth]{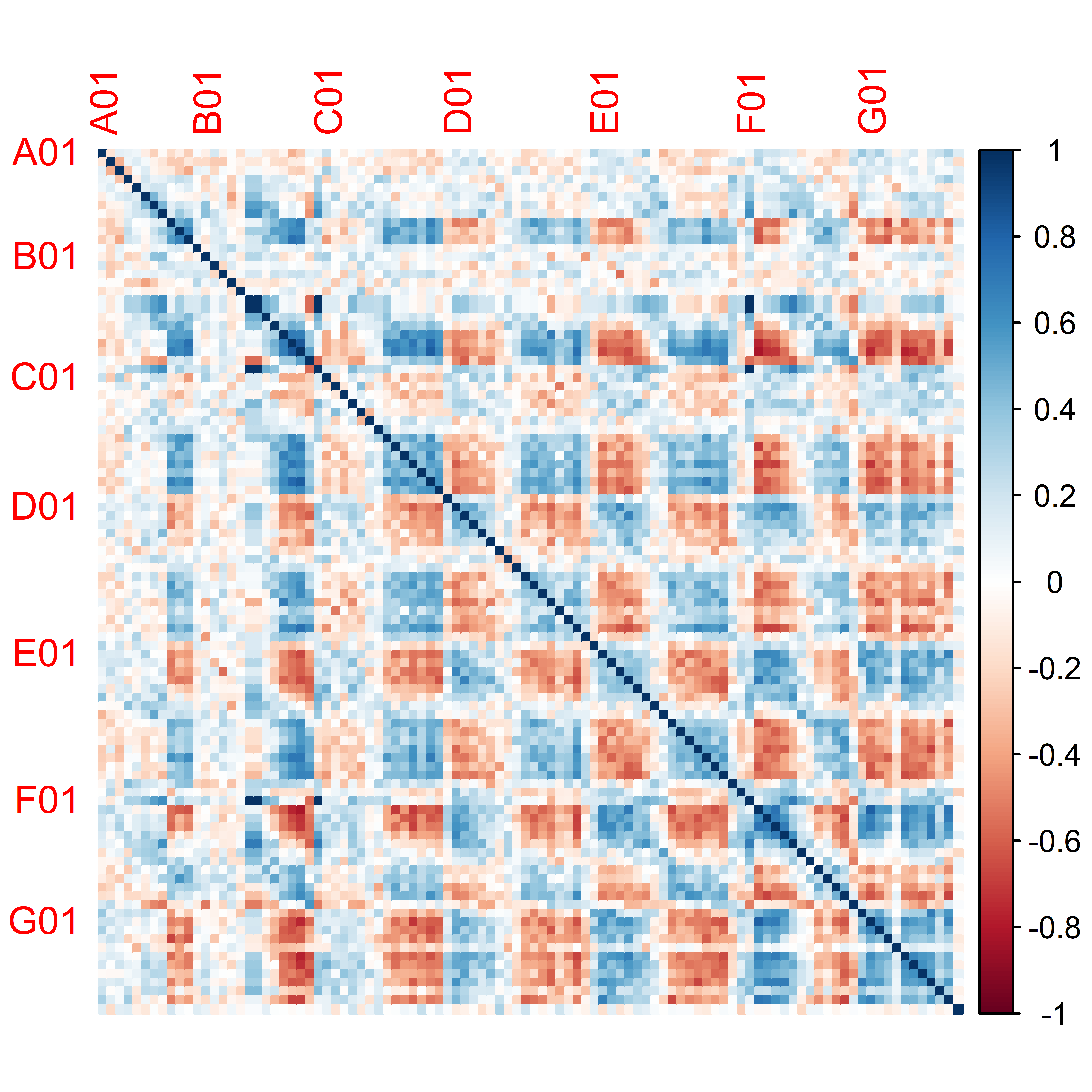}
		\includegraphics[width=0.32\textwidth]{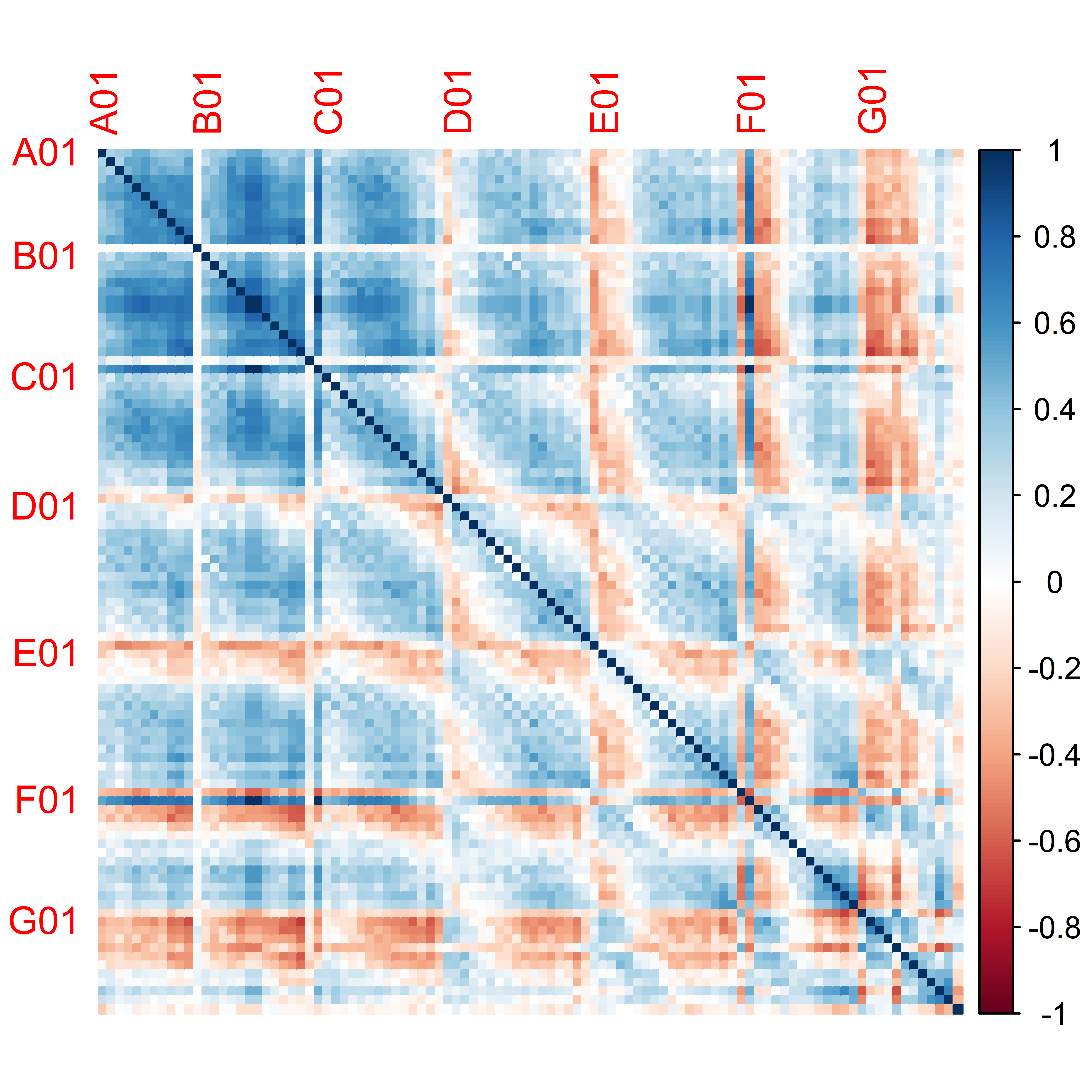}
		\includegraphics[width=0.32\textwidth]{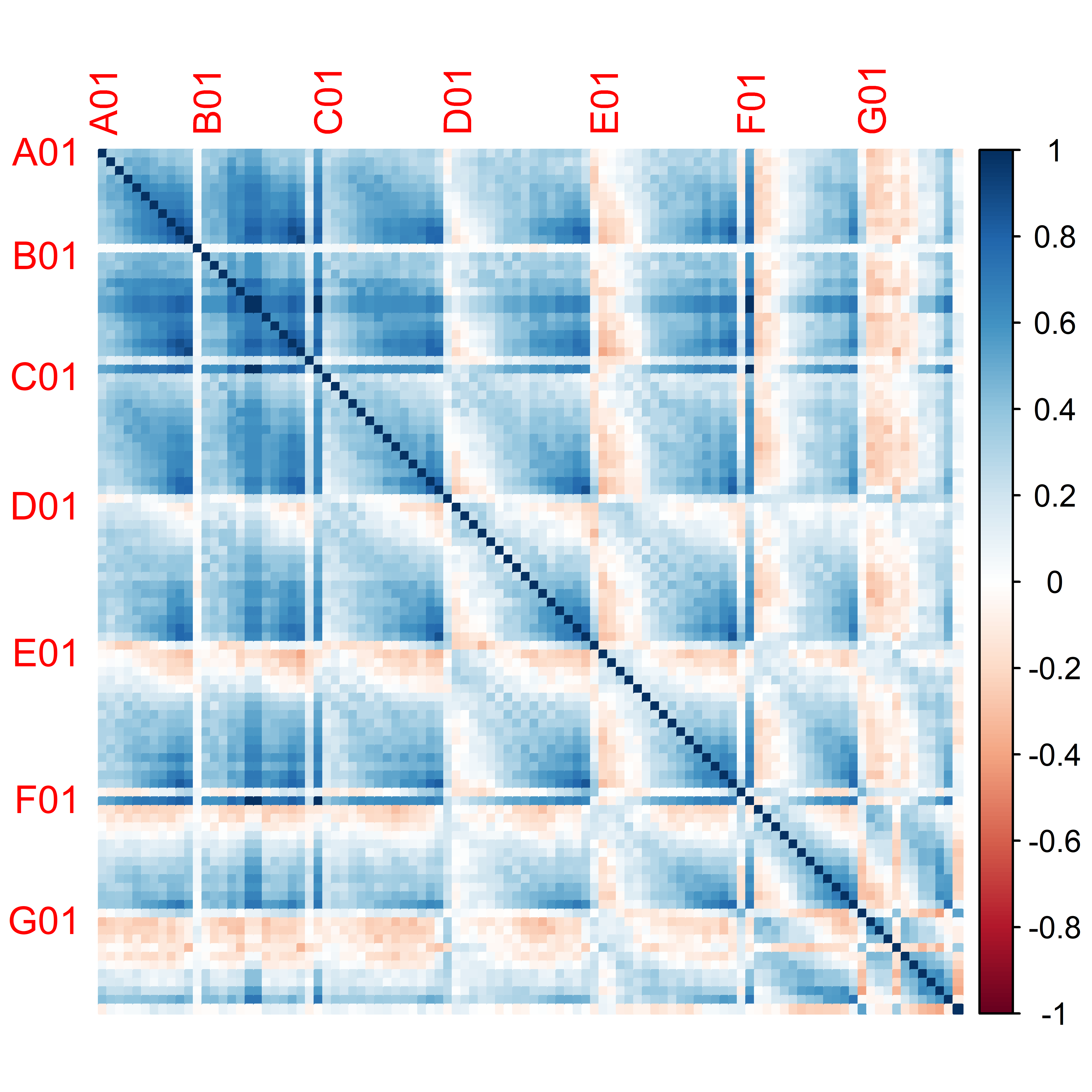}
		\includegraphics[width=0.32\textwidth]{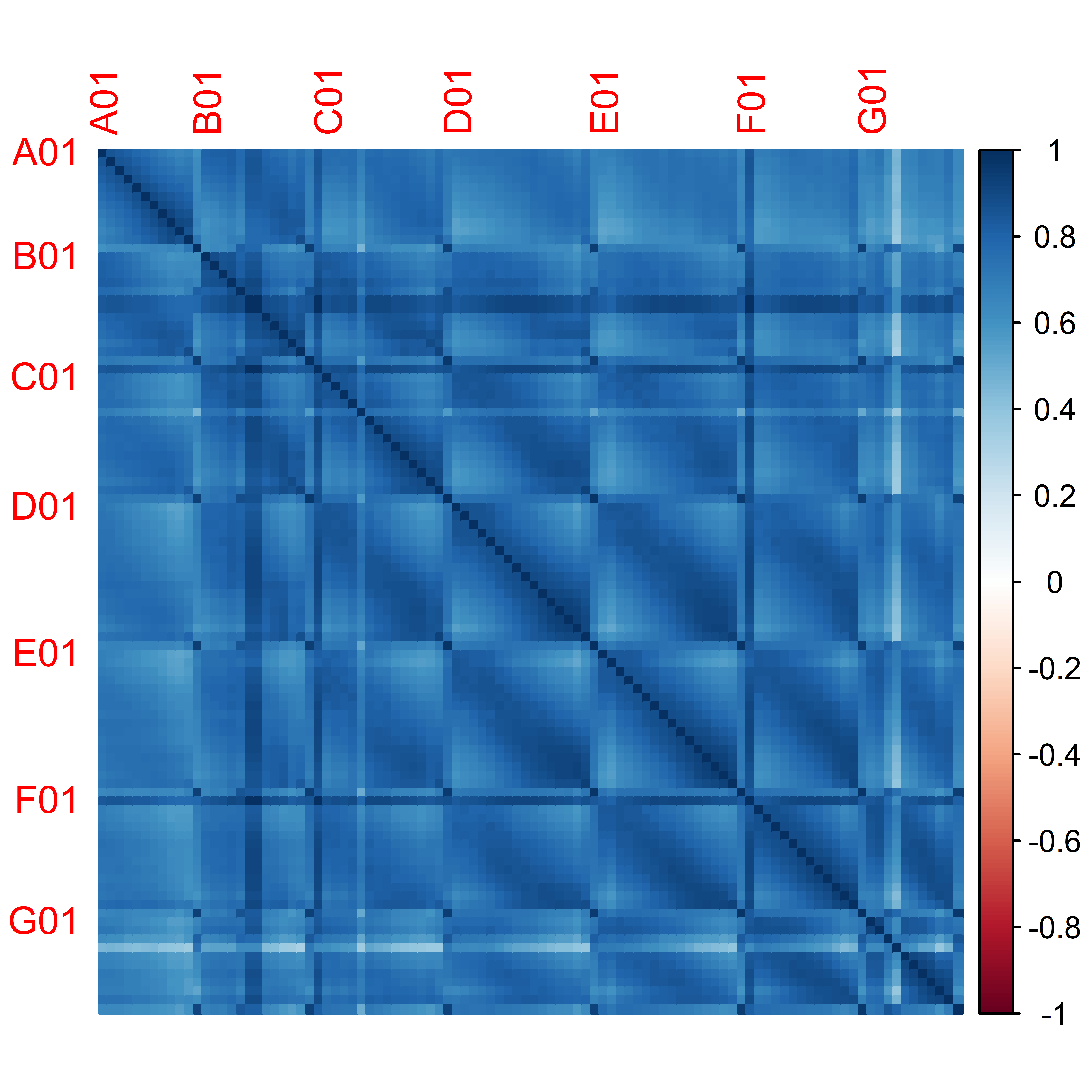}
		\includegraphics[width=0.32\textwidth]{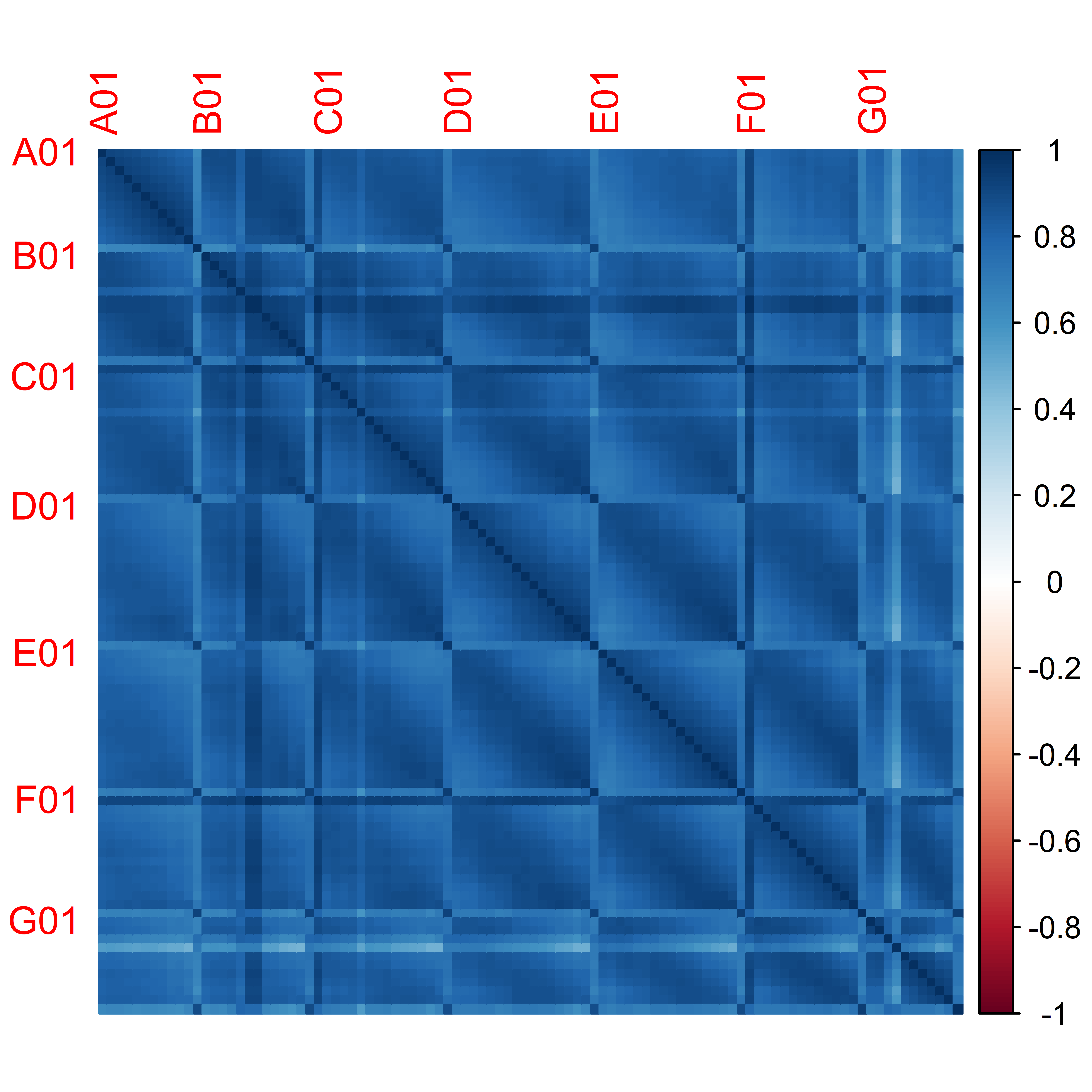}
		\includegraphics[width=0.32\textwidth]{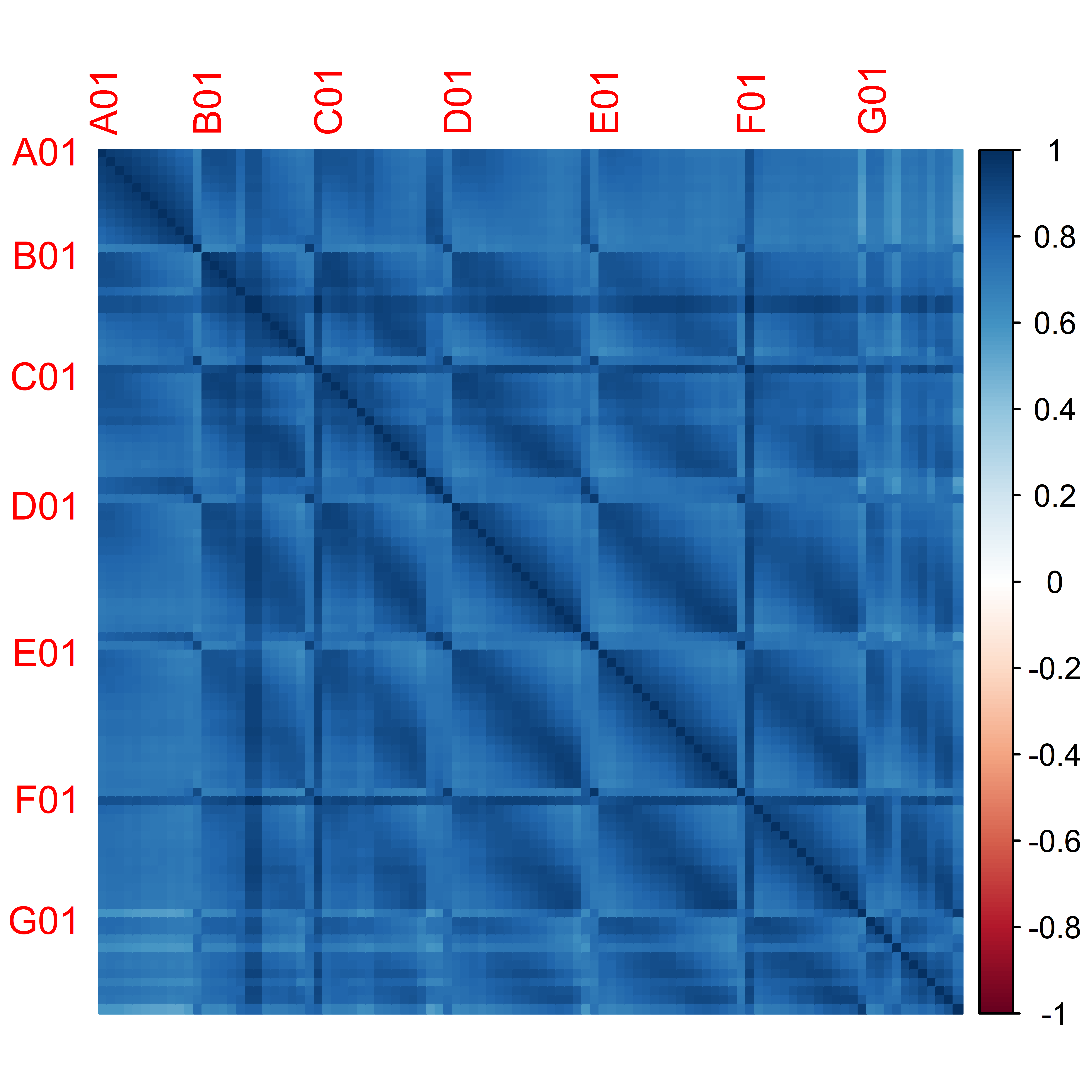}
		\caption{Correlation Matrices for the active power over ten minutes (top left), half an hour (top middle), an hour (top right) six hours (bottom left), half a day (bottom middle) and a day (bottom right) for the wind farm \textsc{Thanet}.}
		\label{fig:Corr_ex_Thanet}
	\end{center}
\end{figure*}

\begin{figure*}[htb]
	\begin{center}
		\includegraphics[width=0.6\textwidth]{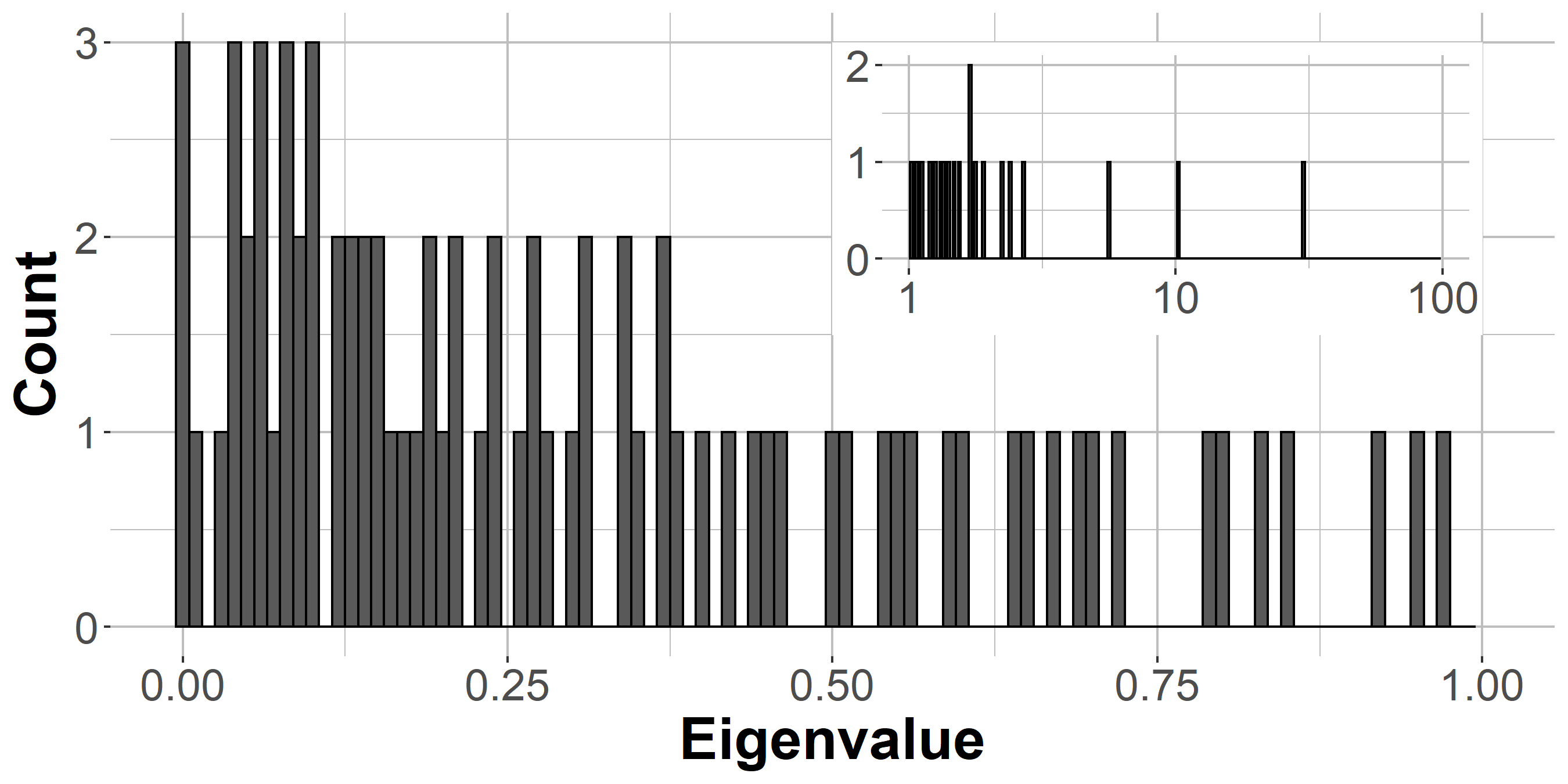}
		\includegraphics[width=0.9\textwidth]{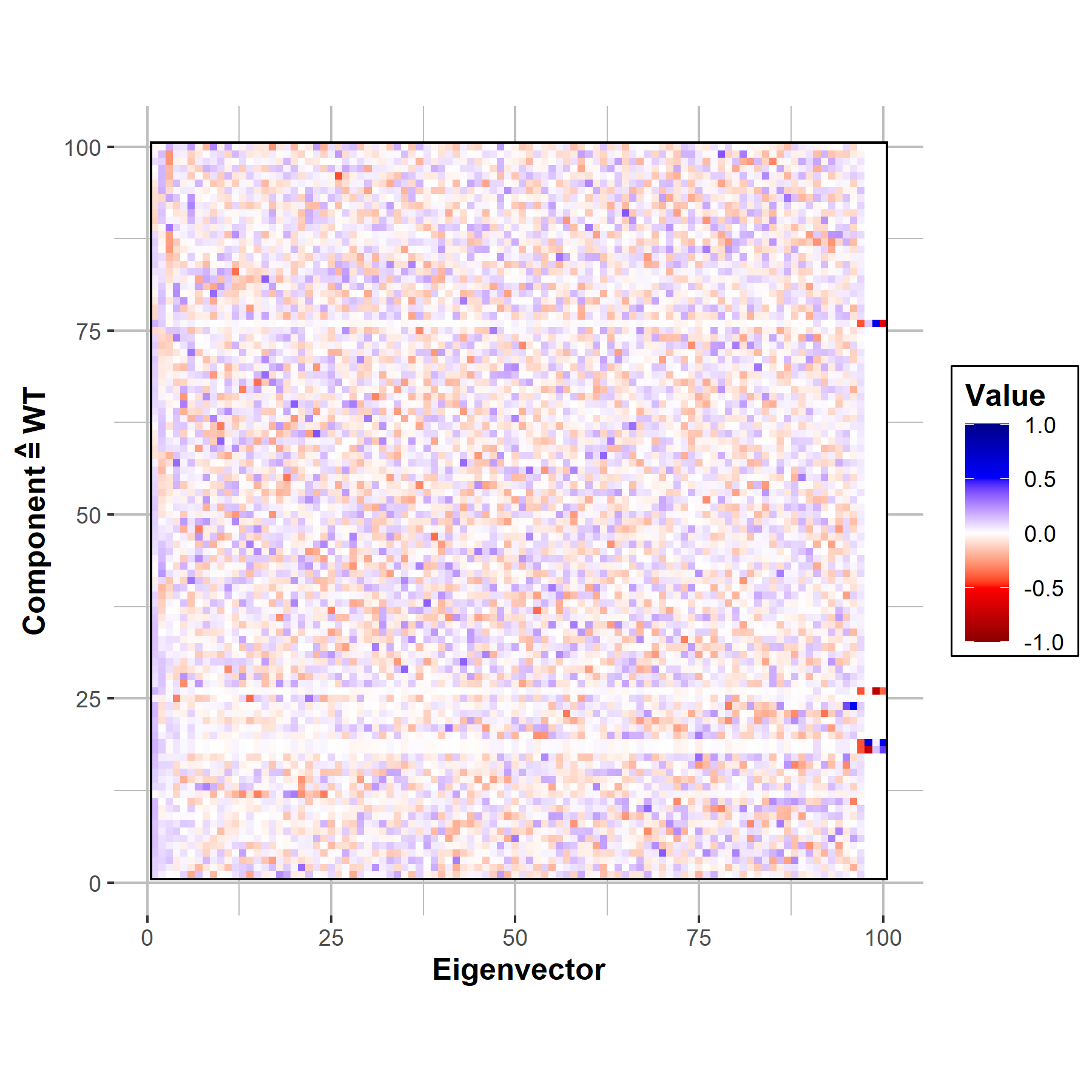}
		\caption{Top: Eigenvalue spectrum as histogram counts of the correlation matrix for the active power for half an hour. The inlay shows the large values, while the main plot is zoomed in on small eigenvalues. Bottom: Corresponding eigenvectors of the correlation matrix for the active power. Each column represents an eigenvector. From left to right the corresponding eigenvalue decreases. For numerical values of the entries, see color code.}
		\label{fig:Eigen_Thanet}
	\end{center}
\end{figure*}

\begin{figure*}[htb]
	\centering
	\includegraphics[width=0.24\textwidth]{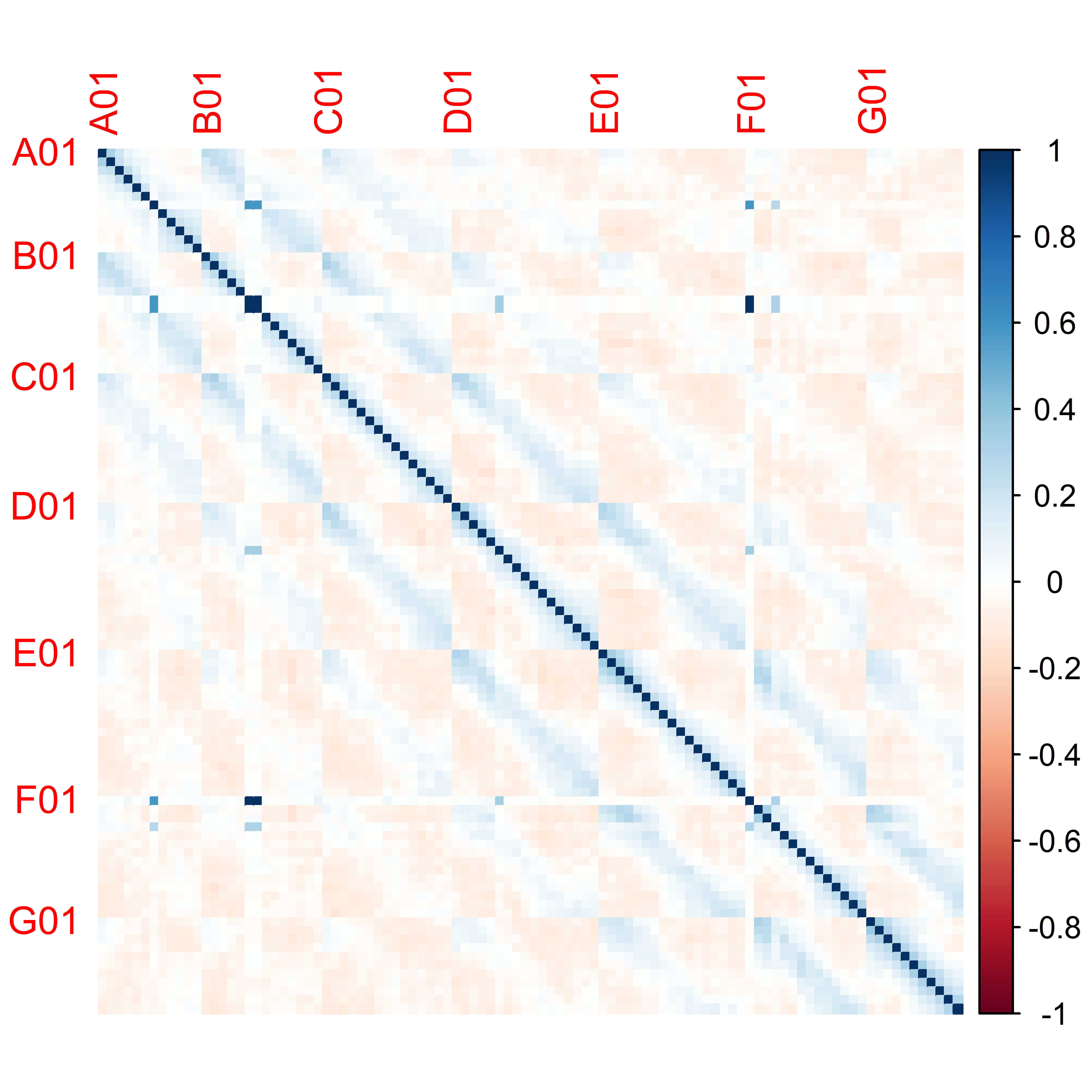}
	\includegraphics[width=0.24\textwidth]{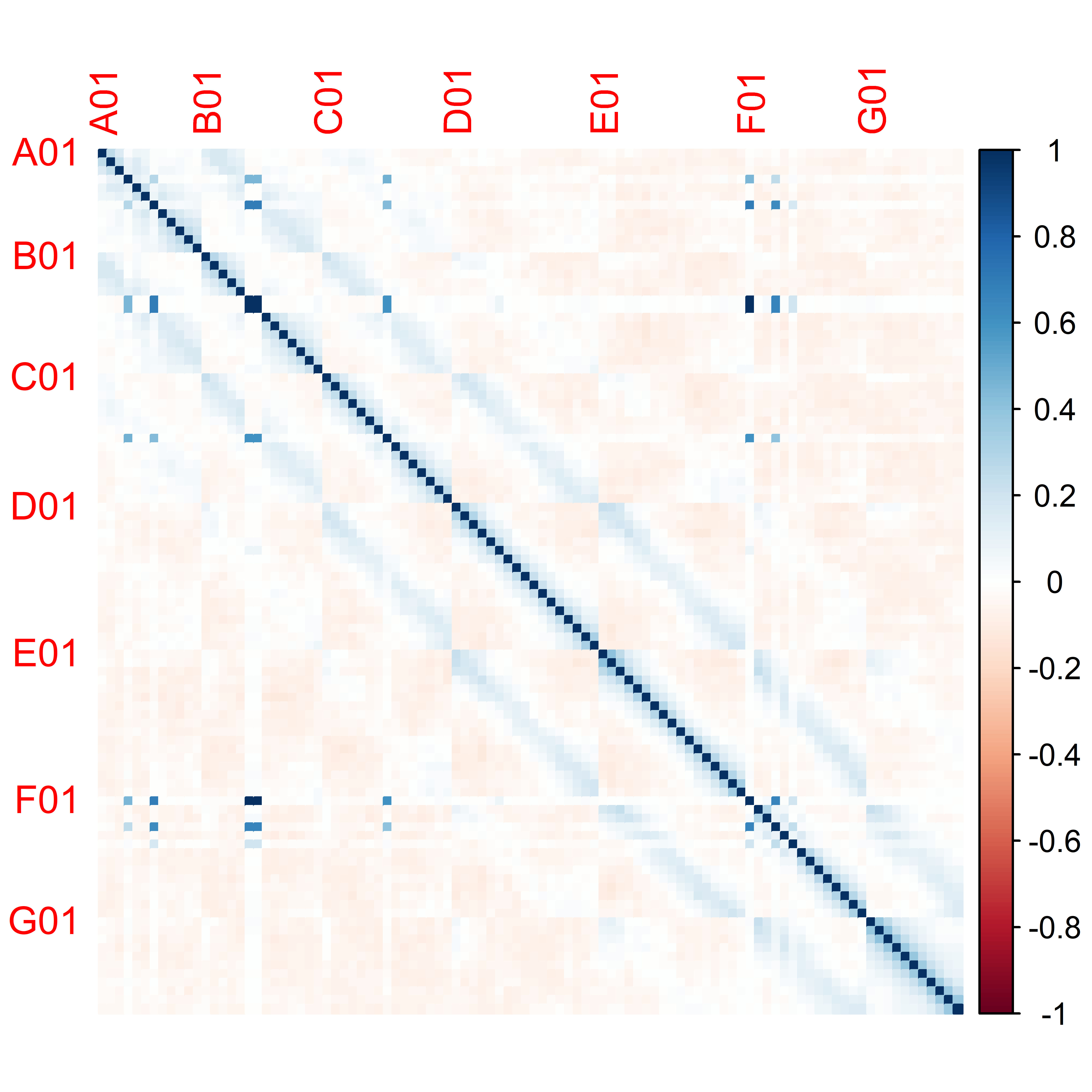}
	\includegraphics[width=0.24\textwidth]{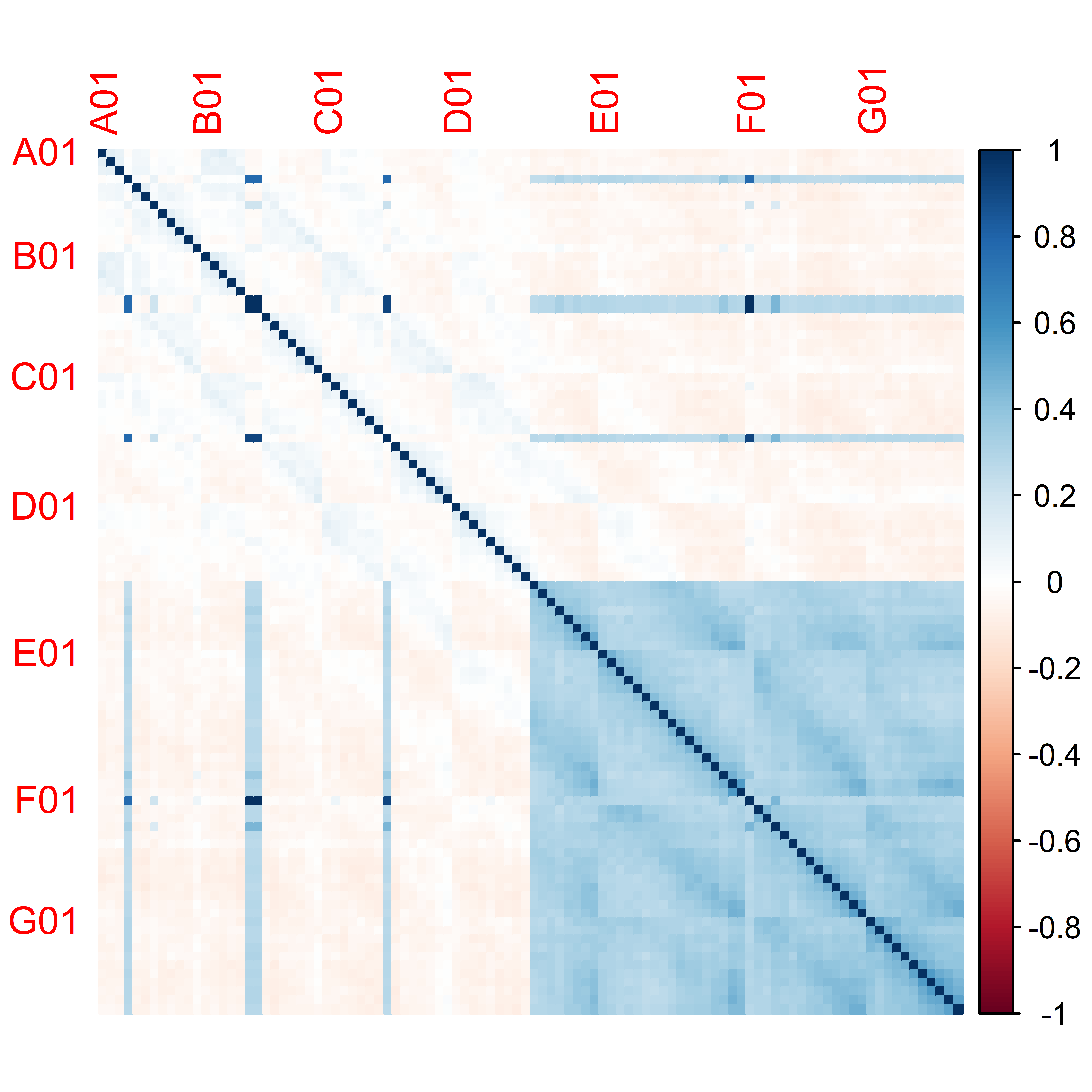}
	\includegraphics[width=0.24\textwidth]{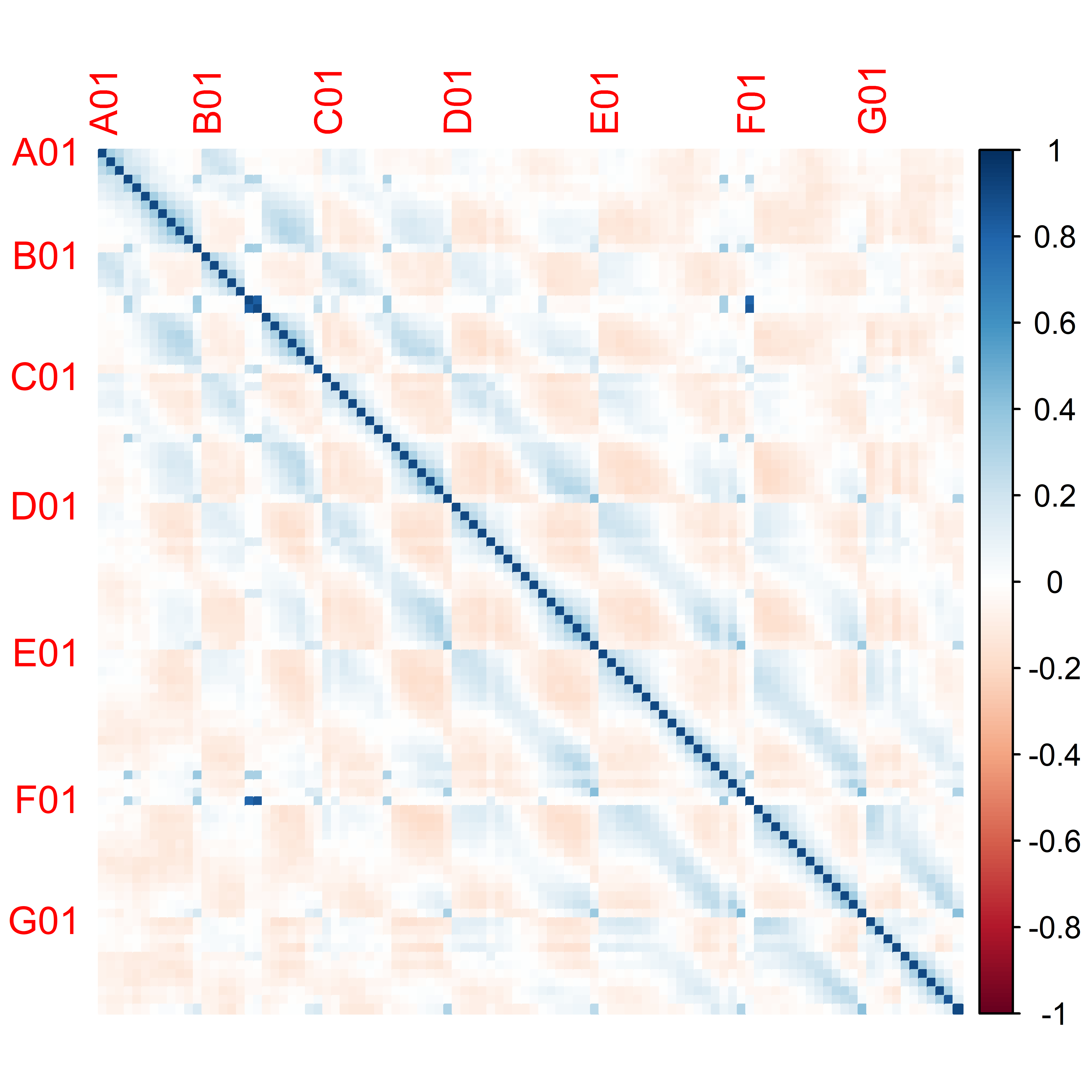}
	\includegraphics[width=0.24\textwidth]{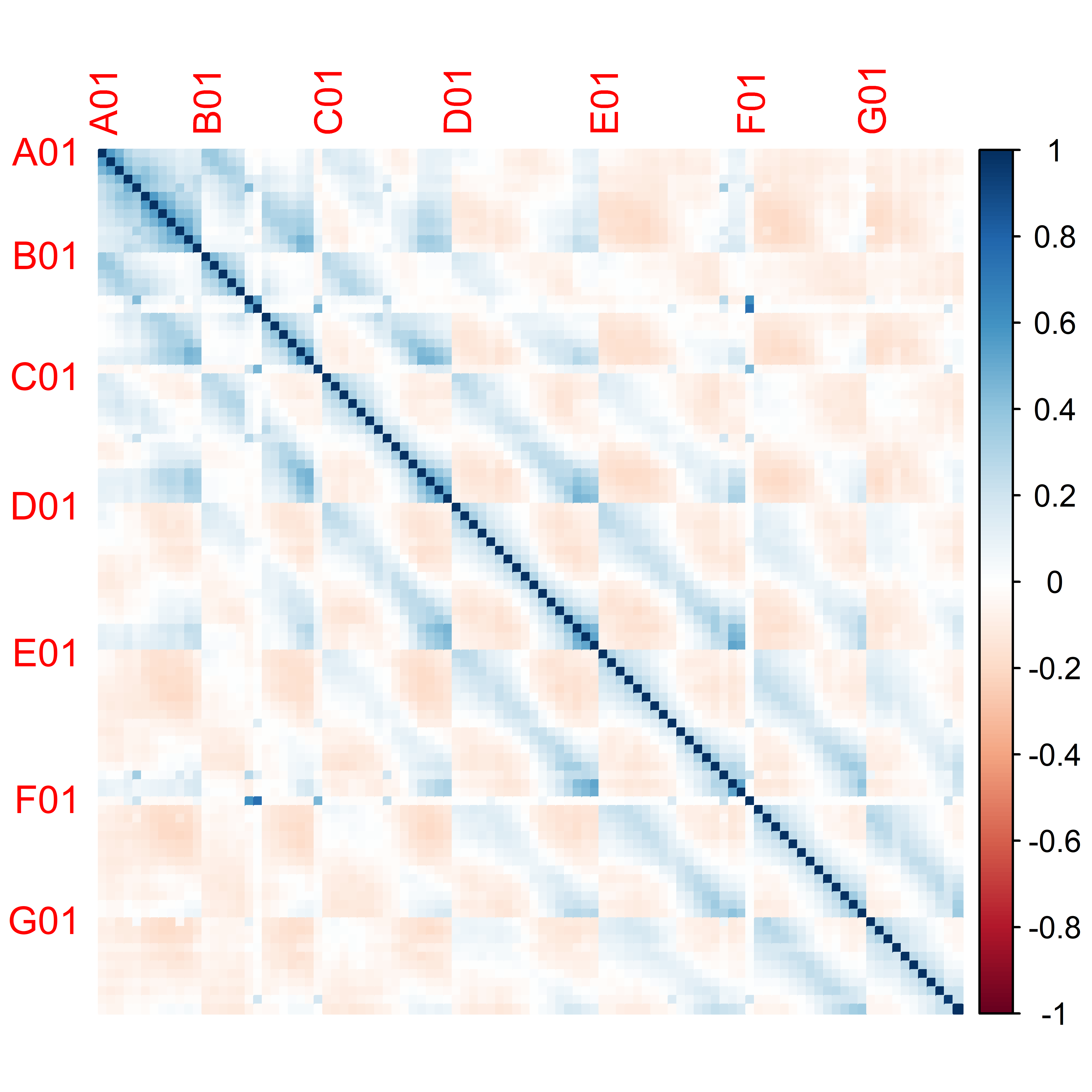}
	\includegraphics[width=0.24\textwidth]{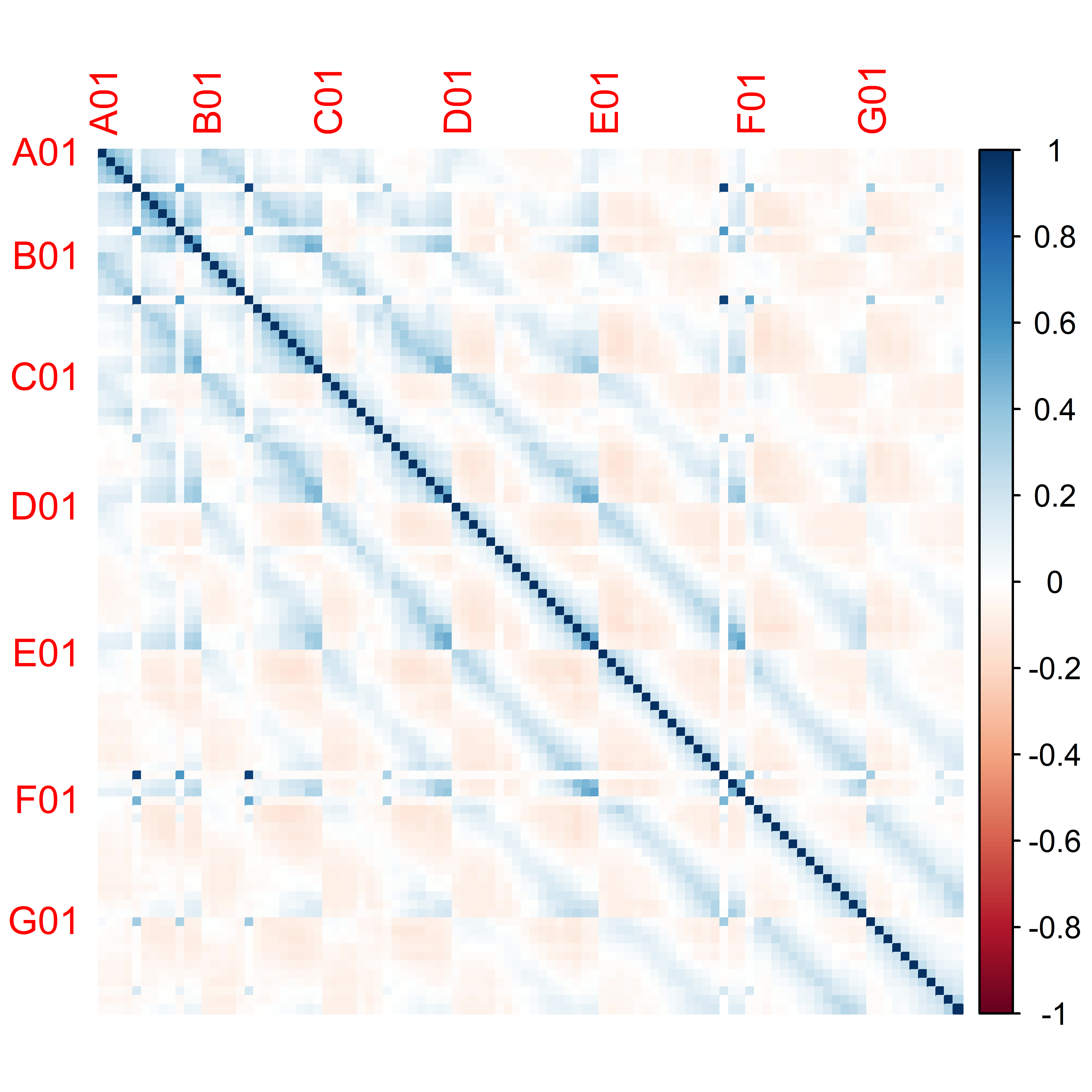}
	\includegraphics[width=0.24\textwidth]{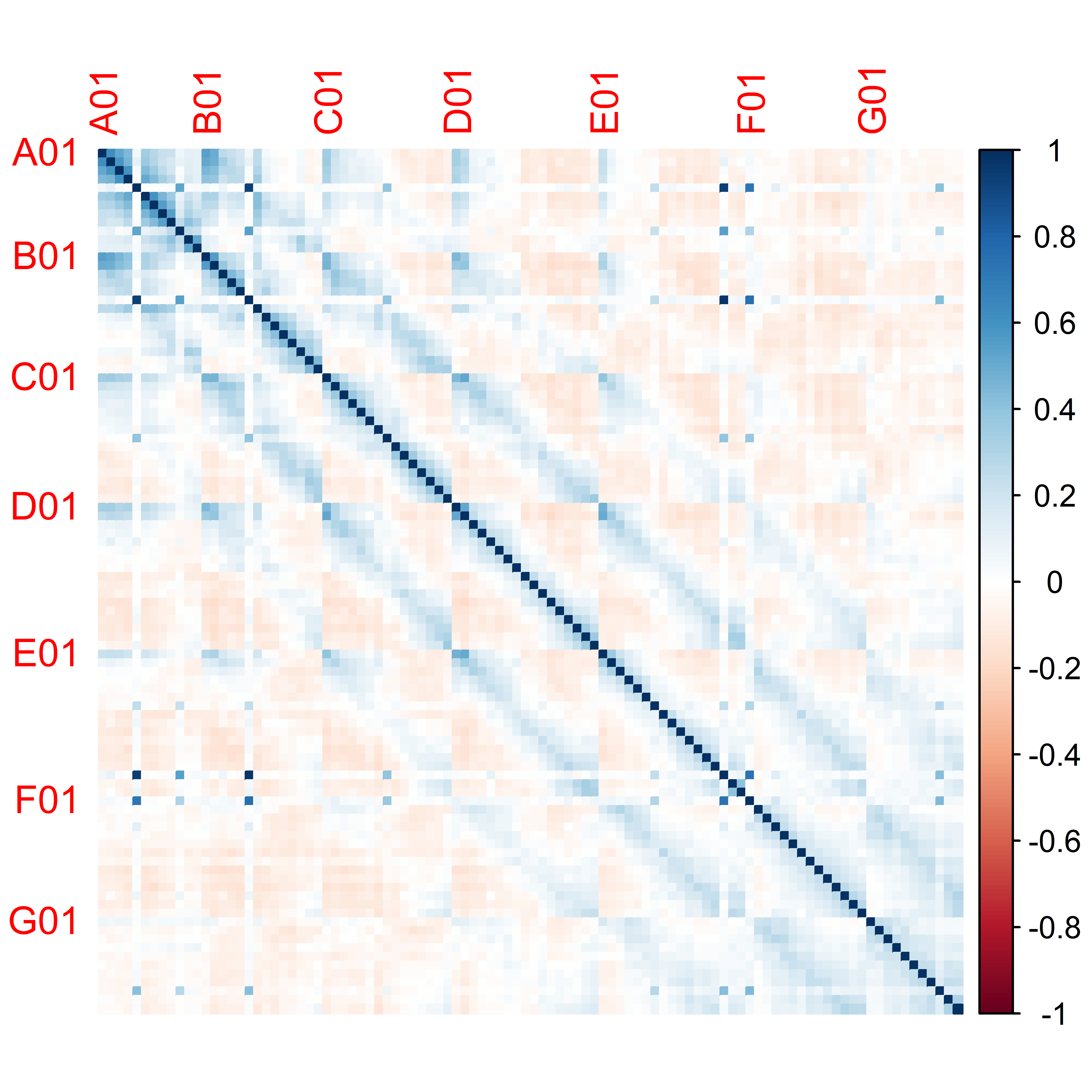}
	\includegraphics[width=0.24\textwidth]{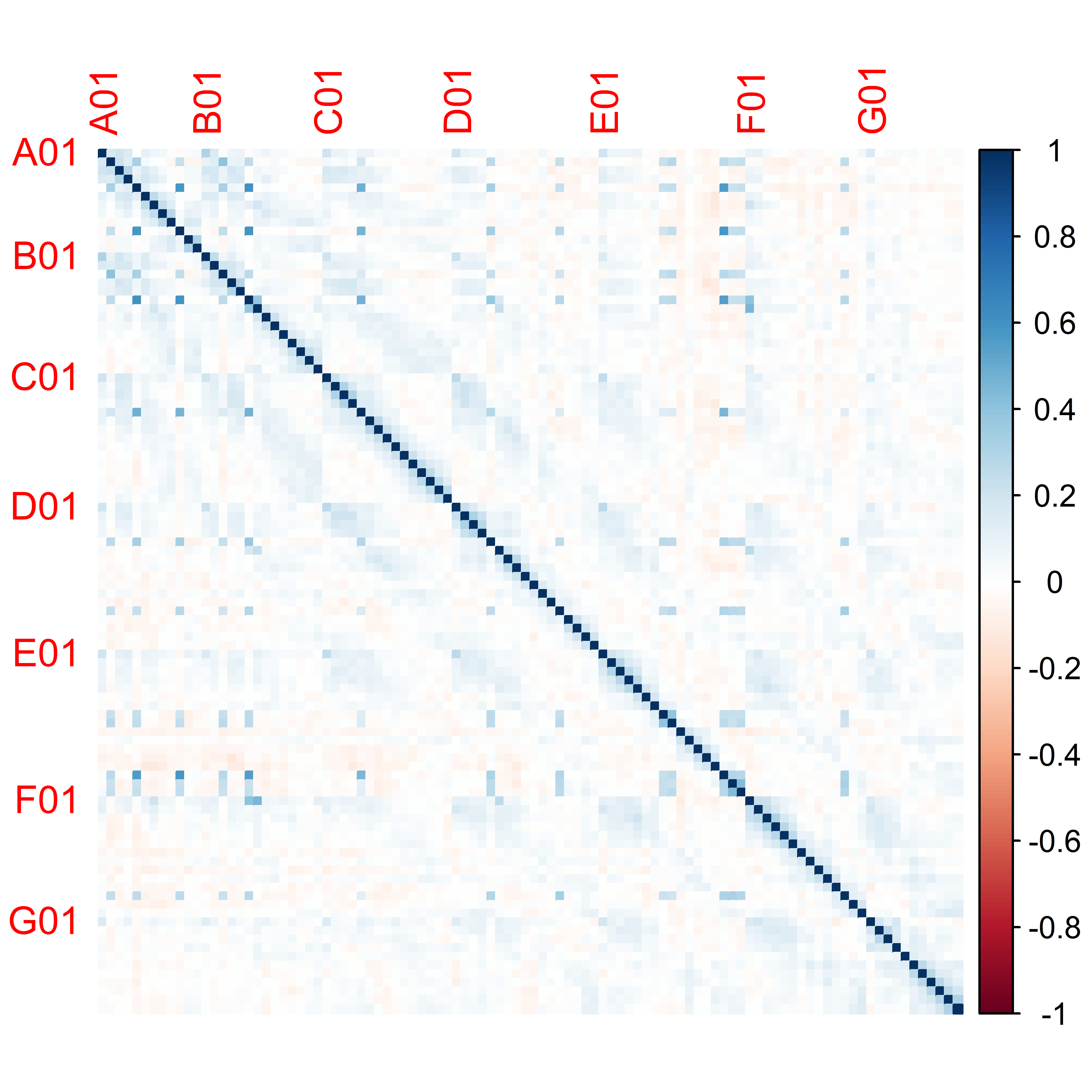}
	\caption{Average correlation matrices for the deviations, $\Psi$, of the individual wind turbines' active power from the mean active power for the different $45^\circ$ ranges for the wind farm \textsc{Thanet}. The wind directions are from left to right N, NE, E, SE and S, SW, W, NW for the two rows.}
	\label{fig:Corr_Red_Dir_Thanet}
\end{figure*}

We first examine the correlation structures of the active power for all wind turbines of each wind farm. Since these are strongly dominated by collective effects, we analyze the correlation structure for the active power without the collective part afterwards. Finally, we relate the correlation structure to the wind farm structure.
We perform the same analysis on both wind farms. In Sec.~\ref{Data analysis Riffgat} we present our results for the wind farm \textsc{Riffgat}. Afterwards, in Sec.~\ref{Data analysis Thanet} we present our results for the wind farm \textsc{Thanet}.

\subsection{\textsc{Riffgat}}
\label{Data analysis Riffgat}

To obtain a first impression of the correlation structure, correlation matrices for the active power for periods of half a day, one day and one week are given in Fig.~\ref{fig:Corr_ex_Riffgat}. We do not use shorter time windows at this point, as the data basis is too small to obtain reliable results for these due to the coarse temporal resolution of the data. For the correlation matrices shown, it is striking that, besides for Turbine $27$ in the matrix over half a day, only positive correlations occur. Along the main diagonals, blockwise associated structures of slightly increased correlations are formed. However, single wind turbines show significantly lower correlations to all other wind turbines, forming stripes of lower correlations. Overall, the correlation matrices are nevertheless dominated by a uniform positive correlation.
The mean correlation increases with the length of the observation period. This is plausible if one considers the common dependence of each power series on one external trend: the wind speed. For long time scales, the wind speed is not constant and the power of all turbines changes strongly in dependence of this. Thus, strong positive cross-correlations are created on these time scales due to trends in the wind speed. With shorter time scales the wind speed varies less strongly allowing for individual variations of the wind turbines to be observed in the correlations. Consequently, the overall correlation for the active power of the wind turbines decreases and shows more interesting structure.
For our further analysis of the \textsc{Riffgat} data we only consider correlation matrices over half a day.

The collective behavior observed for the correlation matrices can be illustrated by an analysis of their eigenvalues. In Fig.~\ref{fig:Eigen_Riffgat}, the eigenvalue spectrum and the corresponding eigenvectors of the correlation matrix are shown.
In the eigenvalue spectrum, a particularly large eigenvalue with a value of $24$ stands out. The associated eigenvector has, with some exceptions, approximately the same value in each component (for each wind turbine). This provides a uniform correlation and indicates a collective behavior for the active power of the wind turbines.

To remove the influence of the collective dynamics, we perform a singular value decomposition of the correlation matrices and take out the contribution of the first singular value/eigenvalue (see Sec.~\ref{Method}).
Correlation matrices for the active power reduced by the contribution of the first eigenvalue are shown in Fig.~\ref{fig:Corr_Red_Riffgat}.
Now, also negative correlations appear in the correlation matrices. In the first matrix, there are positive correlations in blocks along the main diagonal and otherwise mixed positive and negative correlations. The other two correlation matrices show structures that can be interpreted based on the spatial structure of the wind farm. Positive correlations are found between the first and last ten wind turbines (rows), respectively, in conjunction with negative correlations towards the remaining wind turbines.
For correlation matrices other than the ones shown here, positive correlations between the first three wind turbines of each row ($1$, $2$, $3$; $11$, $12$, $13$; $21$, $22$, $23$) together with strong negative correlations of these with the last wind turbines of each row ($10$; $20$; $30$) also occur.
These structures can be interpreted based on the wind farm structure. Depending on the wind direction, different wind turbines form the foremost line, which is the front of the wind farm in the wind. The fluctuations of the active power of these wind turbines are strongly correlated and anti-correlated to the fluctuations of the active power of the wind turbines of the other rows.
It should be noted, however, that the wind direction measurement is slightly different for all wind turbines, and can also change significantly over the observation periods of half a day. Thus, it is only approximately possible to assign a wind direction to each correlation matrix.

In order to investigate whether for these long time periods a dependence of the correlation structure on the wind direction can be found in the correlation matrices, we grouped them as a function of the mean wind direction of all wind turbines over the respective observation period. The mean wind direction is calculated as the circular mean of all wind directions.
The wind direction is divided into eight ranges of $45^\circ$ each. The ranges are divided so that they are centered around the wind directions that are orthogonal on the wind farm rows.
The mean correlation matrices for the $45^\circ$ ranges of the wind direction, calculated elementwise, are shown in Fig.~\ref{fig:Corr_Red_Dir_Riffgat}.
There are clear structural differences between the mean correlation matrices of the wind directions.
For the north direction, the northern wind turbines $1$-$10$ show a strong positive correlation. The correlation for the first and second half of the row is stronger among each other than across the row. The correlation with the other wind turbines is significantly lower, often even negative. Thereby, wind turbines $1$-$5$, $11$-$15$, $21$-$25$ and wind turbines $6$-$10$, $16$-$20$, $26$-$30$ are respectively more strongly (positively) correlated with each other. Wind turbines $11$-$20$ and wind turbines $21$-$30$ show a higher correlation with each other than with wind turbines $1$-$10$.
An analogous structure is even more pronounced for the south direction. Here, wind turbines $21$-$30$ show exclusively negative correlations with the other wind turbines.
For the east direction, wind turbines $10, 20, 30$ are strongly correlated with each other and show anti-correlations with the other wind turbines.
Similarly, wind turbines $1, 11, 21$ are correlated with each other for the west direction.
Overall it can be seen that the wind turbines that are directly downwind have high correlations with each other and have a weaker correlation or anti-correlation with the other wind turbines. Thus, information about the spatial structure of the wind farm can be derived from the correlation structure in combination with the wind direction. The observed correlation structures display the total interaction between turbines as an aggregation of complex dynamics such as the wake effect.
In all these considerations, it should be noted that the wind direction is still an average over half a day and all wind turbines. The emergence of such a clear pattern despite this fact is noteworthy.
It is interesting to investigate to what extent these effects can be found in data with a higher temporal resolution. This will be done in the analysis of the \textsc{Thanet} dataset in the following section.

\subsection{\textsc{Thanet}}
\label{Data analysis Thanet}

The high temporal resolution of the \textsc{Thanet} dataset makes it possible to study effects on a short time scale. Compared to the \textsc{Riffgat} dataset, the electrical power measurements of the \textsc{Thanet} dataset possess a $60$ times higher time resolution. Thus, much smaller observation periods can be chosen with a quantitatively equal data basis. This makes it possible to analyze individual aspects about which no statement could otherwise be made due to a lack of meaningfulness of the data.
For the calculation of correlations for the active power, the length of the observation period (the length of the time series of active power) plays a significant role. For short observation periods, the correlation is masked by noise due to individual statistical fluctuations \cite{Giada_2001, Guhr_2003}.

Therefore, we first discuss the influence of the length of the observation period using different time windows for an exemplary day. In Fig.~\ref{fig:Corr_ex_Thanet} the correlation matrices of the active power over periods of ten minutes, half an hour, one hour, six hours, half a day, and a day are shown.
In the correlation matrix over ten minutes, clear correlations and anticorrelations can be seen, forming a recognizable structure. The proportions of correlations and anticorrelations are balanced.
The correlation matrix over half an hour holds significantly more correlations than anticorrelations. Interrelated structures of correlations and anticorrelations become larger. They are also more pronounced.
The correlation matrix over one hour has essentially the same structure as the matrix over half an hour but weaker anticorrelations.
In the correlation matrix over six hours there are only positive correlations. Here, however, large differences for the different observation times emerge. For the example shown all wind turbines are strongly positively correlated with each other. There is only a slight block and stripe structure in the correlation matrix. For another day the correlation is less strong. A stripe structure clearly shows up for it. In addition, there are also some neutral correlations. Yet for another day the correlation matrix consists to the largest part still of neutral correlations (see appendix Fig.~\ref{fig:Corr_ex2_Thanet} and Fig.~\ref{fig:Corr_ex3_Thanet}).
The correlation matrix over half a day does not show much change for the example shown. However, for the other examples mentioned before the matrix is now also dominated by strong positive correlations (appendix Fig.~\ref{fig:Corr_ex2_Thanet} and Fig.~\ref{fig:Corr_ex3_Thanet}).
Finally, the correlation matrix over a day shows mainly strong positive correlations.
For even larger time periods, the correlation increases further and the correlation matrices become more homogeneous.
We observe that the correlation matrices all tend towards a uniform structure with increasing length of the observation periods. However, this happens for different observation periods on different time scales. These differences emerge from the range of electrical power production (and wind speed) values within each oberservation period. If the wind speed is similar during the whole observation period, the active power has a narrow range of values and individual fluctuations of the wind turbines dominate. Correlations are therefore much smaller. For large changes in the wind speed during an observation period, the active power has a much broader range of values. The values for all wind turbines also show the same timely variation. Hence, a collective behaviour emerges and dominates the individual fluctuations of the wind turbines. Correlations are therefore much stronger.

We again illustrate the collective behaviour observed for the correlation matrices by an analysis of their eigenvalues. In Fig.~\ref{fig:Eigen_Thanet}, the eigenvalue spectrum and the corresponding eigenvectors of the correlation matrix over half an hour are shown.
The correlation matrix over ten minutes was not chosen at this point because the associated time series include only $60$ measurements. This is less than the number of wind turbines considered ($100$). The rank of the correlation matrix is thus limited by the number of measurements. For this matrix, $41$ eigenvalues have a value of zero \cite{heckens2020uncovering}. For the correlation matrix over half an hour, the associated time series include $180$ measurements. Accordingly, the correlation matrix is a full-rank matrix.
The eigenvalue spectrum shows many eigenvalues with values smaller than one. Seventeen eigenvalues have values between one and three. However, there are three significantly larger eigenvalues with values of $5.6$, $10.1$ and $30.5$.
In the components of the first three corresponding eigenvectors, structures can be recognized that indicate a collective behavior of small groups of wind turbines. The first eigenvector has mostly positive entries of a similar magnitude, but also some neutral and even negative entries. In the second and third eigenvectors, several small coherent groups of positive and negative entries form. The entries of the remaining eigenvectors do not show a clear structure, they seem to have randomly distributed positive and negative values. The last three eigenvectors each have entries in only four components. However, these are negligible because the associated eigenvalues have a value of zero within machine precision.

Wind speed and wind direction change on short time scales. Long observation periods thus may include phases of different wind speeds and wind directions. This leads on the one hand to stronger correlations between wind turbines and on the other hand to the mixing of possible wind speed and wind direction dependencies in the correlation structure. In order to be able to resolve effects caused by wind speed or direction, the observation period should therefore be chosen as short as possible.
The correlation matrices over ten minutes are based on a very small data set and do not have a full rank. For that reason, the correlation matrices over $30$ minutes are used for the further analyses.

Despite the short observation period of half an hour, the structures of the correlation matrices are in some cases still strongly characterized by collective effects.
To remove the contribution of the collective behavior of all wind turbines, we now consider the time series $\Psi$ of deviations for the individual wind turbines' active power from the mean active power of all wind turbines. A similar approach was used in Ref.~\cite{borghesi2007emergence}.
At this point, the active power is not reduced by the contribution of the first singular value/eigenvalue as in Sec.~\ref{Data analysis Riffgat}, because the eigenvalue spectra of the correlation matrices show huge differences for the different observation periods. While for some correlation matrices the first eigenvalue is very large and describes the collective dynamics, for other correlation matrices it is smaller and does not describe the collective dynamics. This indicates that sometimes there are influences that are stronger than the collectivity. Taking out the contribution of the first eigenvalue would thus have strongly different effects for the different correlation matrices.

To investigate again a dependence on the wind direction, the correlation matrices are grouped as a function of the mean wind direction of all wind turbines over the respective observation period.
For this purpose, the wind direction is divided into eight ranges of $45^\circ$ each. The ranges are divided so that they are centered around the wind directions that are orthogonal on the wind farm rows.
The mean correlation matrices for the $45^\circ$ ranges of wind direction are shown in Fig.~\ref{fig:Corr_Red_Dir_Thanet}.
There are clear structural differences between the mean correlation matrices of the wind directions.
For the structure to be formed in a manner analogous to the observations for the \textsc{Riffgat} dataset (see Fig.~\ref{fig:Corr_Red_Dir_Riffgat}), the first wind turbines of each row should be correlated with each other for the north-west direction. This can be seen in the mean correlation matrix. However, the correlations are only very weak.
For the south-east direction, the last wind turbines of the rows should be correlated with each other. This can be found in the mean correlation matrix.
Finally, for the south-west and north-east directions, the first and last rows, respectively, should be correlated. For the south-west direction, an increased correlation of the first rows can be seen, but for the north-east direction, no increased correlation of the last rows can be found.
The stronger expression of the structure for the south-east to south-west direction probably stems from the fact that this is the main wind direction for the observation period. Both the most and the largest wind speed measurements are available for this direction.
A transition between the structures of the south-east and south-west directions can further be seen in the correlation matrix of the south direction.
The correlation matrix of the east direction shows strong correlations between the second half of the wind turbines. The respective wind turbines are in a state of failure for a large part of the observation period for that wind direction. The high correlations therefore emerge from our data filling (see Sec.~\ref{Data Pre}).

The spatial structures are less distinct for the \textsc{Thanet} dataset than for the \textsc{Riffgat} dataset.
On the one hand, this may be related to the fact that the distances of the wind turbines for the \textsc{Thanet} wind farm are larger in relation to the rotor diameter of the wind turbines (see Sec.~\ref{Des Farms}). As a result, the wake effects by the wind turbines are less strong and the wind turbines do not influence each other as much. In particular, the distances between rows are much larger in the wind farm \textsc{Thanet} and are in a range for which the influence of the wind turbines can be considered small \cite{mckay2013wake, annoni_2014_eval, crespo_1999_survey, builtjes_1978_cons}.
On the other hand, the difference between the wind farms may also be due to the different time resolutions of the data and the different lengths of the observation periods. By reducing the temporal resolution of the \textsc{Thanet} data to ten minute averages and calculating the correlation matrices over a time window of half a day we are able to reveal spatial structures more clearly (see appendix Fig.~\ref{fig:Corr_Red_Dir_Thanet_12h}). However, reducing the temporal resolution has again the disadvantage that wind direction and other external circumstances can change significantly within the observation periods, making it impossible for us to draw clear conclusions about their influence.

\section{Conclusion}
\label{Conclusion}

We analyzed the operational data of the offshore wind farm \textsc{Riffgat} for the period from 01.03.2014 to 28.02.2015 and the operational data of the offshore wind farm \textsc{Thanet} for February 2017.
In both cases, a significant dependence on the spatial structure of the wind farm could be found for the correlation structure of the fluctuations of the active power for the wind turbines. Wind turbines that have spatial proximity to each other show stronger correlations for the electrical power fluctuations with each other.
For the \textsc{Riffgat} dataset, a connection between the structures of the correlation matrices and the prevailing wind direction was found. This is surprising since the time periods considered for the correlation matrices were half a day. Over such a long period of time, the wind direction is not constant and can only be approximated to a specific direction.
For the high resolution \textsc{Thanet} dataset, this problem was circumvented by considering shorter time periods. It was expected that the structures would be even more pronounced as a function of wind direction for this case. This could not be confirmed in the analysis. Increased correlations due to the spatial proximity of the wind turbines were again found, but a clear correlation with wind direction could only be found for a certain range of wind directions. Using the same time resolution and time periods for the \textsc{Thanet} dataset as previously used for the \textsc{Riffgat} dataset resulted in an enhancement of the structures as a function of wind direction. This is due to the fact that a high temporal resolution leads to stronger noise. Therefore a trade-off between resolution and noise has to be made.

One possible cause for the differences between wind farms is the wind turbine spacing within the wind farms. Especially in relation to the rotor diameters of the wind turbines, the distances between the wind turbines in the wind farm \textsc{Thanet} are significantly larger than in the wind farm \textsc{Riffgat}.
For the \textsc{Riffgat} wind farm, the wind turbines within rows have a spacing of $550\,$m ($4.6$ rotor diameter). The distance between rows is $600\,$m ($5$ rotor diameter). For the \textsc{Thanet} wind farm, the wind turbines within the rows have a spacing of $470\,$m ($5.2$ rotor diameter). The spacing between rows is $720\,$m ($8$ rotor diameter).
Thus, the interactions and wake effects of the wind turbines, especially between rows, are much weaker for the \textsc{Thanet} wind farm than for the \textsc{Riffgat} wind farm. Hence, structures depending on the wind direction are also significantly weaker pronounced. While the influence of wake effects and their complex dynamics are the subject of many studies \cite{PorteAgel2020}, we have shown that they are intimately connected to the collective behavior. Their influence is measurable without any further input in the cross-correlations of wind farms once the collective behavior is accounted for.

The methods of this paper are suitable to analyze a large number of existing wind farm layouts with reasonable effort and input-free beyond the measured data. The presented statistical approach complements existing methods to evaluate which layouts minimize turbine interactions. Of course, it cannot replace complex analysis such as fluid dynamics simulations for detailed analysis.

Within the paper, attempts have already been made to quantify differences in the correlation structure, caused by the length of the observation period for the individual correlation matrices. However, for a better understanding of the emergence of collective dynamics, further investigations are necessary. These should consider not only the lengths of the observation periods, but also the changes in external parameters such as wind speed and wind direction within them.

\section*{Acknowledgements}
We are most grateful to Vattenfall AB for providing the data. We acknowledge fruitful conversations with Joachim Peinke and Matthias Wächter. This study was carried out in the project Wind farm virtual Site Assistant for O\&M decision support – advanced methods for big data analysis (WiSAbigdata) funded by the Federal Ministry of Economics Affairs and Energy, Germany (BMWi). One of us (H.M.B.) thanks for financial support in this project.

\bibliography{bibliography_revised_HB}

\onecolumngrid
\appendix
\clearpage
\section{Further Correlation Matrices \textsc{Thanet}}
\begin{figure*}[h]
	\begin{center}
		\includegraphics[width=0.32\textwidth]{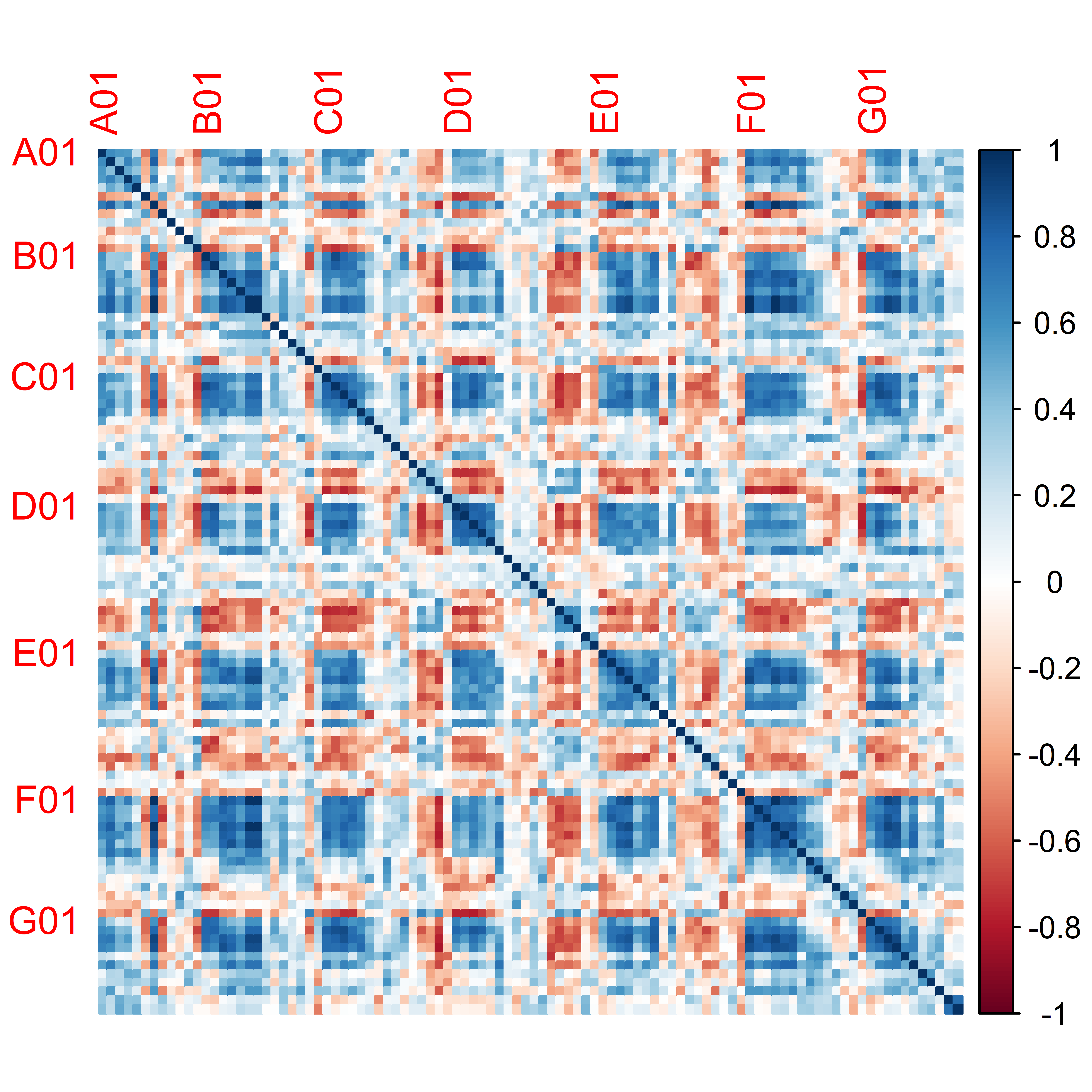}
		\includegraphics[width=0.32\textwidth]{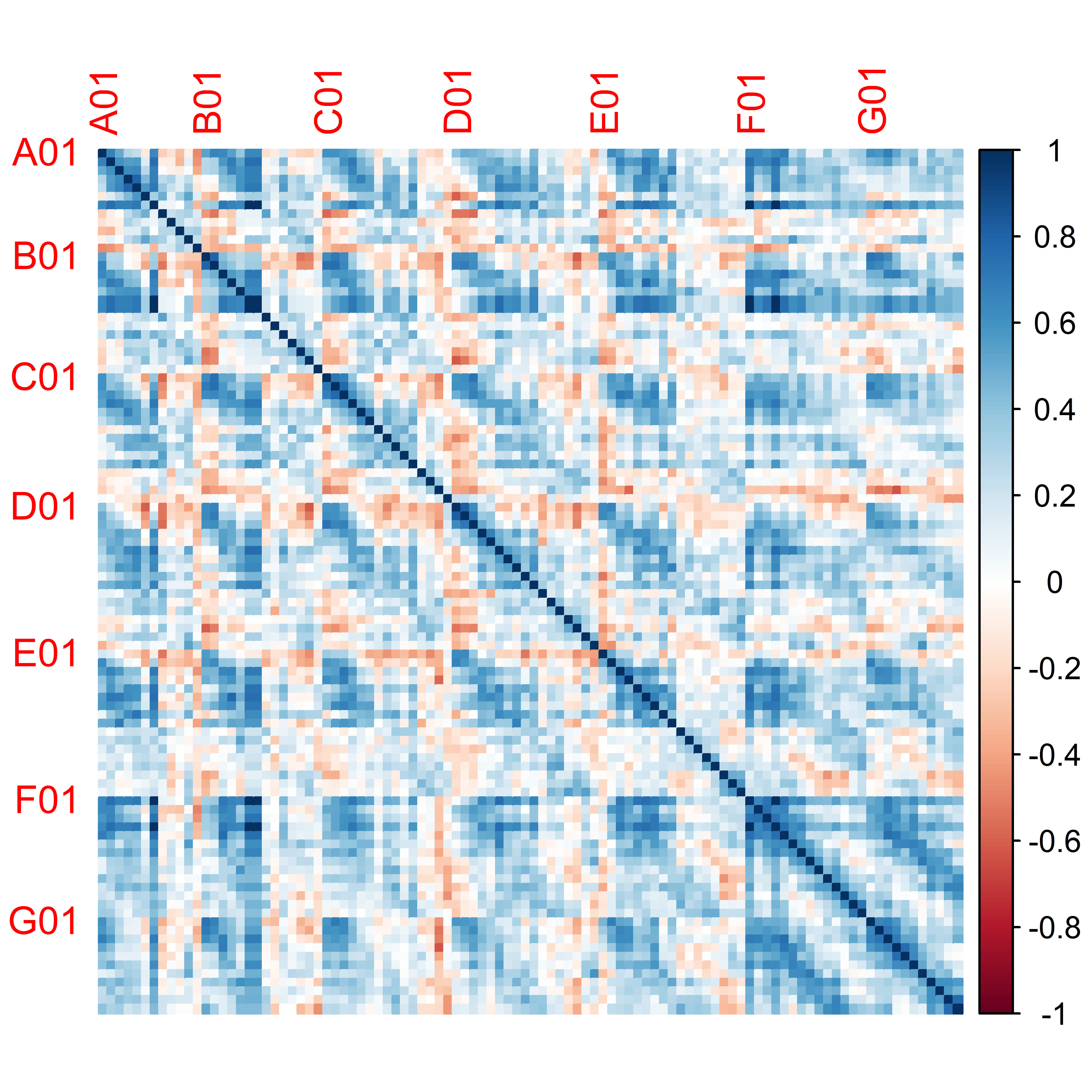}
		\includegraphics[width=0.32\textwidth]{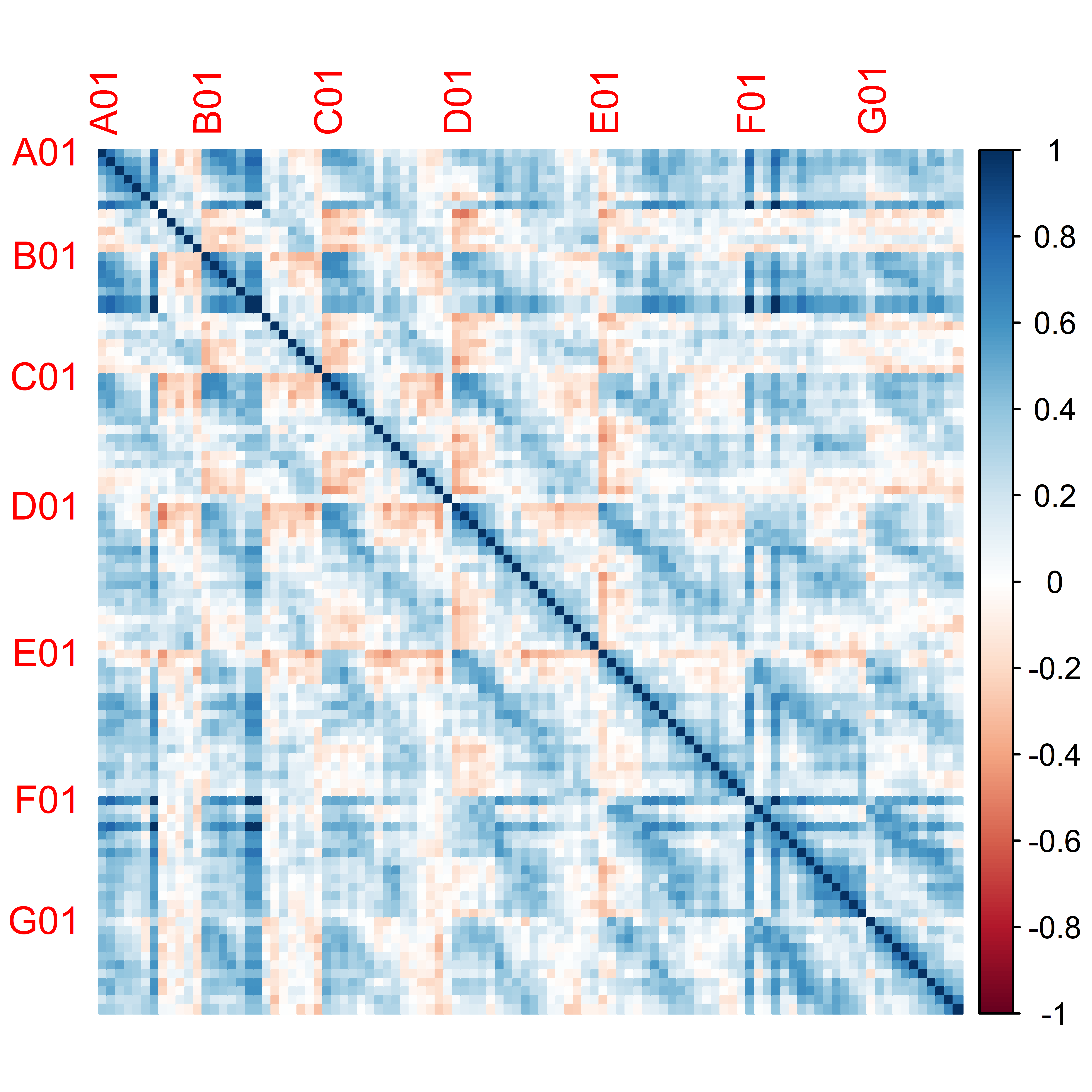}
		\includegraphics[width=0.32\textwidth]{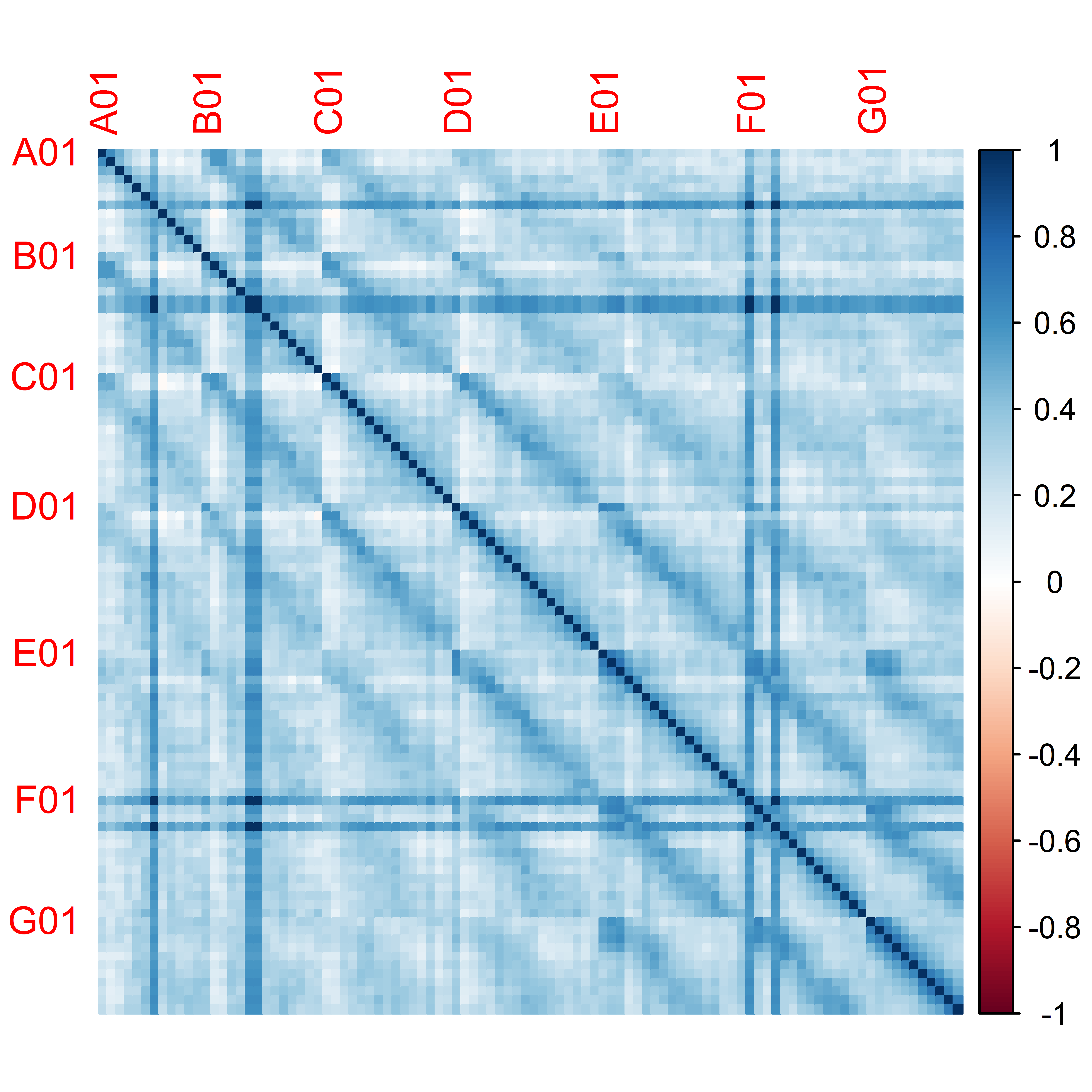}
		\includegraphics[width=0.32\textwidth]{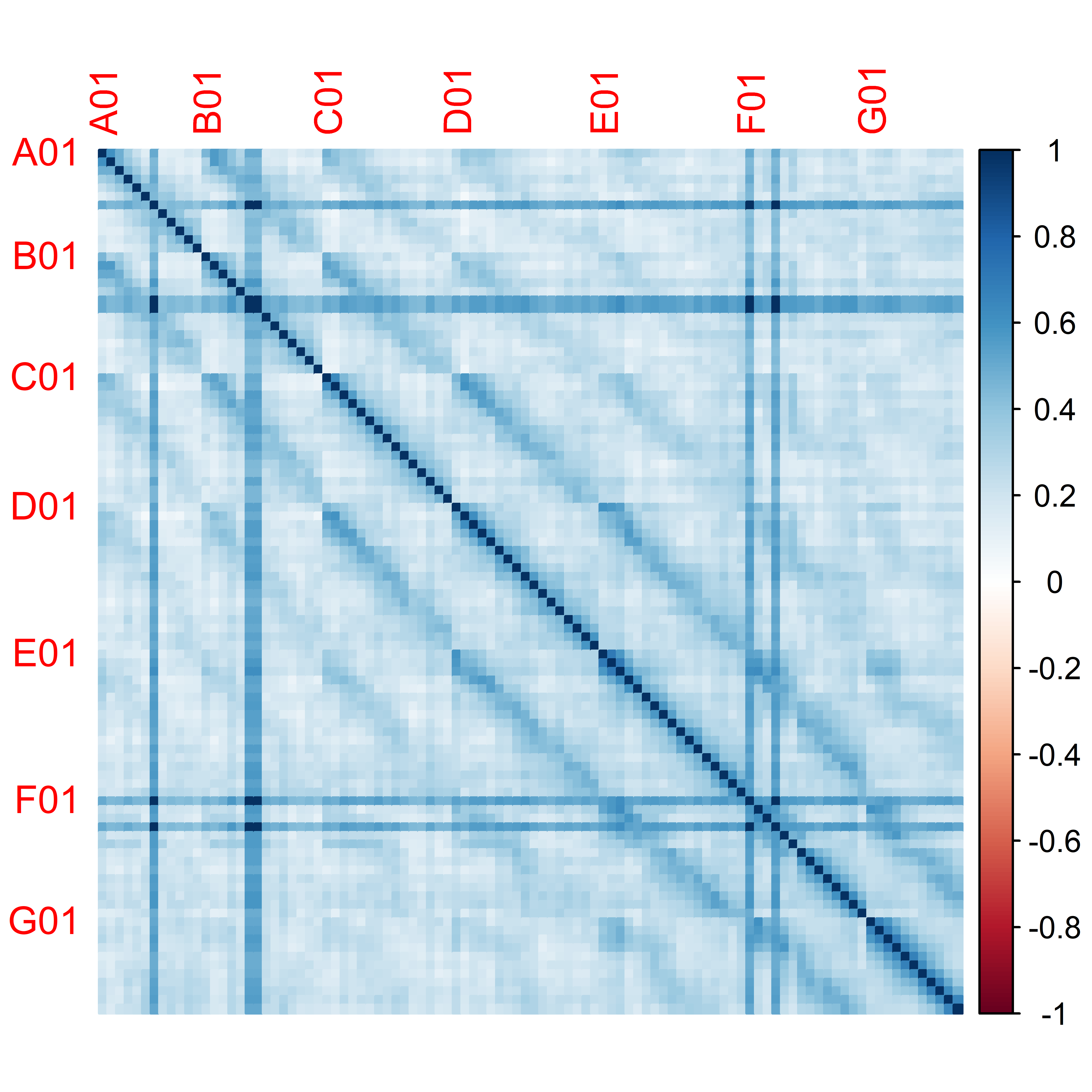}
		\includegraphics[width=0.32\textwidth]{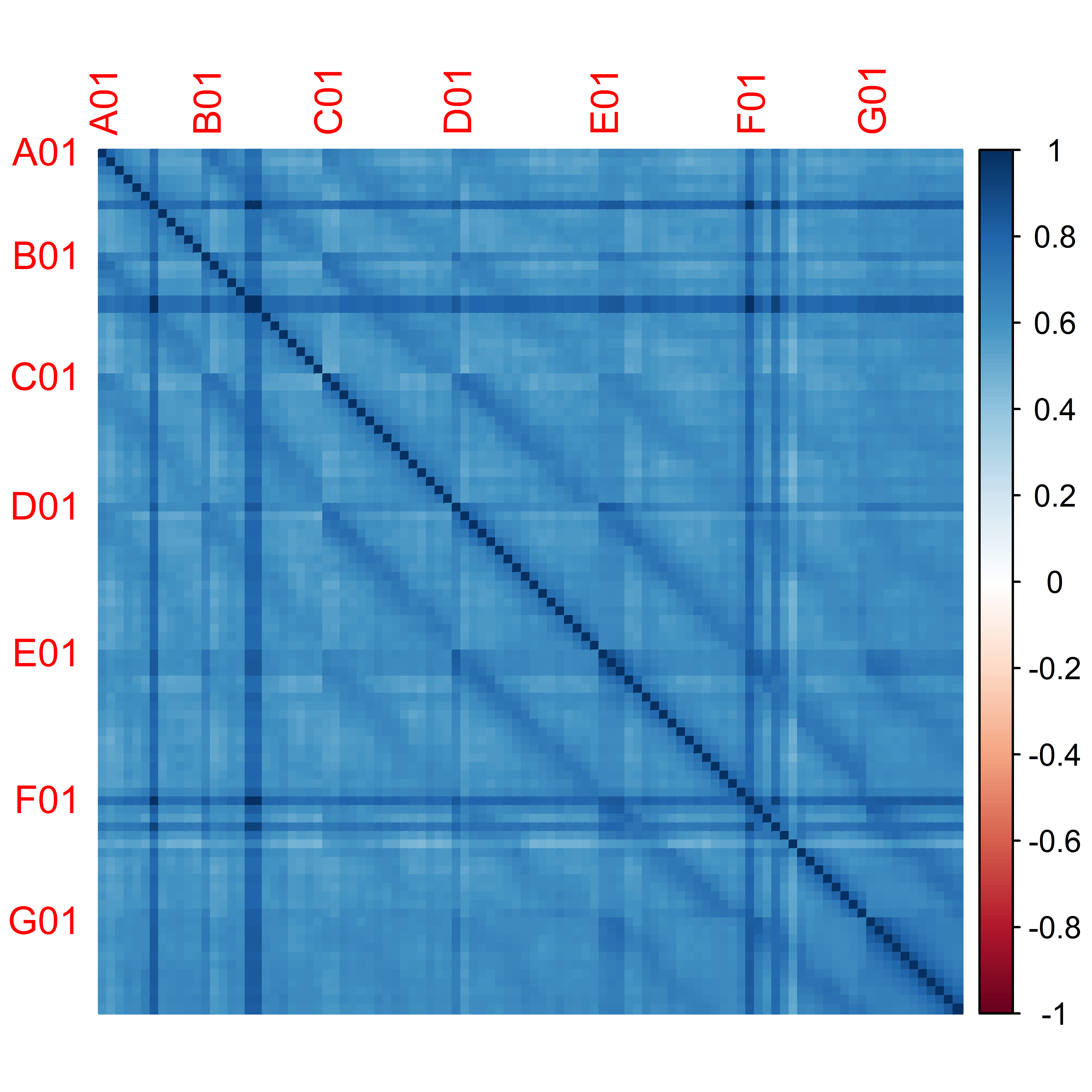}
		\caption{Correlation Matrices for the active power over ten minutes (top left), half an hour (top middle), an hour (top right) six hours (bottom left), half a day (bottom middle) and a day (bottom right) for the wind farm \textsc{Thanet}, February 8.}
		\label{fig:Corr_ex2_Thanet}
	\end{center}
\end{figure*}
\begin{figure*}[h]
	\begin{center}		
		\includegraphics[width=0.32\textwidth]{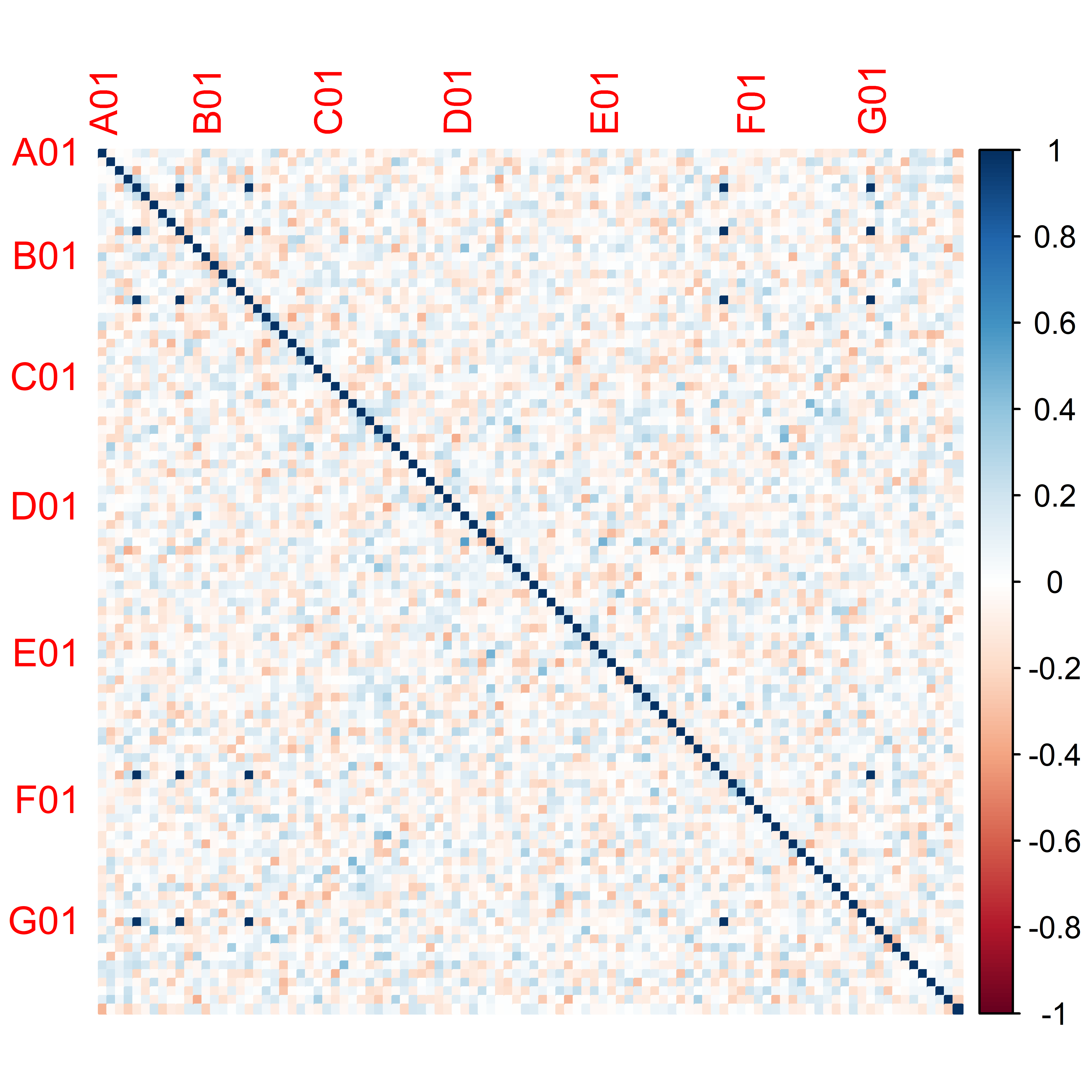}
		\includegraphics[width=0.32\textwidth]{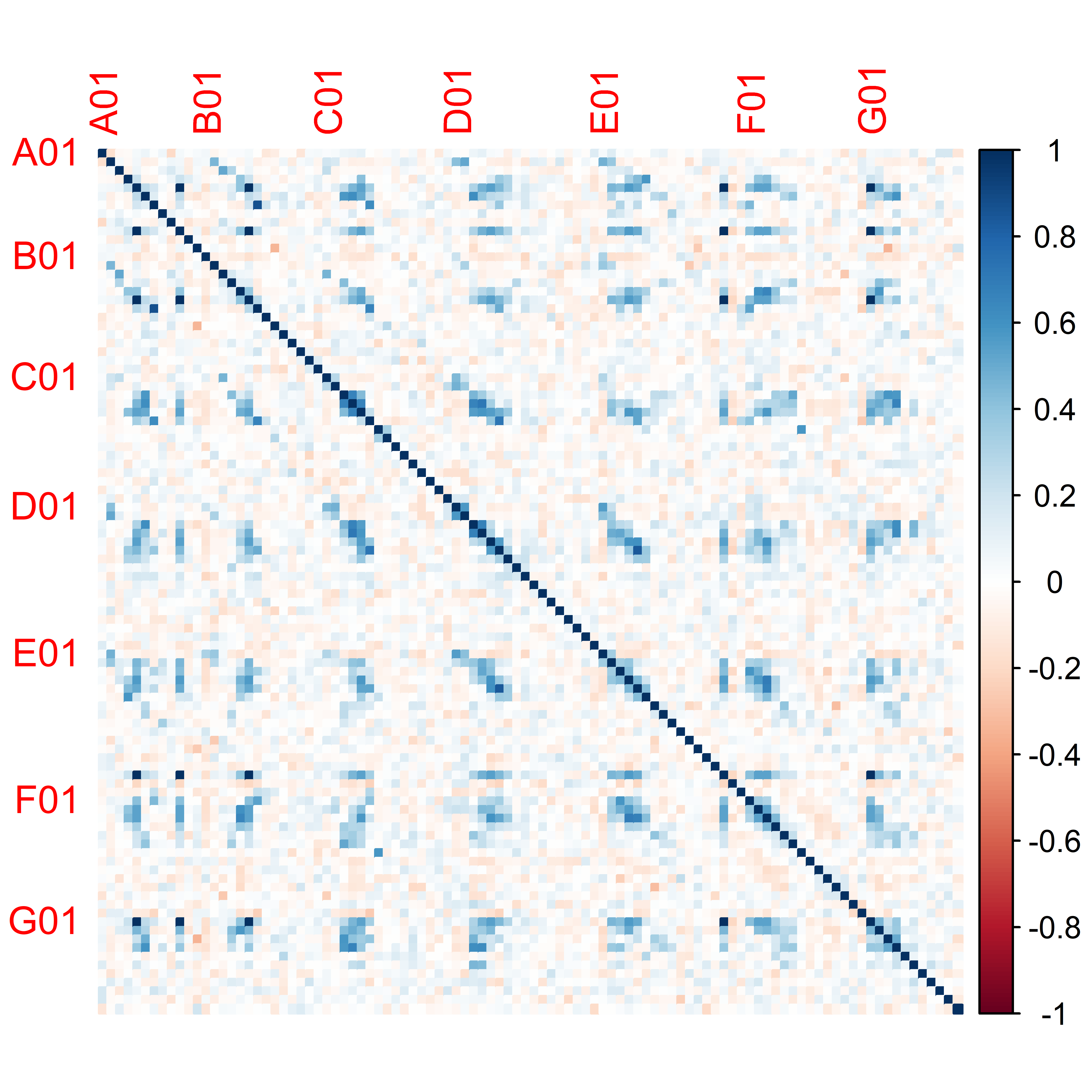}
		\includegraphics[width=0.32\textwidth]{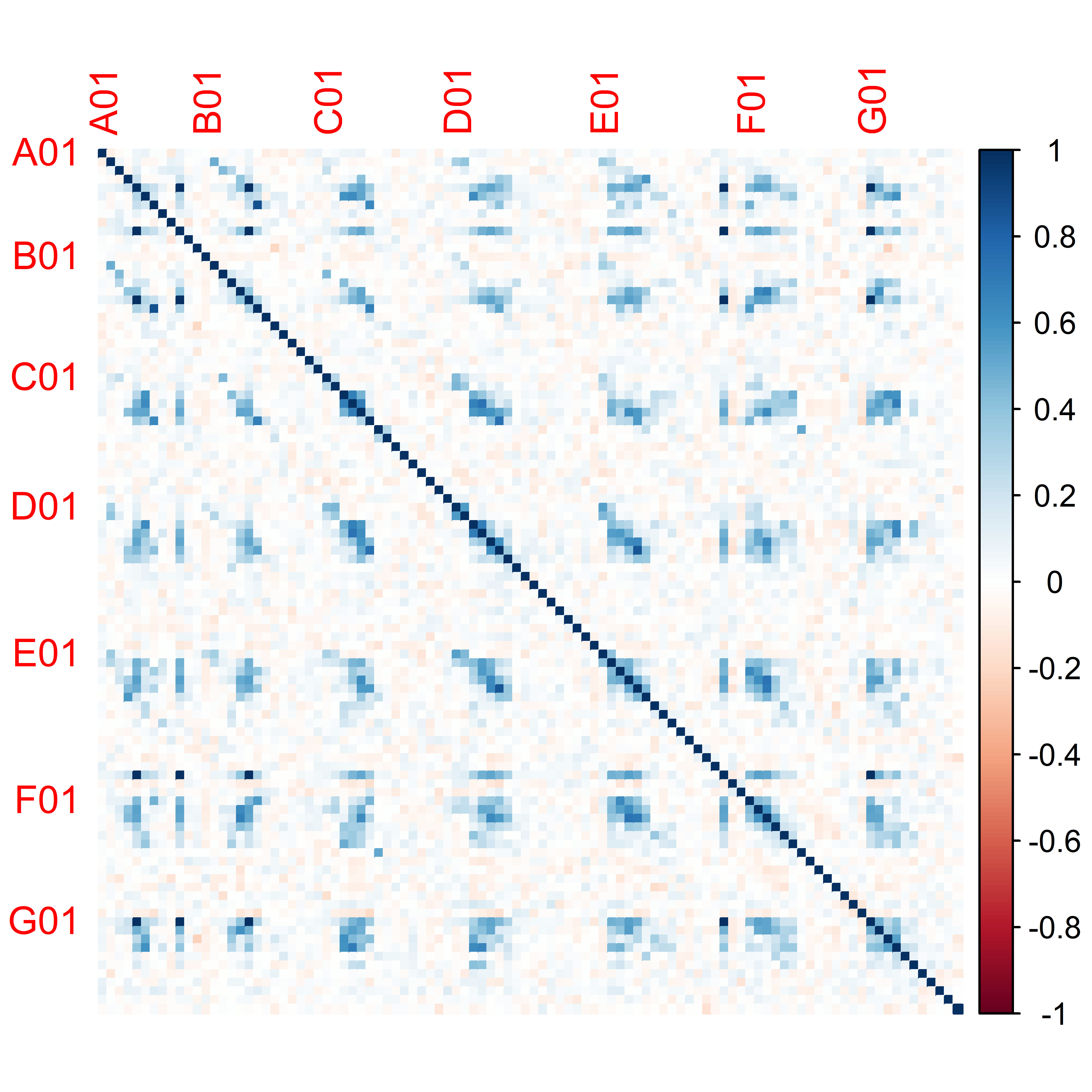}
		\includegraphics[width=0.32\textwidth]{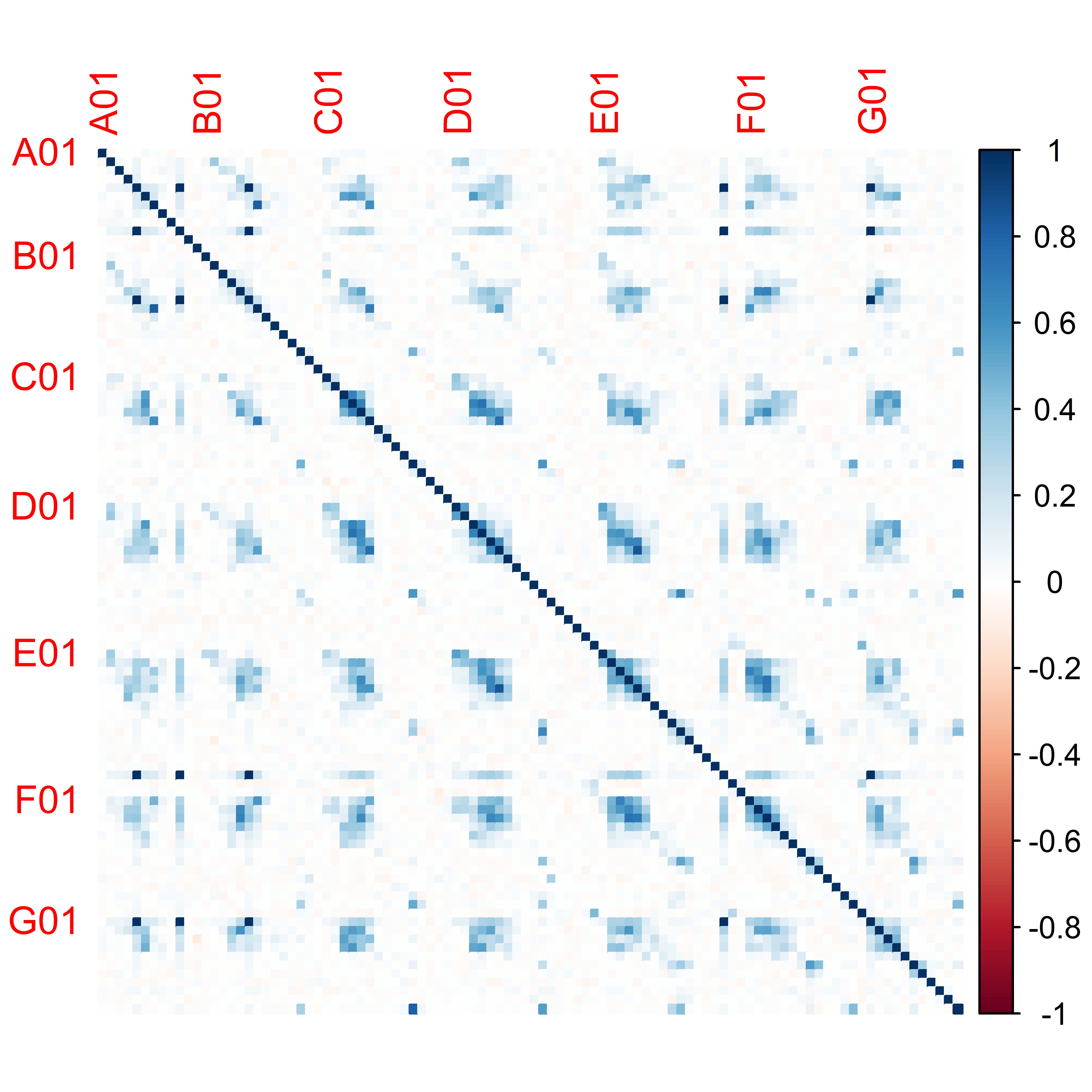}
		\includegraphics[width=0.32\textwidth]{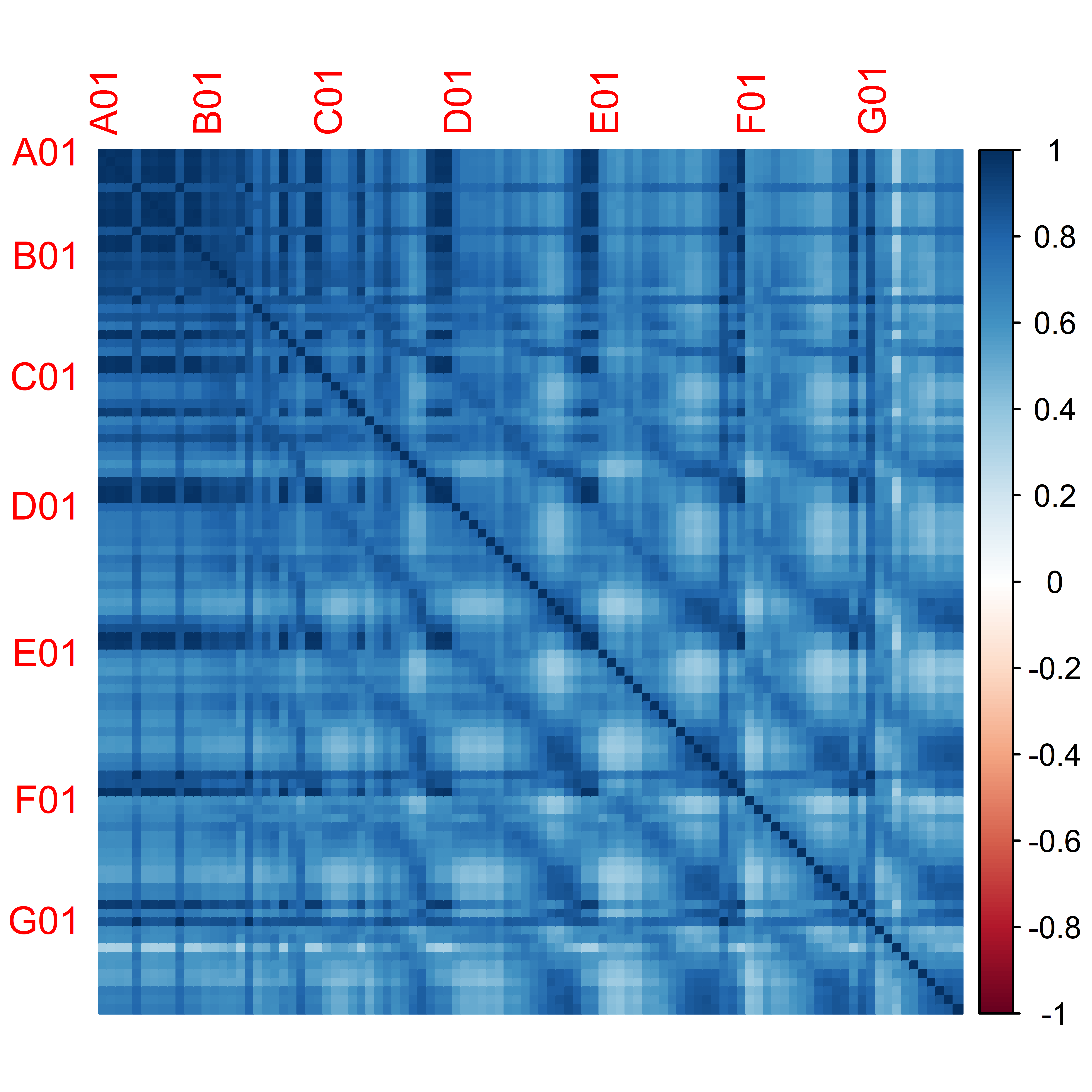}
		\includegraphics[width=0.32\textwidth]{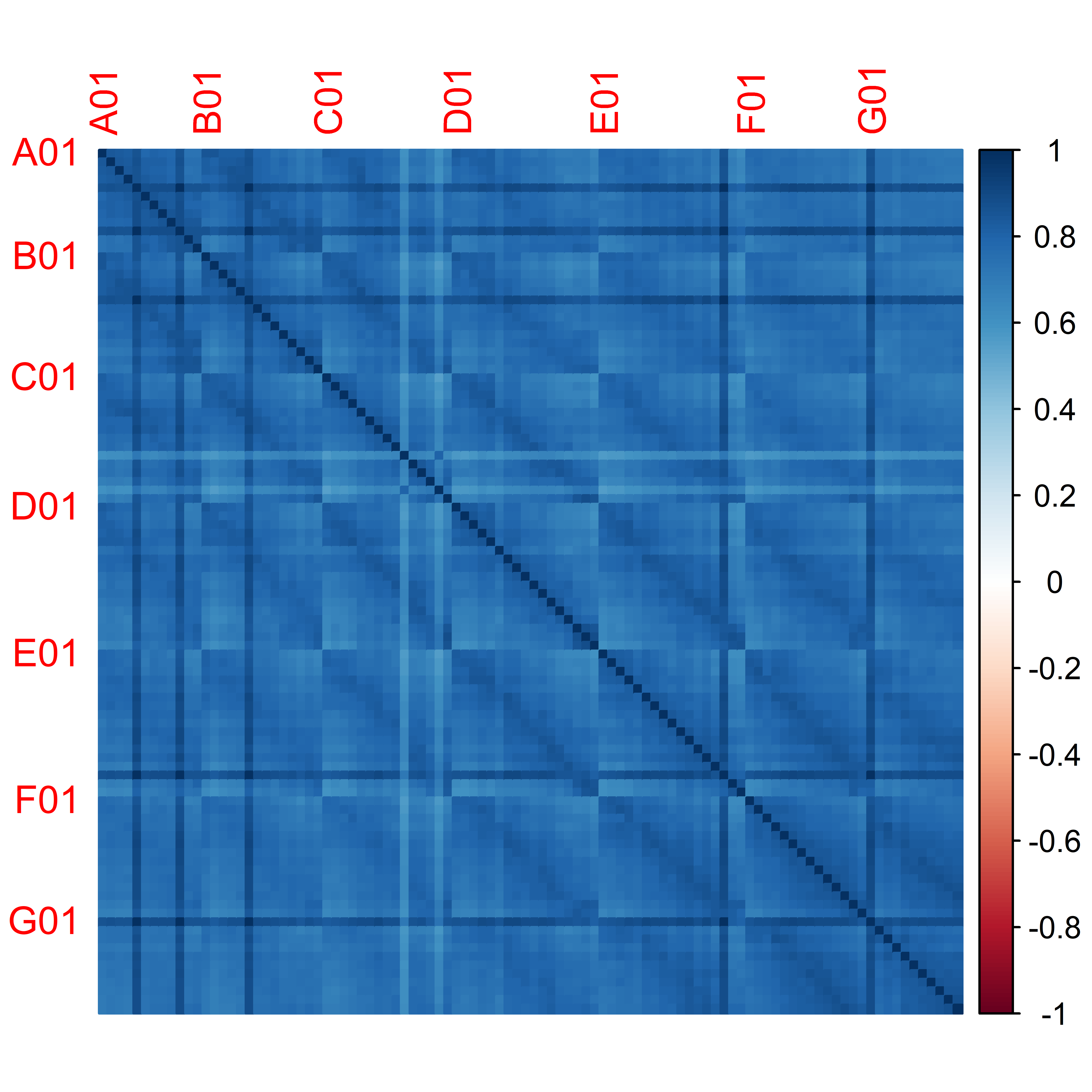}
		\caption{Correlation Matrices for the active power over ten minutes (top left), half an hour (top middle), an hour (top right) six hours (bottom left), half a day (bottom middle) and a day (bottom right) for the wind farm \textsc{Thanet}, February 27.}
		\label{fig:Corr_ex3_Thanet}
	\end{center}
\end{figure*}

\FloatBarrier
\clearpage
\section{Correlation Matrices \textsc{Thanet} for half a day}
\begin{figure*}[h]
	\centering
	\includegraphics[width=0.24\textwidth]{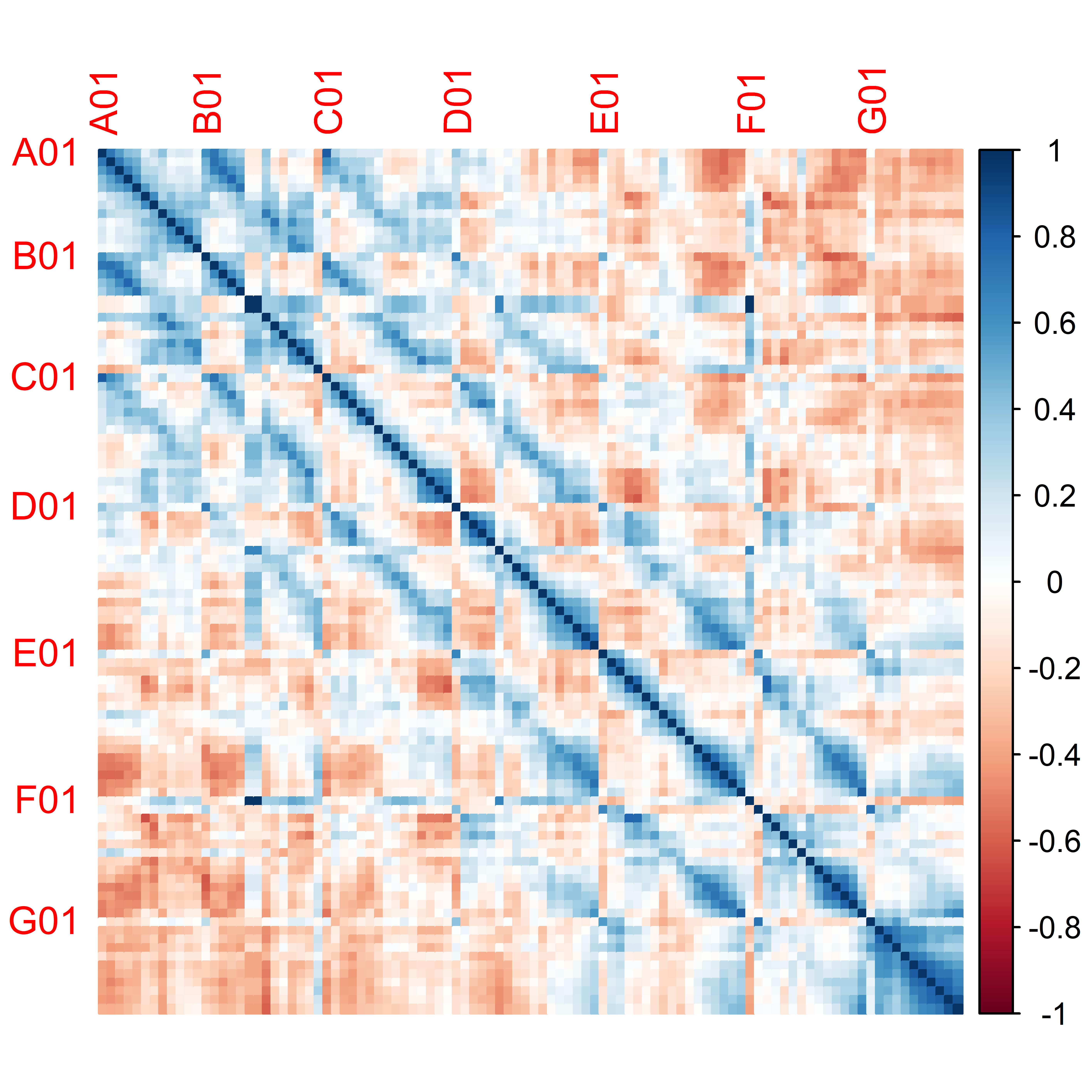}
	\includegraphics[width=0.24\textwidth]{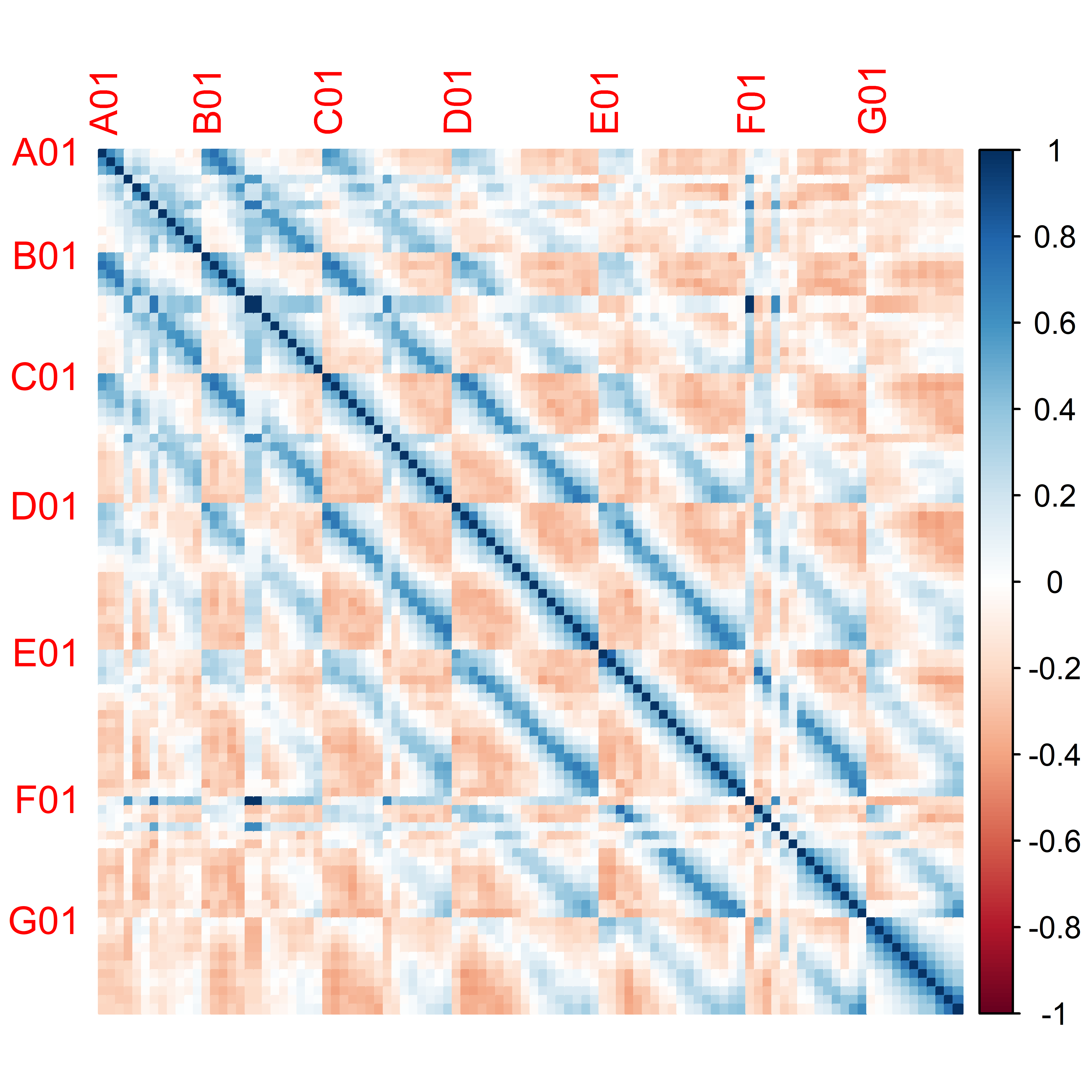}
	\includegraphics[width=0.24\textwidth]{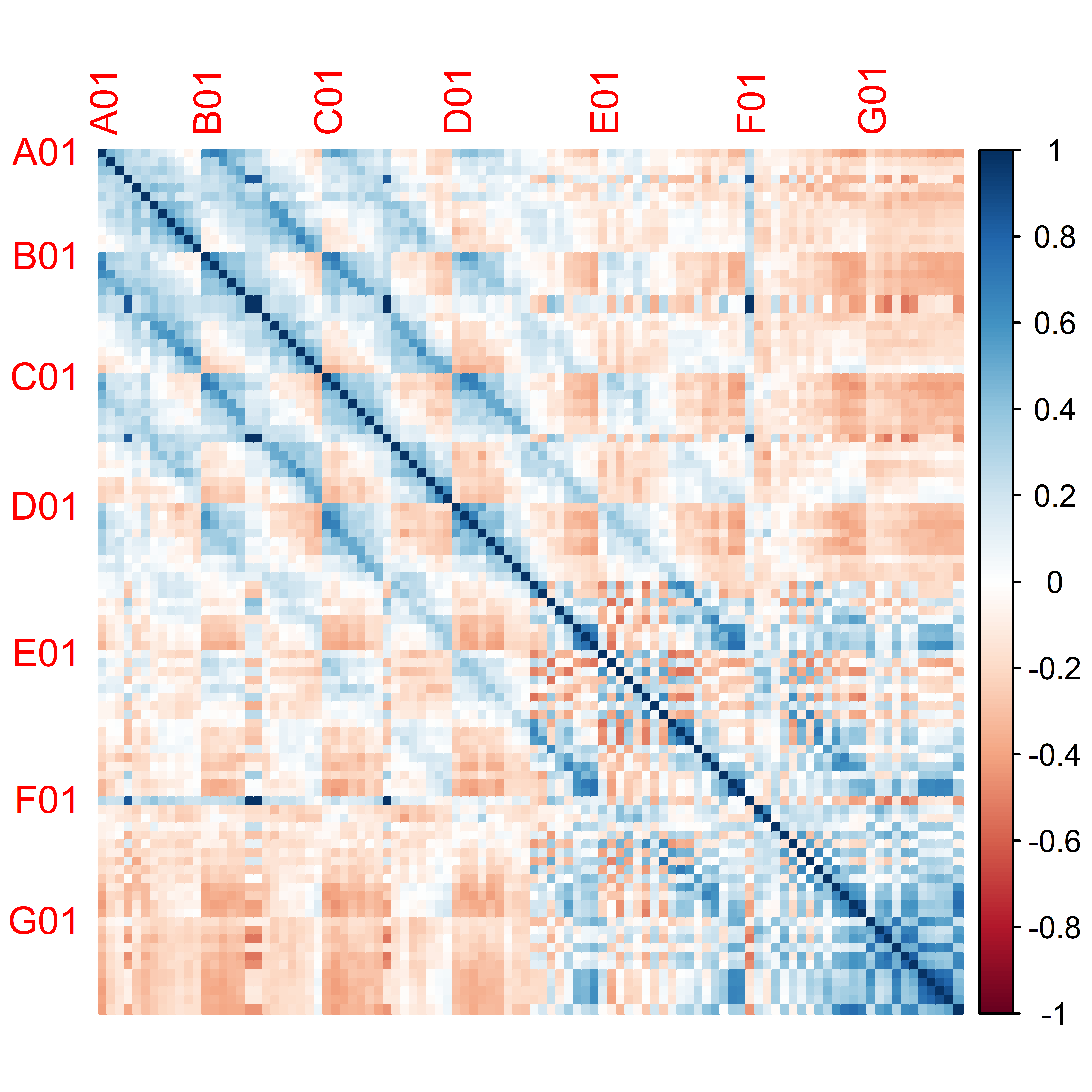}
	\includegraphics[width=0.24\textwidth]{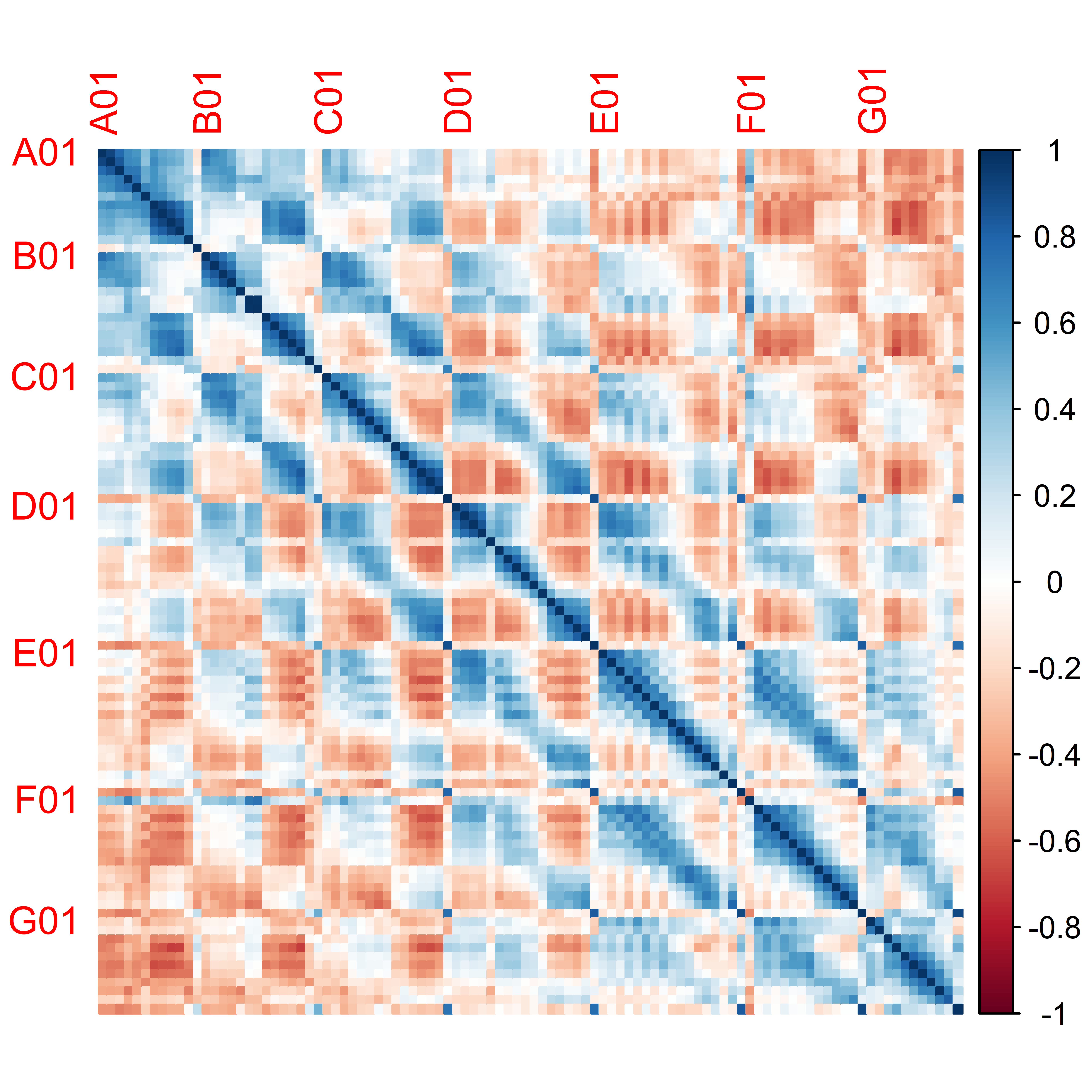}
	\includegraphics[width=0.24\textwidth]{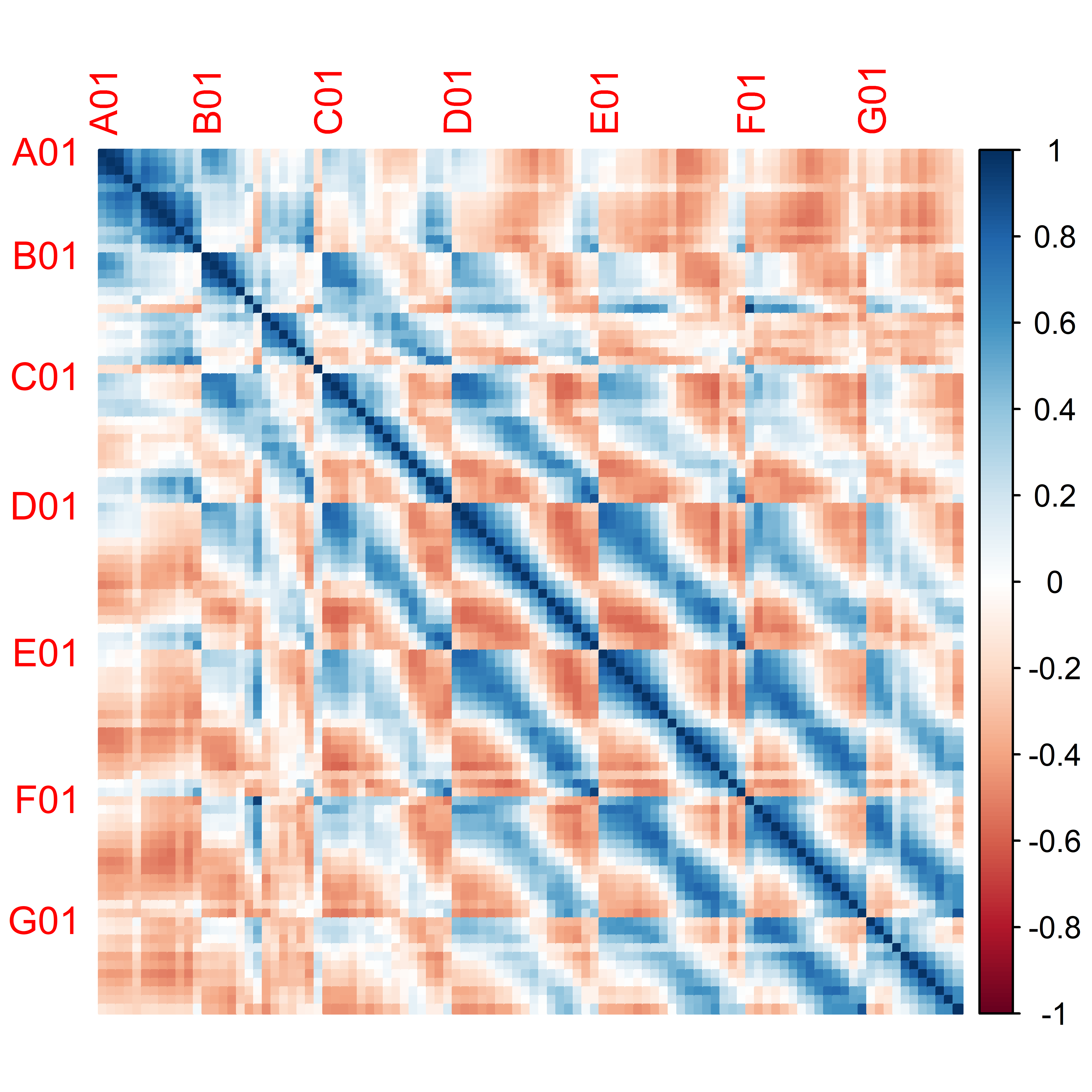}
	\includegraphics[width=0.24\textwidth]{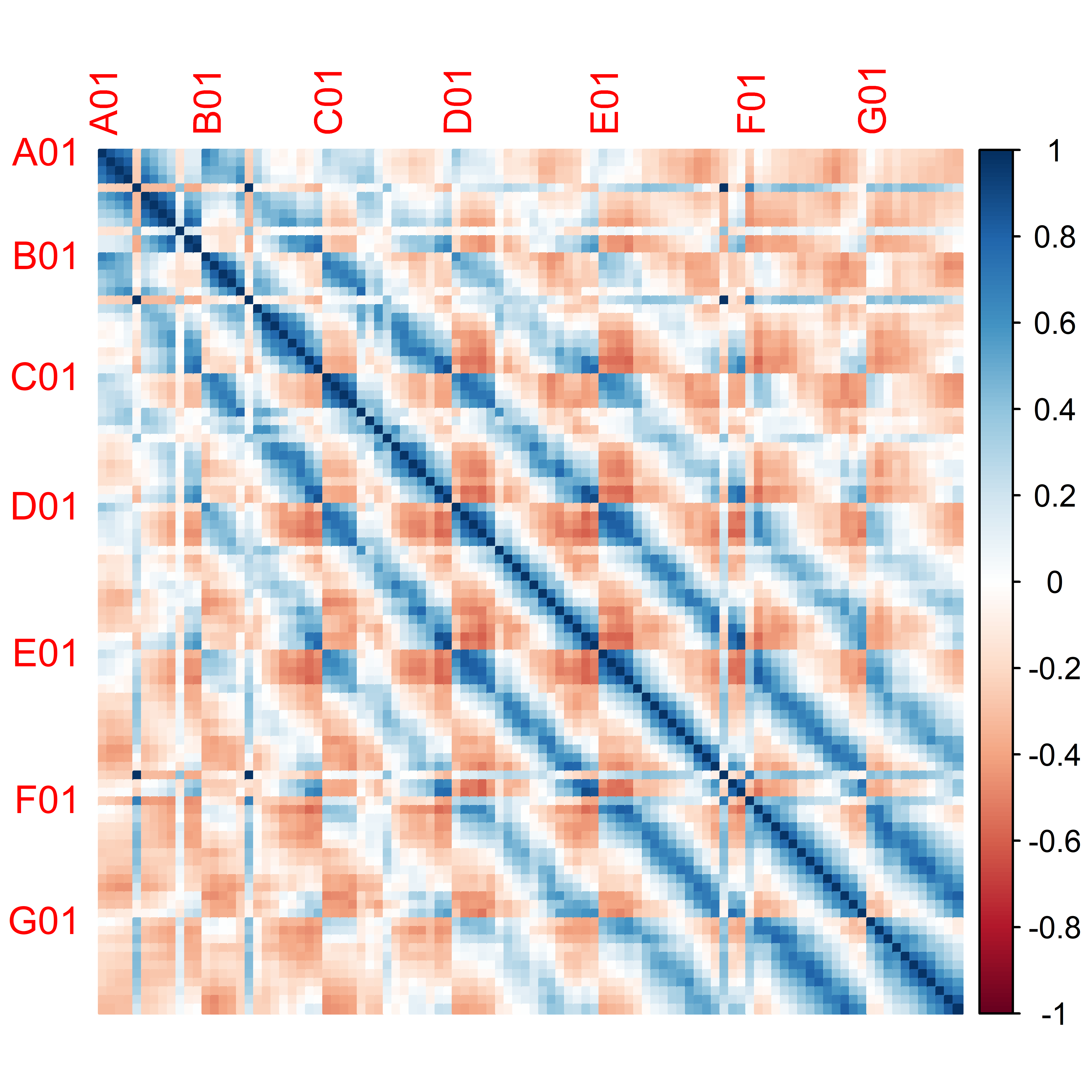}
	\includegraphics[width=0.24\textwidth]{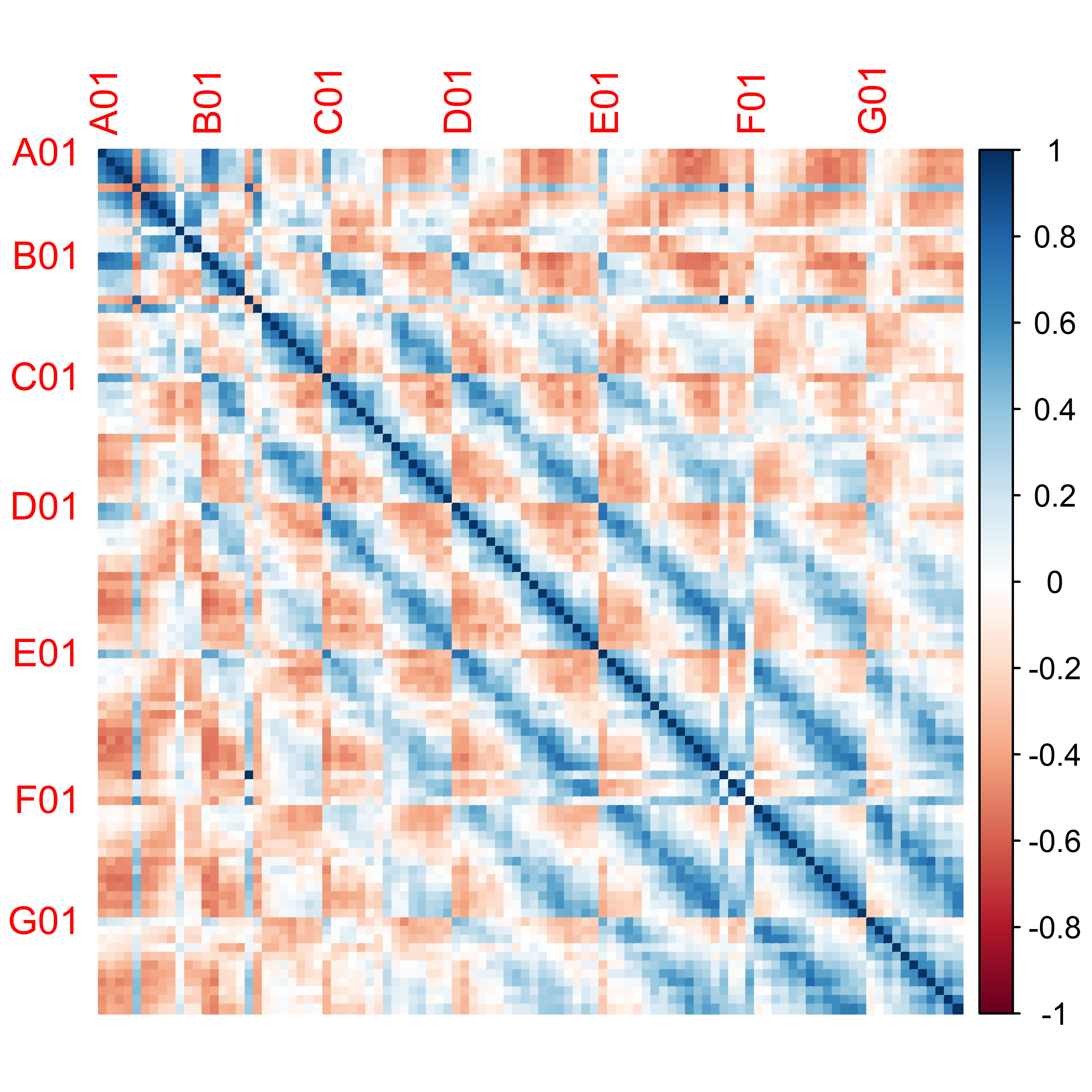}
	\includegraphics[width=0.24\textwidth]{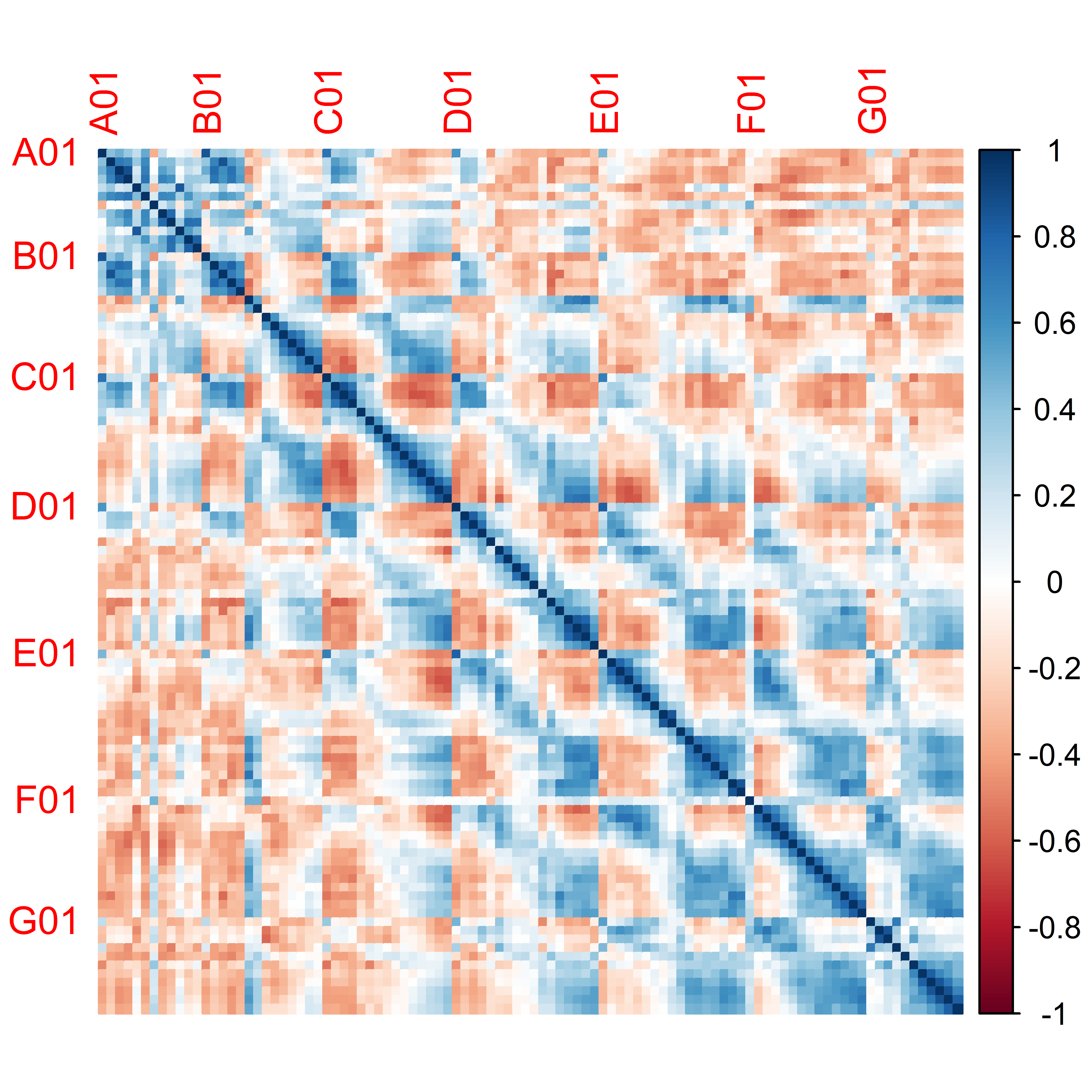}
	\caption{Average correlation matrices for the active power without the contribution of the first eigenvalue for the different $45^\circ$ ranges for the wind farm \textsc{Thanet} over half a day. The wind directions are from left to right N, NE, E, SE and S, SW, W, NW for the two rows.}
	\label{fig:Corr_Red_Dir_Thanet_12h}
\end{figure*}

\end{document}